\documentclass[10pt]{article}
\usepackage[utf8]{inputenc}
\usepackage{multicol}
\usepackage{amsmath}
\usepackage{amssymb}
\usepackage{graphicx}
\usepackage[dvipsnames]{xcolor}
\usepackage[normalem]{ulem}
\usepackage{hyperref}
\usepackage{caption}
\usepackage{subcaption}
\hypersetup{colorlinks, urlcolor=blue, allcolors=blue}
\newcommand{\norm}[1]{\left\lVert#1\right\rVert}

\usepackage[affil-it]{authblk}
\usepackage{algorithm}
\usepackage[noend]{algpseudocode}
\usepackage[semicolon]{natbib}
\setcitestyle{square}
\usepackage{geometry}
\usepackage{graphicx}
\usepackage[outdir=Images/]{epstopdf}
\graphicspath{ {Images/} }
\usepackage{bm}
\usepackage{dirtytalk}
\usepackage[title]{appendix}
\usepackage{listings}
\usepackage{csquotes}

\providecommand{\keyword}[1]{{\textit{Keywords:}} #1}

\title{A graph-based probabilistic geometric deep learning framework with online enforcement of physical constraints to predict the criticality of defects in porous materials}
\author[1,*]{Vasilis Krokos}
\author[2]{Stéphane P. A. Bordas}
\author[3,*]{Pierre Kerfriden}

\affil[1]{Cardiff School of Engineering, Cardiff University, UK}
\affil[2]{Institute for Computational Engineering, Faculty of Science, Technology and Communication, University of Luxembourg, Luxembourg}
\affil[3]{Centre des Matériaux, Mines Paris - PSL, Evry, France}
\affil[*]{Corresponding authors: Vasilis Krokos, KrokosV@cardiff.ac.uk; Pierre Kerfriden, pierre.kerfriden@minesparis.psl.eu}

\date{29 October 2023}

\begin{document}

\maketitle

\begin{abstract}%
    Stress prediction in porous materials and structures is challenging due to the high computational cost associated with direct numerical simulations. Convolutional Neural Network (CNN) based architectures have recently been proposed as surrogates to approximate and extrapolate the solution of such multiscale simulations. These methodologies are usually limited to 2D problems due to the high computational cost of 3D voxel based CNNs. We propose a novel geometric learning approach based on a Graph Neural Network (GNN) that efficiently deals with three-dimensional problems by performing convolutions over 2D surfaces only. Following our previous developments using pixel-based CNN, we train the GNN to automatically add local fine-scale stress corrections to an inexpensively computed coarse stress prediction in the porous structure of interest. Our method is Bayesian and generates densities of stress fields, from which credible intervals may be extracted. As a second scientific contribution, we propose to improve the extrapolation ability of our network by deploying a strategy of online physics-based corrections.
    Specifically, we condition the posterior predictions of our probabilistic predictions to satisfy partial equilibrium at the microscale, at the inference stage. This is done using an Ensemble Kalman algorithm, to ensure tractability of the Bayesian conditioning operation. We show that this innovative methodology allows us to alleviate the effect of undesirable biases observed in the outputs of the uncorrected GNN, and improves the accuracy of the predictions in general.
\end{abstract}

\keyword{Multiscale stress analysis, Graph neural network, Surrogate modelling, Bayesian machine learning, Physics-corrected neural network, Porous media}

%#######################################################################################################################################

\section{Introduction}

    Multiscale simulations are common in areas of science and engineering such as geo-engineering (heterogeneous soil layers and subterranean cavities), biomechanics (bones and soft tissues) and mechanical engineering (composite and architectured materials). Finite Element Analysis (FEA) of multiscale structures is notoriously challenging since the finite element mesh required to fully resolve features at the fine scale has to be sufficiently dense, which  often results in intractable computational analyses. 
    
    Common approaches to tackle multiscale problems usually fall either in the homogenisation or concurrent scale coupling approaches. In both cases the multiscale problem is split into a macroscale and a microscale problem. The macroscale problem diffuses the overall stress field over the entire structure, whilst local microscale corrections are required to fully account for the effect of the microstructure. In homogenisation it is assumed that the gradients of the macroscale displacement are homogeneous over the entire material sample (\textit{i.e.} the representative volume element, RVE) \citep{sanchezpalencia74, zohdiwrigger}. If that assumption does not hold, for instance due to the existence of fast macroscale gradients induced by sharp macroscopic geometrical features, then concurrent scale coupling methods are employed. In concurrent scale coupling, the results of homogenisation are applied to the boundary of regions of interest where concurrent microscale corrections are to be performed \citep{Ghosh_2004, Oden_2006, Kerfriden_Allix_2009, Multiscale_Zhu, Paladin_Kerfriden_2016, krokos2021bayesian} which may be performed in a purely downward approach \citep{ODEN19993}, which is often called submodelling, or with 2-way coupling \citep{gendre:hal-00437023}. Concurrent scale coupling is usually more demanding in terms of computing power and implementation than homogenisation approaches.

    In this work we propose a Neural-Network-based (NN) surrogate modelling approach to replace 3D linear elastic simulations in multiscale structures. The method is largely inspired by submodelling techniques. Specifically, we focus on stress predictions in linear elastic porous media with random spatial distribution of pores and with no separation of scales (\textit{i.e.} the interaction between the structure's geometry and the microscopic pores cannot be faithfully accounted for using a homogenisation strategy). The distribution of pores may or may not be known to the analyst. In the former case, which may arise when CT-scans of the porous structure are available, one single finite element analysis is needed, for a known spatial distribution of micro constituents and/or voids. We may need to perform that prediction fast, with limited computer resources that are not compatible with direct FE simulations. Offline however, direct FE simulations may be ran many times using super computers. In the latter case, when only statistics of the pore network is known to the analyst, a comprehensive Monte-Carlo analysis may be required (scales cannot be separated by assumption) to investigate the various possible interactions between microscopic pores and structural geometrical features. In that case, direct FE simulations may be ran tens to hundreds of times. Further predictions would gain from being accelerated using the data contained in the set of full microscale FE simulations. Following our previous work on multiscale stress CNNs presented in \citep{krokos2021bayesian}, we will use a convolution neural network as surrogate model for the direct microscale stress analysis of porous structures. Our approach boasts a multi-fidelity element, as we assume that macroscale simulations, whereby the network of pores is ignored, may always be performed inexpensively, and that the results of such simplified analysis may be used as an input to the NN. Subsequently, the fine scale geometry of the material and that of the structure being properly encoded, the NN will correct the macroscopic predictions over regions of interests in such a way that the corrected microscale stress fields emulate the output of the direct FE simulation.
    
    Neural Networks have been successfully employed in reproducing the output of simulation codes in a variety of fields such as fluid dynamics \citep{Lye2020DeepLO, Karniadakis_HiddenFluidDynamics}, solid mechanics \citep{ResCNN, Nie2020, deshpande2021fembased, krokos2021bayesian} and computational biology \citep{DeepMind_Protein_2019, DeepMind_Protein_2020}. Most of the publications that focus on learning the response of FE models employ Convolutional Neural Networks (CNNs), that have initially been developed to perform AI tasks on 2D arrays of pixels, i.e. images, \citep{ResCNN, Cotin1, sun2020predicting}, but have been extended to tasks on 3D arrays of voxels, for instance in the context of volume segmentation \citep{BUDA2019218, LU2019422} and classification \citep{3D_class_2018, VU2020117328}. The majority of the work dedicated to physics-based simulation deal with 2D problems only. One of the reasons, in our opinion, is that 3D CNNs are  computationally expensive to train, requiring specialist hardware and large amounts of CPU/GPU memory and time. Consequently, the literature dedicated to physics-based simulations using 3D CNNs is rather limited (see e.g. \citep{Cotin1, RAO2020109850}).
    Incidentally, representing 3D geometries using binary voxels is inefficient because the resolution of the image is homogeneous while the geometry of the object can be better described with a variable resolution. Specifically, for multiscale problems, this means that the voxel size needs to be very small to capture the details of the microstructure. This results in a computationally wasteful data representation that uses a large number of voxels to represent areas with no microscale features that could be represented using a coarser functional representation. This is also true for CNNs applied to 2D problems, but the computational cost is disproportionately larger in 3D problems. In this paper, we propose to follow a different approach, that of geometric learning.
    \par
    
    Another general problem with current deep learning models is their poor generalisation ability in cases where the training dataset is small or far from the cases for which predictions are required in the online -inference- stage. 
    This difficulty impedes the straightforward application of deep learning methodologies, which are usually associated to the \say{big data} paradigm, not to physics-based modelling. Indeed, replacing heavy physics-based simulators, which can only be run a few hundreds of times (or sometimes a few tens of times for industrial scale FEA models), by deep learning models, is challenging (hence the enduring superiority of Model Order Reduction, which involves solving some of the equations of the high-fidelity model online, over deep learning \citep{HOANG2016121, ROCHA2020103995}). 
    The problem is even more acute when no physical model is available and data needs acquiring through cumbersome series of experiments \citep{ma13153298, Goetz_2022}. 
    An interesting option that is being actively investigated, but is outside the scope of our current investigations, is that of deploying transfer learning methodologies \citep{Ayan_2022}, where a pretrained model for a more general task is fine-tuned for the specific task of interest.
     
    Our paper addresses the aforementioned difficulties (inefficiency of 3D deep learning architectures for physics-based simulations and lack of generalisation ability) by extending our previous work on 2D Bayesian multiscale stress CNNs presented in \citep{krokos2021bayesian}. The two main sources of novelty introduced in this extended deep learning approach are as follows. Firstly, we will address 3D problems efficiently using geometric learning. Instead of working with voxelised images of the computational domain, we will learn the output of the simulation code of interest using Graph Neural Networks that can directly handle the type of unstructured meshes used to perform FE simulations. Secondly, we propose a novel way to introduce physics-based online corrections to improve the generalisation ability of the NN. Both steps are dedicated to the surrogate modelling of linear elastic stress analysis in porous structures. At the present stage of our investigations, the extension to stress prediction in general heterogeneous materials and structures, and/or to the context of non-confined plasticity and damage is unclear.
    
    Geometric learning can be seen as an extension of previously established NN techniques from Euclidean, grid structures to non-Euclidean, graph and manifold structures. Graph Neural Networks (GNNs) may be utilised to perform geometric learning, by representing the manifold of interest using a graph. GNNs in general may be used to overcome the constant voxel resolution problem by operating on an unstructured grid. In the context of our current investigations, this mesh may be the mesh used to solve the FEA problem of interest, or an auxiliary mesh created specifically for the deep learning task. The mesh used by the GNN may have refined triangles or tetrahedra in areas where the geometry is complex and/or large mechanical gradients are to be expected, and a smaller number of triangles or tetrahedra in the remainder of the computational domain. In principle, this adaptive resolution strategy seems to be a better starting point for efficient use of computational resources than voxel-based CNNs. In this paper, the deep learning task will be performed over the surface of the structure of interest, including that of the microscopic pores. This surface will be represented by a mesh of triangles, typically the trace of the FEA mesh. We choose the Von Mises stress as quantity of interest for this paper, which, incidentally, tends to peak at the pores' surface [\ref{vol_vs_surf}] (this will be shown numerically in the core of the paper). Therefore, the mapping between geometry and quantity of interest can be encoded without loss of information using the graph-based geometrical learning structure mentioned above. Convolutions are only performed on surface nodes, thereby limiting computational expenses to a minimum.

    We will use an Encoder-Processor-Decoder architecture that employs message-passing GN blocks as convolution blocks as introduced in \citep{battaglia2018relational}. This type of well-established neural network architectures that has been shown to be state-of-the-art for a variety of applications \citep{sanchez2018graph, sanchezgonzalez2020learning, pfaff2021learning, mylonas2021bayesian}.
    The novelties we bring in terms of GNNs are specific to the multiscale nature of the surrogate modelling task at hand. Firstly, instead of operating on the entire structure of interest we operate on patches extracted from the structure. This is in accordance with our previous work that employs CNNs for multiscale stress predictions \citep{krokos2021bayesian}. Secondly, since we perform geometric learning on the surface mesh of the porous structure of interest we employ dual geodesic and Euclidean convolutions to diffuse information across disconnected regions of the surface mesh. The use of dual convolutions was firstly proposed from \citep{schult2020dualconvmeshnet} for 3D semantic segmentation tasks. In the context of the current work we provide numerical experiments to support these two choices.
    
    Our GNN is made Bayesian by using the Bayes-by-Backpropagation variational inference framework introduced in \citep{blundell2015weight}, consistently with our previous work \citep{krokos2021bayesian}. Bayesian NNs output densities of posterior mechanical quantities, from which credible intervals (CIs) may be computed. When the BGNN makes a prediction for a case that is different from the cases seen in the training set, then the CIs will be broad, reflecting the fact that the prediction is uncertain. Conversely, tight CIs indicates trust in the prediction. Active learning strategies may be developed based on such probabilistic predictions. This has been discussed in our previous work and is not investigated further here \citep{krokos2021bayesian}. However, the probabilistic formalism of our deep learning framework is the corner stone of the online correction method that is the second novel methodological element of the present paper.
    
    The geometric deep learning architecture that we deploy outputs all stress tensor components at every vertex of the mesh used to represent the surface of the structure of interest, including the surface of the pore network. In order to improve the predictions of the NN, we propose to enforce the natural boundary conditions on the stress tensor at inference stage (i.e. the homogeneous Neumann boundary conditions), for every vertex belonging to the Neumann boundary. This is done using a Bayesian data assimilation approach. To do this, we consider the posterior density of stress fields generated by the BGNN to be a prior for the online correction step. Then, we formally consider the Neumann conditions as \say{data} resulting in a new posterior density following the application of Bayes' theorem. Sampling this corrected posterior is difficult, as retraining the BGNN through a modification of the loss function would result in intractable inference stages. Instead, we use an Ensemble Kalman sampler \citep{Evensen1994_EnKF}, which is a standard algorithm used in the data assimilation literature to perform recursive inference in large-dimensional settings (e.g. weather forecasting). Samples are drawn from the posterior of the BGNN, which is assumed to be Gaussian. From this assumption and the linearity of the Neumann condition to the predicted surface stress fields, the online posterior update becomes explicit (Gaussian as well), and samples from the BGNN are inexpensively and robustly transformed into samples of the posterior using the usual law-rank update techniques associated with Ensemble Kalman methods. We show that this simple approach produces surprisingly good results. Indeed, while the BGNN tends to underestimate critical (maximum) values of equivalent stresses, the corrected result does not suffer from this downward bias, thereby granting reliability to the overall approach.
    
    We stress that the online correction approach described in the previous paragraph does not yield a Physics Informed Neural Network (PINN) as introduced in \citep{RAISSI2019686}. PINNs are full Neural-Network-based PDE and inverse problem solvers. Our offline-online correction approach is more closely related to Model Order Reduction \citep{Ryckelynck2009, HOANG2016121}, whereby a reduced space of possible mechanical fields is mathematically extracted during an offline exploration stage, and prediction are made online by enforcing \say{some} of the equations of physics \footnote{In fact, this link becomes even more apparent when one realises that the Ensemble Kalman sampler utilises a reduced basis to describe the Gaussian prior. Further discussions on the relationship between Model Order Reduction and Ensemble Kalman samplers may be found in \citep{Pereira2021}}
    
    Our paper is organised as follows. In section [\ref{Governing Equations}] we present the multiscale structural model for porous structures that we aim to learn the output of. In section [\ref{GNN section}] we discuss the GNN architecture and the choice of the input features and engineering quantity of interest. Next, in section [\ref{BGNN subsection}] we explain the strategy that we follow to convert the deterministic GNN into a BGNN. In section [\ref{EKF section}] we introduce the online correction method developed in this work. Finally, in section [\ref{Numerical Examples}] we apply our multiscale stress GNN method to two different problems, we demonstrate the effectiveness of the GNN on accurately predicting spatial stress distribution and, as a particular case, maximum equivalent stress values. We also demonstrate that the online stress correction technique that we propose improves the GNN predictions, thereby circumventing the problem of systematic underestimation of maximum equivalent stress values observed in the context of training datasets of limited size.

%#######################################################################################################################################
\section{Deep-learning-based meta-modelling for multiscale finite element analysis} \label{Governing Equations}
    
    %-----------------------------------------------------------------------------------------------------------------------------------
    \subsection{Problem statement: linear elasticity}
    
        We consider a 3D body occupying domain $\Omega \in \mathbb{R}^3$ with a boundary $\partial \Omega = S$. The body is subjected to prescribed displacements $\textbf{U}_D$ on its boundary $\partial \Omega_{u}$ and prescribed tractions $\textbf{T}_D$ on the complementary boundary $\partial \Omega_{T}$ = $\partial \Omega \backslash  \partial \Omega_{u}$. We consider a body force \pmb{f} and a displacement $ \textbf{u} : \Omega \rightarrow \mathbb{R}^3$. The boundary value problem of linear elasticity, under the assumption of the small perturbation theory, consists in finding $\textbf{u} = \text{arg } \displaystyle{\min_{\textbf{u}^{*}} E_p(\textbf{u}^{*}) }$
        where the potential energy is defined as
        
        \begin{equation}
            E_p(\textbf{u}) = \int_{\Omega} W(\pmb{\epsilon}) \,d\Omega - 
                     \int_{\Omega} \textbf{f} \cdot \textbf{u} \,d\Omega +
                     \int_{\partial \Omega_{T}} \textbf{T}_D \cdot \textbf{u} \, dS
            \label{eq:PDE}
        \end{equation}
        We consider a linear isotropic material model defined by its strain energy density
        \begin{equation}
            W(\pmb{\epsilon}) = \frac{1}{2} \lambda [\textrm{tr}(\pmb{\epsilon})]^2 + \mu\textrm{tr}(\pmb{\epsilon}^2)
            \label{eq:strain-energy}
        \end{equation}
            
        \noindent    
        In the equations above, $\lambda$ and $\mu$ are the Lamé elasticity parameters and $\pmb{\epsilon}$ the linearised strain tensor defined as
    
        \begin{equation}
            \pmb{\epsilon} =  \frac{1}{2} ( \nabla \textbf{u} + (\nabla \textbf{u})^\top ) 
            \label{eq:linear-strain-energy}
        \end{equation}
        
        \noindent  
        The Cauchy stress tensor $\pmb{\sigma}$ is calculated as follows:

        \begin{equation}
            \pmb{\sigma} = \frac{\partial W(\pmb{\epsilon})}{\partial \pmb{\pmb{\epsilon}}}
            \label{eq:linear-Cauchy-stress}
        \end{equation}
    
    \subsection{Case study: mechanical specimen made of a porous material} \label{multiscale_structure}
        
\begin{figure}[htb]
            \begin{center}
                \includegraphics[width=\linewidth]{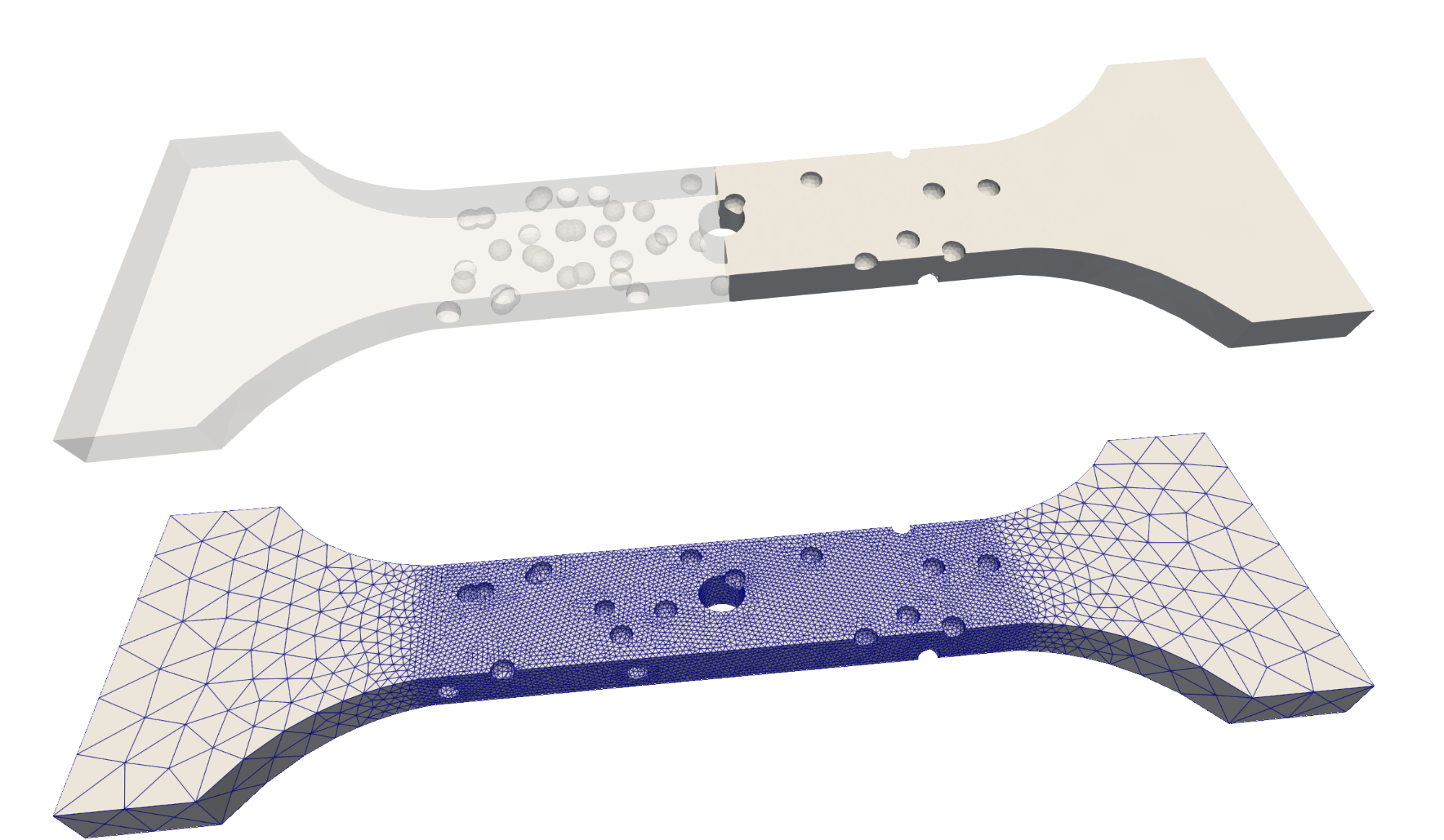}
                \caption{In the top subfigure, we can see a realisation of the dogbone structure. The dogbone is porous and it has a cylindrical hole in the middle. The porous phase is geometrically defined as the union of randomly distributed spheres. The spheres are allowed to intersect with each other, the boundaries of the dogbone and the cylindrical hole. In the bottom subfigure, we see the mesh of the dogbone. We observe that the mesh is denser in the middle of the structure, where the porous phase is present, and coarser everywhere else.}
                \centering
                \label{fig:random_realisations}
            \end{center}
        \end{figure}

        For the purpose of training the GNN we need to create multiple realisations of a multiscale geometry with both macroscale and microscale features. In this work we consider
        a structure that we will refer to from now on as \say{dogbone} [Fig  \ref{fig:random_realisations}]. The dogbone specimen has a cylindrical hole as a macroscale feature. The material is porous, made of randomly distributed spherical pores as microscale features.
        
    \subsection{Mesh and Finite element solver}        
        
        The FE meshes are created using the open-source 3D finite element mesh generator gmsh \citep{gmsh}, using pygmsh \citep{schlomer_nico_2021_5591953} as a Python interface. Unstructured tetrahedral meshes will be used.
        
        The FE problems, comprising the macroscopic surrogate models and the full fine-scale models (as defined in the next subsection [\ref{Global-local framework}]), will be solved using the FEniCS software which is a Python based popular open-source computing platform for solving partial differential equations (PDEs) \citep{fenics, LoggMardalEtAl2012}. We use standard linear tetrahedral elements to solve the elasticity problem.
        
        Meshes are manipulated using the Python package pyvista \citep{sullivan2019pyvista} that allows for easy interaction with VTK objects through Numpy arrays. 
        
    \subsection{Global-local framework} \label{Global-local framework}
    
        The multiscale framework that we use here is a direct extension of our previous work \citep{krokos2021bayesian}. As mentioned in section [\ref{multiscale_structure}], the examples that we show in this work are dedicated to linear elasticity for porous media made of a homogeneous matrix with a random distribution of microscopic spherical voids. 

\begin{figure}[p]
            \begin{center}
                \includegraphics[width=\linewidth]{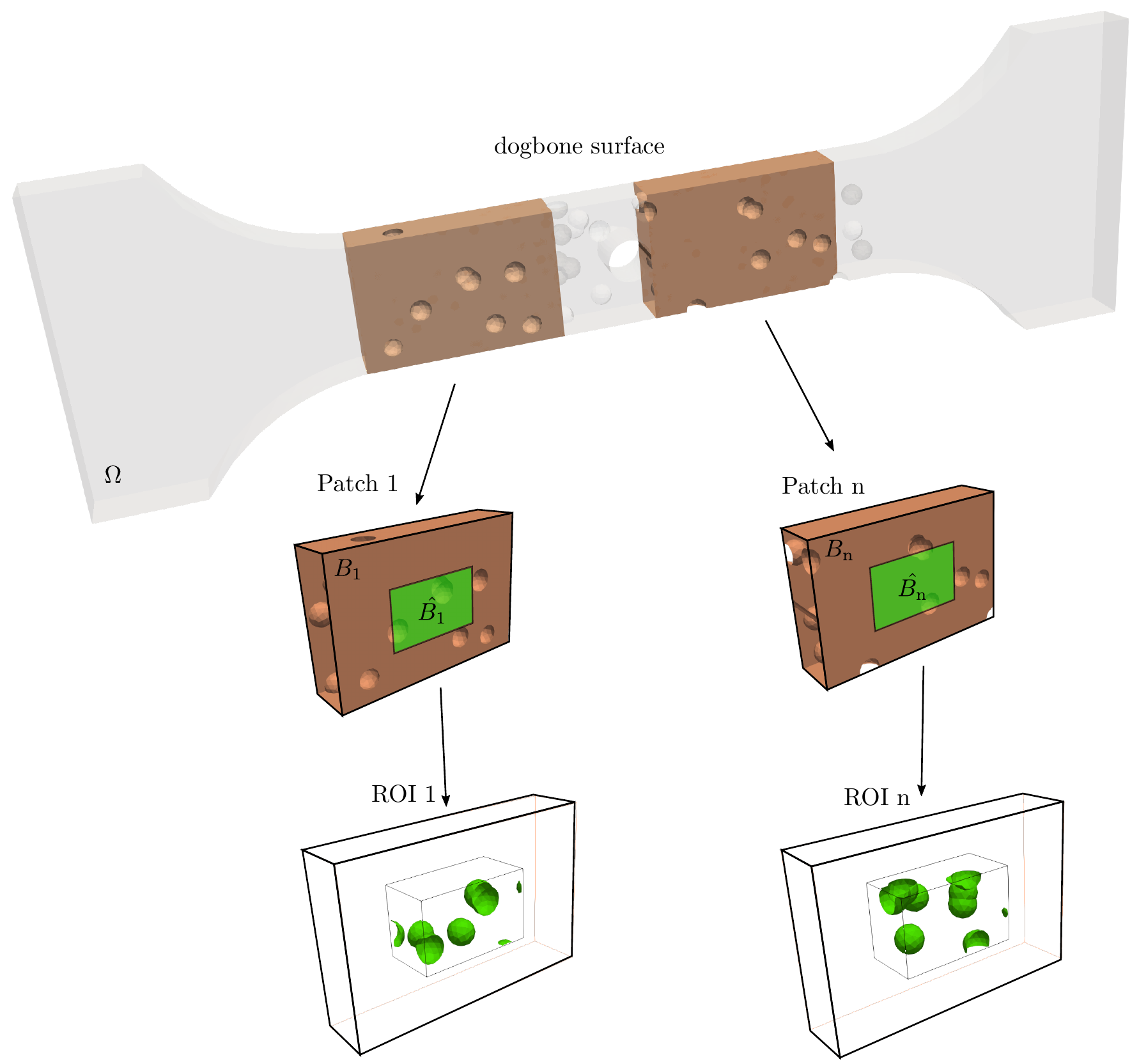}
                \caption{Porous material, $\Omega$, with patches, $B$. We can observe a dogbone structure where we have extracted two patches. With green we can see the ROI of the patches, $\hat{B}$}
                \centering
                \label{fig:patches}
            \end{center}
        \end{figure}

        Our approach is multiscale. At inference time, we first solve for linear elasticity using FEA, in the entire computational domain $\Omega$ but ignoring the voids. This computation is relatively inexpensive as the mesh needs not represent the microscopic elements of our problem, leading to comparatively small numbers of degrees of freedom. The Neural Network will then correct the macroscopic stress field, "guessing" the microscale stress field perturbed by the presence of the microstructural elements. This is done over a sliding window $B \subset \Omega$, which we call patch. The Neural Network described later will take as input the geometry of the patch -both the voids and the boundaries of the structure- the macroscale stress field computed by the coarse finite element procedure, and will output the microscale field in a subregion $\hat{B} \subset B$ of the patch, that we call Region of Interest (ROI), as represented in [Fig  \ref{fig:patches}].
        
        Region $B\backslash \hat{B}$ is called the buffer region. It needs to be large enough so that the ROI is given enough information about the macroscopic mechanical state and about the micro and macro geometry of its surroundings. Following Saint-Venant's principle, we postulate that for sufficiently wide buffer regions, the function that the Neural Network is to approximate is injective.

        Training will be performed by providing input-output pairs for a large number of patches, which will be extracted from appropriate microscale finite element simulations. When required, predictions over regions that are larger than patches may be performed by stitching ROI predictions together, which is demonstrated in our numerical example section.

    \subsection{Quantity of interest and accuracy measure}
    
    We are interested in predicting a mechanical quantity that indicates critical regions that could be responsible for the loss of structural integrity. In this work we choose the Von Mises stress (VM) as indicator of stress intensity, which reads as follows:
    
    \begin{equation}
        \sigma_\textrm{VM} = \sqrt{ \frac{1}{2} [(\sigma_{xx}-\sigma_{yy})^2 + 
                                          (\sigma_{xx}-\sigma_{zz})^2 + 
                                          (\sigma_{yy}-\sigma_{zz})^2] + 
                                          3(\sigma_{xy}^2 + \sigma_{xz}^2 + \sigma_{yz}^2)}
        \label{eq:VM_stress}
    \end{equation}

    We will define accuracy as the fraction of patches for which the maximum Von Mises stress over the RoI is predicted with less than a certain percentage of relative error (which we call threshold). Unless stated otherwise, this threshold is set to 10\%. This procedure is summarised in [Appendix  \ref{appendix:accuracy_algorithm}].

%#######################################################################################################################################
\section{Graph-based geometric deep learning} \label{GNN section}

    In this section we briefly review geometric learning methods reported in literature. Also, we explain the geometric learning framework that we choose to use and finally we discuss the structure of the GNN along with the input and output features.
    
    \subsection{Brief review of earlier work on geometric deep learning}
    
    GNNs have been vastly developed over the last years with various works on segmentation \citep{Hanocka_2019, schult2020dualconvmeshnet, Lei2021PicassoAC} and shape correspondence or retrieval tasks \citep{masci2018geodesic, gong2020geometrically}. One early attempt of deep learning in non-Euclidean spaces is introduced by \citep{qi2017pointnet} where a method for processing point clouds is described. Even though point clouds is a very simple way to describe 3D objects, a more common and informative way is by using a mesh. One common approach for geometric learning with meshes is described in \citep{Hanocka_2019} where a generic framework for processing 3D models for classification and segmentation tasks is described. MeshCNN operates directly on the mesh and not a voxel-based representation of the object, which leads to increased computational cost, or a point cloud, which results in loss of valuable topology information.
    
    GNNs have also been used for various engineering applications. In \citep{Guo_Buehler_2020}, the authors developed a semi-supervised approach to design architected materials using a binary classification GNN. The input to the GNN is the load levels for 1\% of the nodes and the output is the load levels for the rest of the nodes. The load level is a binary label, where the two possible values are the low stress and high stress area. The framework  requires experimental data at test time (the load levels for 1\% of the nodes). The authors mention that such data can be obtained from embedded sparse tensors.
    In the same year, \citep{Vlassis2020} used geometric deep learning to model anisotropic hyperelastic materials. Specifically, they modelled polycrystals using a graph where the nodes represent the monocrystals and the edges their connectivity. This utilization of non-Euclidean data allowed them to take advantage of the rich microstructure information not described by classical Euclidean descriptors like density and porosity. 
    Additionally, \citep{sanchezgonzalez2020learning} used GNNs to simulate particle-based dynamic systems where one or more types of particles, ranging from water to sand, are involved. The physical state of the system is described by a graph of interacting particles and the physics is predicted by message passing through the nodes (particles) of this graph. 
    Later, \citep{pfaff2021learning} used GNNs for mesh based dynamic simulations. In contrast to particle-based problems, in this work the authors took advantage of the connectivity of the mesh to define the edges of the graph. They demonstrated their method in a variety of applications such as deformable bodies, fluid flow around obstacle and air flow around an aircraft wing. Specifically for the air flow around an aircraft wing they show that GNNs have superior performance compared to CNNs. 
    Moreover, \citep{mylonas2021bayesian} used a Bayesian GNN to infer the position and shape of an unknown crack via patterns of dynamic strain field measurements at discrete locations. 
    Recently, \citep{lino2021simulating} developed a multiscale GNN that efficiently diffuses information across different scales making it ideal for tackling strongly nonlocal problems such as advection and incompressible fluid dynamics. 
    Furthermore, in \citep{Perera_Guzzetti_Agrawal_2022} the authors develop a GNN based framework to simulate fracture and stress evolution in brittle materials due to multiple microcracks’ interaction. The GNN predicts the future crack-tip positions and coalescence, crack-tip stress intensity factors, and the stress distribution throughout the domain at each future time-step.
    Lastly, \citep{deshpande2023magnet} introduced a U-Net GNN with mesh pooling and unpooling operations, in accordance with the U-Net architecture for CNNs, that efficiently scales with the size of the problem. MAgNET is used to predict nonlinear force-displacement mappings and the authors provided examples of applying the MAgNET in real-world geometries such as those arising in biomechanics.
    In their latest work, \citep{Deshpande_Sosa_Bordas_Lengiewicz_2023}, the authors compared MAgNET with a novel attention-based architecture called Perceiver IO, proposed  by \citep{Jaegle_et_al_2022}. Perceiver IO was developed with the goal to easily integrate and transform arbitrary information for arbitrary tasks, so the authors use it without adding information about the underlying data structure (such as the mesh connectivity). In terms of training, Perceiver IO requires less parameters but much more training time. Nonetheless, it is much faster in the inference stage. In terms of results, for inputs of small size, the Perceiver IO slightly outperforms MAgNET, but as the size and complexity of the mesh increases it becomes less robust and fails to learn efficiently, which is attributed to the fact that the mesh connectivity is not provided so it has to implicitly learn the nodal data dependencies.
    
    Most of these works are using the formulation described in \citep{battaglia2018relational} which provides a general framework to work with GNNs. We use the same formulation since it allows for a very natural encoding of information on the mesh where absolute information, for instance material properties, can be encoded on the vertices of the mesh while relative information, for example distance between vertices, is encoded on the edges of the mesh.
    
    \subsection{Geometric learning for multiscale stress analysis: assumptions and justification} \label{vol_vs_surf}
    
    As can be seen from the literature review, researchers can take advantage of the flexibility that GNNs offer, to reduce the computational cost by modelling their system in more meaningful ways that stem from the physical understanding of the underlying problem, as for example \citep{Vlassis2020} who modelled the behaviour of a polycrystal by considering every crystal as a node of the graph instead of operating on the volume mesh. Following this paradigm, for reasons we explain in the next paragraphs, we choose to work only on the surface mesh of the porous medium and not the volume mesh.

\begin{figure}[htb]
        \begin{center}
            \includegraphics[width=.6\linewidth]{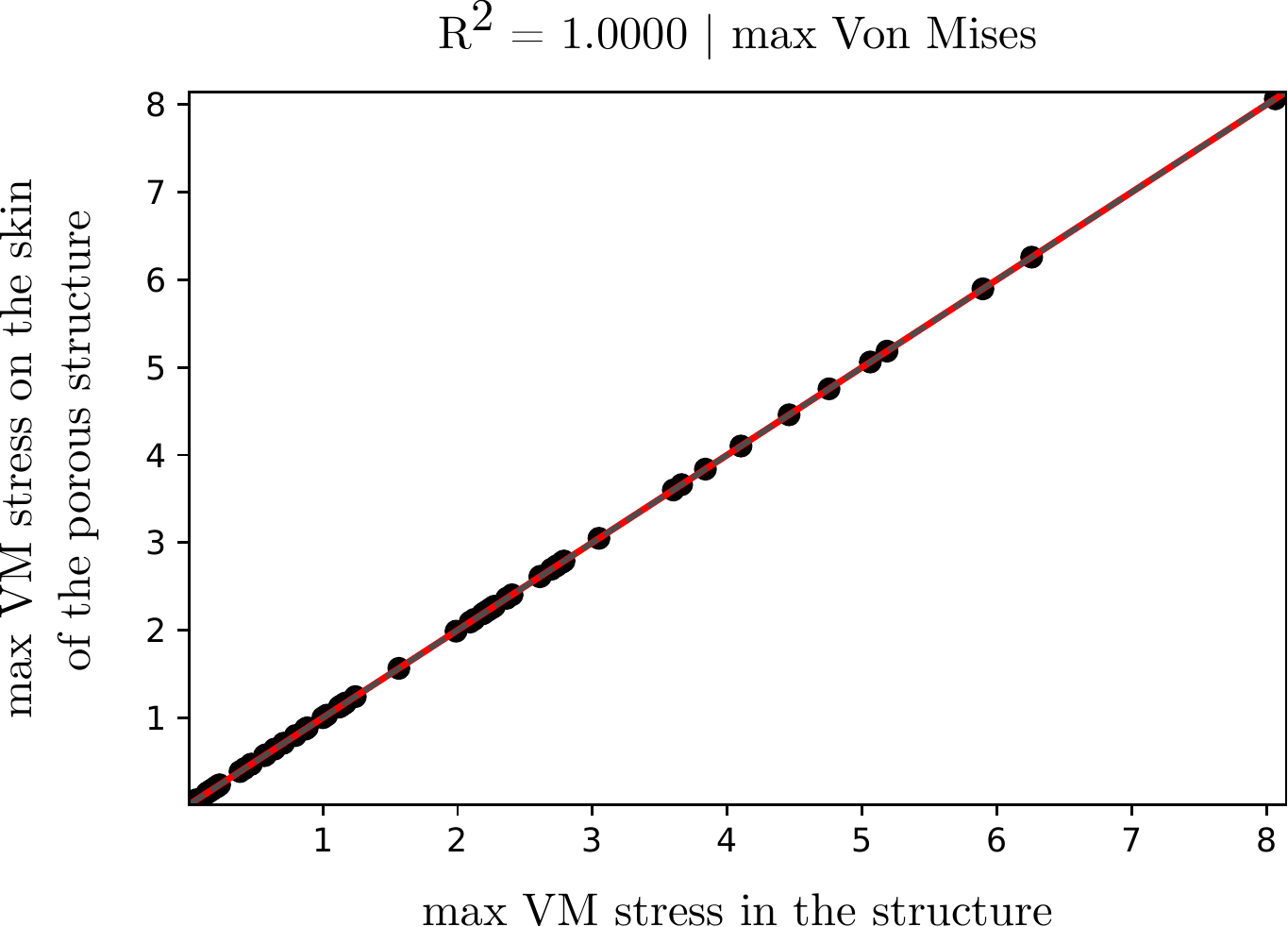}
            \caption{The x axis corresponds to the maximum Von Mises stress on the volume mesh and the y axis to the maximum Von Mises stress on the surface mesh for each of the 100 performed simulations. The red line is the $y=x$ line. We can observe that all the points lay on this line and thus the maximum volume and surface stress values are the same.}
            \centering
            \label{fig:skin_volume}
        \end{center}
    \end{figure}

    First of all, we try to evaluate to what degree the choice to work only on the surface mesh is going to affect the maximum stress in the structure, which is the quantity of interest.
    We conduct an experiment where we perform 100 FE simulations with 100 realisations of the dogbone structure. For the boundary conditions we apply a displacement along the $x$ and $-x$ direction of the same magnitude but with opposite direction. Afterwards, we extract from the volume mesh the surface mesh with the stress tensors encoded on the nodes of the surface mesh (see [Appendix \ref{appendix:Volume_vs_Surface}]). Lastly, we calculate the maximum Von Mises stress both on the volume and surface mesh and we compare the two. From [Fig  \ref{fig:skin_volume}] we can conclude that the maximum volume stress is the same as the maximum surface stress. We do not claim that this would be the case for every possible distribution of pores, for every sample geometry and for every macroscale loading condition but we will make this assumption for computational efficiency reasons.
    
    Additionally, we hypothesise that the surface mesh is a comprehensive representation of the geometry of the specimen. Lastly, we assume that the trace of the, smooth, macroscale stress over the skin mesh is sufficient to inform the GNN of the macroscopic stress state over the patch. We prove these two hypotheses numerically in the context of the current paper.
    
    Consequently, we choose to work only on the surface mesh since we are primarily interested in predicting the maximum stress. This greatly reduces the memory requirements and the training time.
        
    %-----------------------------------------------------------------------------------------------------------------------------------
    \subsection{Dual geodesic/euclidean convolution operators for disconnected graphs} \label{neighbourhood}
    
    To perform geometric learning on the surface mesh of our porous structures, we consider a graph $G$ that can be described by a set of nodes $\pmb{V}:\{V_1, V_2, ... V_N\}$ where $N$ is the number of nodes of $G$ and a set of edges $\pmb{E}:\{E_1, E_2, ... E_M\}$ where $M$ is the number of edges of $G$. 
    
\begin{figure}[htb]
        \begin{center}
            \includegraphics[width=\linewidth]{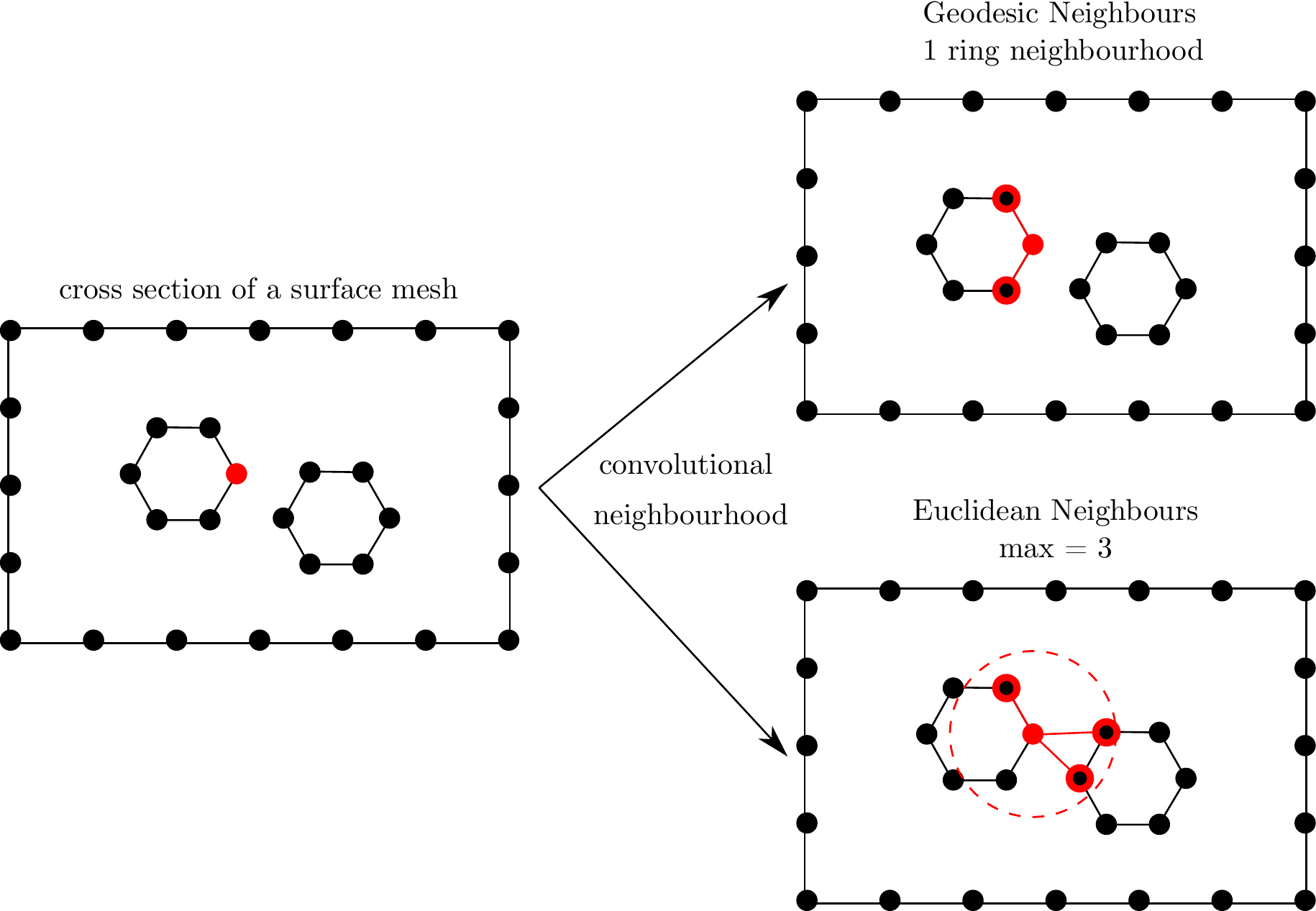}
            \caption{On the left we show a cross section of the surface mesh depicting a rectangle with 2 spherical features. With red with see the node that is considered the central node for the convolutional neighbourhood. On the right we can see two possible neighbourhoods for the central node. The Geodesic (1-ring) neighbourhood on top and the Euclidean neighbourhood at the bottom with a threshold for the maximum number of neighbours equal to 3. In the Geodesic neighbourhood there is no edge between the two spherical features and thus message passing between the two is not possible.}
            \centering
            \label{fig:Neighborhood_2}
        \end{center}
    \end{figure}

    We assume in this work that the nodes of the graph $\pmb{V}$ coincide with the nodes of the surface FEA mesh. The reason is that the computational mesh is optimised by the mesh generation software to best describe the geometry of the structure. However, the proposed methodology can easily be extended to cases where the nodes of the graph used for geometric learning do not coincide with the surface FEA mesh. The edges of the graph are used to pass messages between the nodes, and they define the connectivity of the graph. The edges of the graph cannot coincide with the edges of the surface FEA mesh since the surface mesh has disconnected areas. This means that it is possible for areas of the surface mesh that are close to each other not to share any edges, for example in the case of two interacting but non intersecting defects. This does not allow passing of information through networks of pores. To overcome this problem we define the edges of our graphs via two different types of nodal neighbourhoods [Fig \ref{fig:Neighborhood_2}].
    
    \begin{itemize}
    
        \item Geodesic neighbourhood. The Geodesic, or 1-ring neighbourhood, includes the nodes that share an edge with the central node in the FEA surface mesh. This is the neighbourhood that is directly implied by the connectivity of the potentially disconnected surface mesh.
    
        \item Euclidean neighbourhood. Any node $V_i$ is connected to every node of the FEA surface mesh that falls within a sphere with a predefined radius $r$ and with centre $V_i$. The predefined radius $r$ that is used for this work is 4 times the radius of the spherical pores, as suggested in \citep{krokos2021bayesian}. In practice that results into a very high number of neighbours and thus into a very expensive computational problem. To tackle this problem we need to restrict the number of neighbours for each node. We do this by setting a fixed integer threshold and randomly choosing neighbours within the euclidean neighbourhoods so that the total number of neighbours of each node is either equal to the threshold, or is the number of neighbouring nodes located in the sphere if this number is smaller than the threshold. In [Appendix \ref{appendix:Number of neighbours}], we discuss the most appropriate value for this threshold.
              
    \end{itemize}
    
    In this work we use either Euclidean or a combination of Euclidean and Geodesic neighbours, as proposed by \citep{schult2020dualconvmeshnet}, both of which allow the modelling of interactions between non-intersecting parts of the mesh. A typical dual convolution graph is represented in [Fig \ref{fig:GNN_graph}], where we see that edges created by considering euclidean neighbourhoods allow to create connections between nearby pores.
    
    The edges of the graph are bidirectional so that the information can be exchanged both ways between two connected nodes. This means that for every edge $E_{k,l}$ passing a message from node $V_k$ to $V_l$ there exists the edge $E_{l,k}$ passing a message from node $V_l$ to $V_k$.
    
    %-----------------------------------------------------------------------------------------------------------------------------------
    \subsection{Input and Output features}
    
    The framework we use requires the input and output graphs to be of the same structure. Consequently, we choose to interpolate the 3D FE solution from the macroscale mesh, i.e. mesh without spherical voids, to the microscale mesh, i.e. mesh with the spherical voids.
        
\begin{figure}[htb]
        \begin{center}
            \includegraphics[width=0.8 \linewidth]{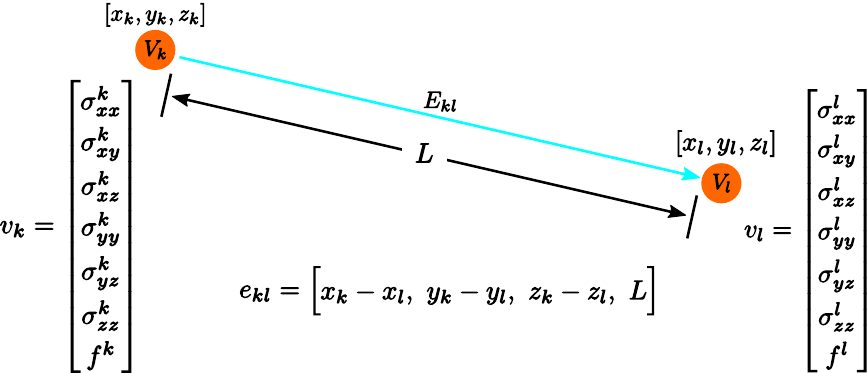}
            \caption{Example of input node features, $u_\text{k}$ and $u_\text{l}$, and edge features, $e_\text{kl}$. With orange we see the nodes and with light blue the directed edge. $f$ is the micro indicator, a single integer that indicates if the node is a macroscale or microscale feature. $\sigma_{i,j}$ corresponds to the i,j component of the macroscale stress tensor. $L$ is the distance between the two nodes.}
            \centering
            \label{fig:Inputs}
        \end{center}
    \end{figure}

    For the input of the GNN we encode node features, $v_i$ where $i=1,N$, on the nodes of the graph and edge features, $e_{i,j}$ where $j=1,M$, on the edges of the graph. The input node features are the independent components of the macroscale stress tensor along with the microscale feature indicator, a single integer indicating if the node corresponds to a microscale feature or not. The input edge features are the relative position between the two nodes that the edge is connecting along with the distance between the two nodes. This is summarised in [Fig \ref{fig:Inputs}].
    
    The output of the graph contains information only on the nodes of the mesh. Specifically, the output node features are the components of the microscale stress tensor.
    
\begin{figure}[!h]
        \begin{center}
            \includegraphics[width=.95\linewidth]{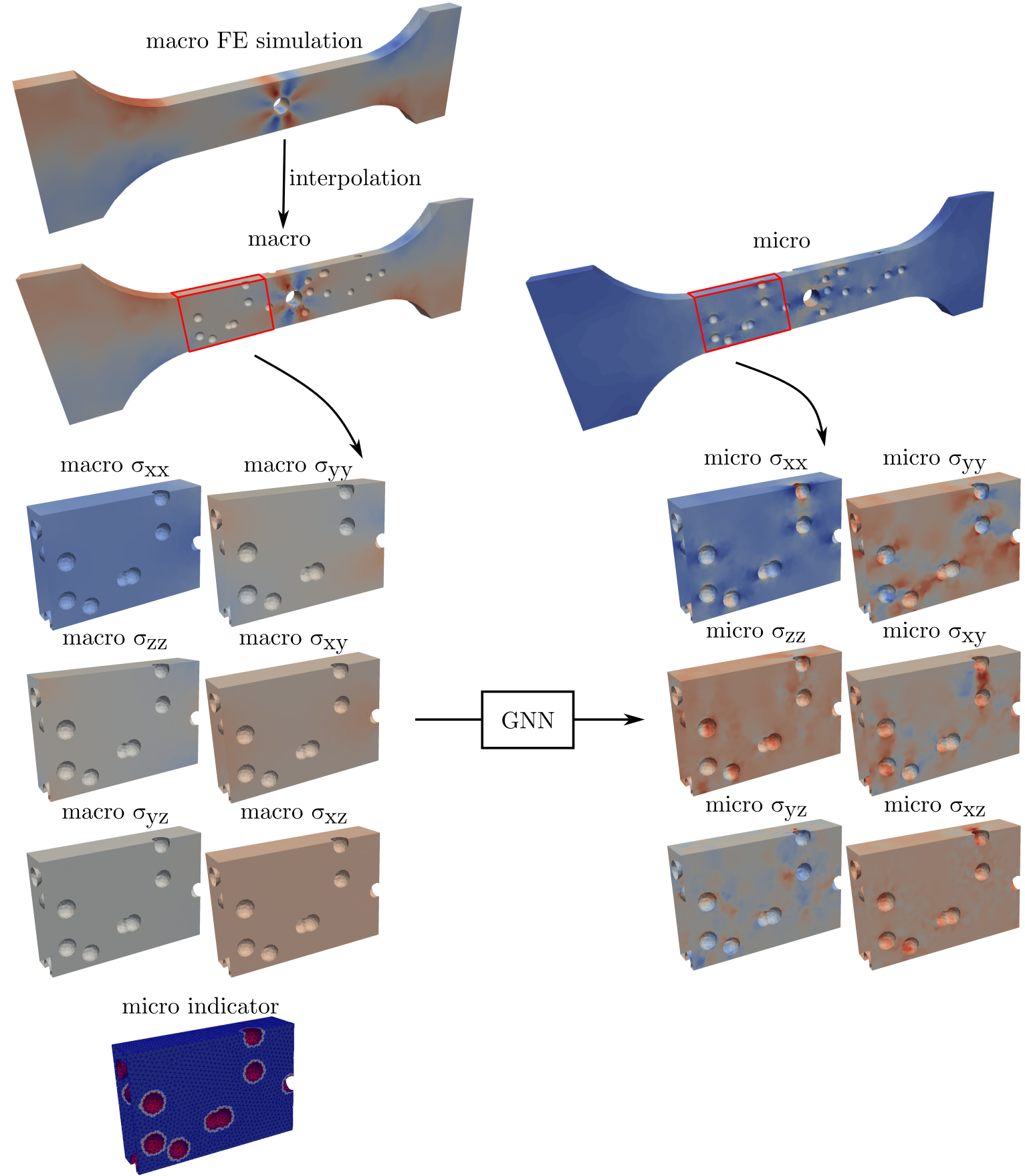}
            \caption{In this figure we can see the input and output of the GNN. The input consists of the 6 stress components of the macroscale stress tensor along with the micro indicator, a single integer per node that determines if the node belongs to a microscale or a macroscale geometrical feature. For the patch corresponding to the micro indicator the FE mesh is visible. The output of the GNN is the 6 stress components of the microscale stress tensor. At the top of the figure, we can see the full structures from which the patches are extracted both for the input and the output. As we can see, the macro stress is computed without the pores, i.e. stress gradients in the gage section are solely created by the cylindrical hole, and it is projected onto the microscale mesh to be provided as input to the GNN.}
            \centering
            \label{fig:in_out}
        \end{center}
    \end{figure}

    We sum up in [Fig \ref{fig:in_out}] the input and output of the GNN. We have only included the node features since the edge features are calculated as a pre-processing step from the connectivity of the mesh.
    
    %-----------------------------------------------------------------------------------------------------------------------------------
    \subsection{Loss function}
        
        The loss function used for the training of all the networks in this work is the Mean Squared Error (MSE) between the stress tensor predicted by the GNN and the one calculated using direct microscale FEA.
          
    %-----------------------------------------------------------------------------------------------------------------------------------
    \subsection{GN block}
    
\begin{figure}[htb]
        \begin{center}
            \includegraphics[width=\linewidth]{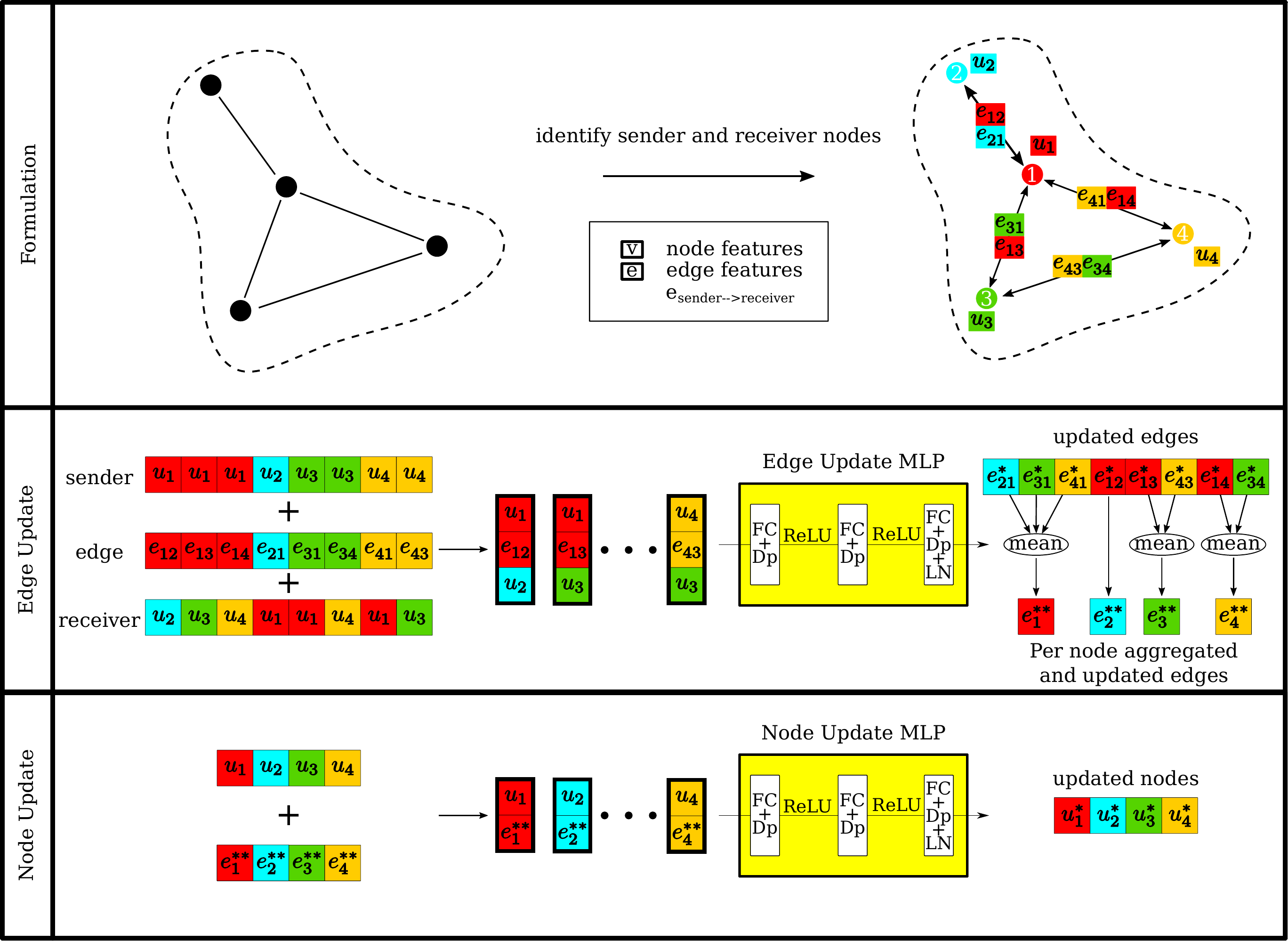}
            \caption{Structure of the GN block used in this work. The update of node and edge features happens through a 2-step procedure as described in this figure. FC stands for fully connected layer, Dp for dropout and LN for layer normalisation.}
            \centering
            \label{fig:GN_Block}
        \end{center}
    \end{figure}

    The GN block is the basic building block of the GNN, and it is used to map an input graph to an output graph with the same nodes and edges but with updated node and edge features. The node and edge features are jointly passed through an edge update Multilayer Perceptron (MLP) and a node update MLP in a 2-step procedure described in \citep{battaglia2018relational} that sequentially updates the edge features first and the node features later. A sketch of this procedure can be found in [Fig \ref{fig:GN_Block}]. 
    
    For the edge update step, if we consider an edge $E_{k,l}$ with edge features $e_{kl}$ that sends a message from the sender node $V_k$ with node features $v_k$ to the receiver node $V_l$ with node features $v_l$ then the updated edge feature, $e_{kl}^*$, is calculated by passing through the edge MLP the concatenation of these features, $e_{kl}^* = \text{MLP(concatenate}(e_{kl}, v_k, v_l$)). If $M$ is the number of edges of the graph, $NNF$ is the number of node features and $NEF$ is the number of edge features then the input of the edge update MLP is of size $[M \times (2 NNF + NEF)]$
    
    For the node update step, the updated edge features, $e^{*}$, are aggregated per node. In this work we use mean aggregation although elementwise summation, maximum or any other symmetric operation can be used. After the aggregation step, we have a single edge feature, $e^{**}$, corresponding to each node feature $v$. The updated node features, $v^*$, are calculated by passing through the node MLP the concatenation of the original node features with the updated and aggregated edge features corresponding to this node, $v^* = \text{MLP(concatenate}(v, e^{**}$)). If $N$ is the number of nodes of the graph, $NNF$ is the number of node features and $F$ is the dimensionality of the updated edge features then the input of the node update MLP is of size $[N \times (NNF + F) ]$.
    
    %-----------------------------------------------------------------------------------------------------------------------------------
    
    \subsection{Residual Blocks}

       Training NNs with many layers is numerically challenging due to problems like the exploding and vanishing gradients. This causes the network to not be able to learn simple functions like the identity function between the input and the output \citep{chapter-gradient-flow-2001, sussillo2015random}. In practise this means that adding a lot of layers might, counter intuitively, deteriorate the performance of the network. To tackle this problem residual blocks where introduced that allow the network to skip some of the convolutional blocks \citep{he2015deep, kim2016deeplyrecursive, zagoruyko2017wide, lim2017enhanced}. This is achieved by using a skip (also called residual) connection that connects the input with the output of the convolutional block. This means that if the input to the convolutional block is denoted by $x$ and the output by $f(x)$ then the output of the residual block is $f(x) + x$. In the case that the network decides that the best strategy is to completely ignore a convolutional block ($f(x)=0$) the input to this block is successfully passed to the rest of the network (the expression of the output would be simplified to: $f(x) + x = 0 + x = x$). This strategy was first applied to CNNs but it seamlessly extends to GNNs \citep{sanchezgonzalez2020learning, pfaff2021learning, mylonas2021bayesian}.
           
    \subsection{Model} \label{Model}
    
    In this work we deploy an Encode-Process-Decode GNN as described by \citep{battaglia2018relational}. The encoder encodes the node and edge features in a latent space of higher dimensions. The processor updates these node and edge features and finally the decoder, decodes these features from the latent space to the output space.
    
    \begin{itemize}
    
        \item \textbf{Encoder}: The node and edge features are independently encoded into a latent space of 128 dimensions using 2 distinct MLPs with the same structure. The MLP has 3 hidden layers followed by a Dropout layer \citep{hinton2012improving} and ReLU activation function each and a Layer Normalization layer after the output layer, which improves training stability and decreases training time in recurrent NNs \citep{ba2016layer}.
        
        \item \textbf{Processor}: The processor is composed of 5 residual GN Blocks. The structure of the MLPs of the Processor is the same as the structure of the encoder MLPs. The processor combines information from edges and nodes and passes messages between nodes that share an edge.
        
        \item \textbf{Decoder}: The decoder operates only on the node features as the output of the GNN is the microscale stress field on the nodes. A single MLP is used to decode the node features from the latent space to the output space. The decoder MLP is similar to the MLPs used in the encoder and the processor but there is no Layer Normalization layer applied after the output layer.
        
    \end{itemize}
    
\begin{figure}[htb]
        \begin{center}
            \includegraphics[width=\linewidth]{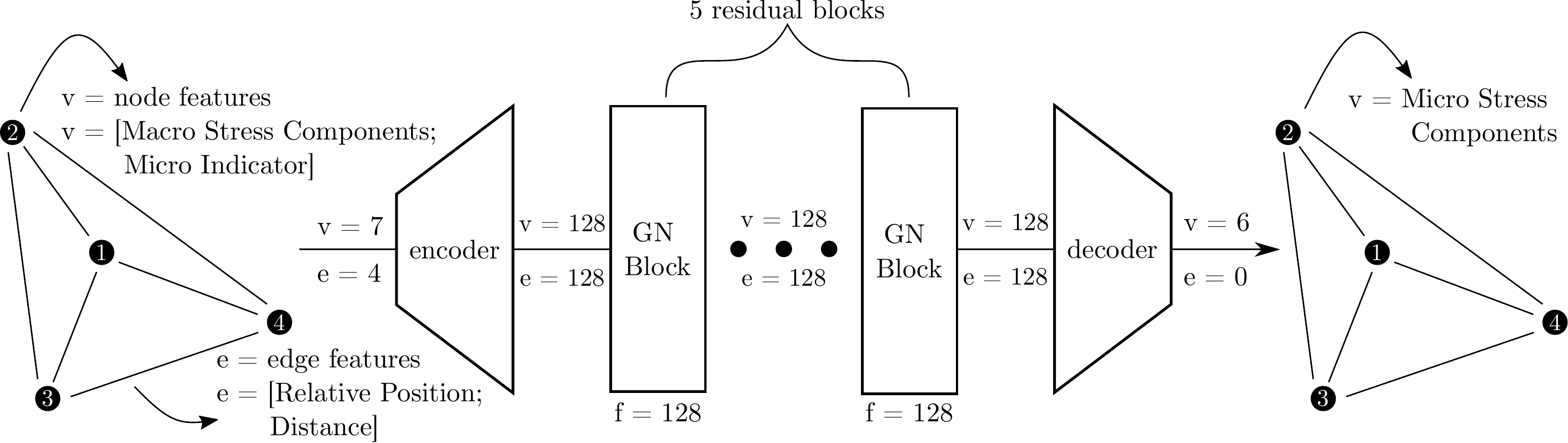}
            \caption{Architecture of the GNN}
            \centering
            \label{fig:GNN}
        \end{center}
    \end{figure}

    There is a skip connection that connects the macroscale stress tensor, from the input, with the output of the Decoder. That way the network can learn how the microscale stress field diverges from the macroscale stress field. In [Appendix \ref{appendix:Skip Connection}], we demonstrate that this leads to smoother and faster convergence. A sketch of the GNN can be found below [Fig \ref{fig:GNN}].
    
%#######################################################################################################################################
    
\section{Physics-based corrections of the NN predictions : enforcing Neumann conditions online via an ensemble Kalman approach} \label{BGNN section} 

    In this section we introduce an online hierarchical Bayesian method to constrain the prediction of the GNN. This method allows us to improve the network prediction without creating more data nor retraining the network. We impose Neumann conditions (which are homogeneous in all our numerical examples), which are the only equations that may be enforced on the stress field after restricting it to the surface of the porous structure. These conditions are not imposed while training the network, even though the training examples satisfy them up to the FE accuracy. This is in contrast to semi-supervised and self-learning learning approach that penalise the violation of some of the physical constraints during training (through specific terms in the loss function), but do not enforce them at inference stage.
    
    To apply this method we will construct a statistical distribution for the GNN predictions. To this end, we convert the deterministic GNN to a Bayesian GNN (BGNN). The posterior estimation of the BGNN for the stress field over the surface of the structure will then serve as a prior for the online correction step, which is formulated as a standard Bayesian data assimilation problem. More precisely, the Neumann conditions are applied statistically by considering them as partial information of the stress field, yielding a Bayesian posterior update. We further use a Gaussian approximation of the posterior in the form of the classical ensemble Kalman method, which is widely used to solve the filtering problems associated with weather forecasting. Alternative approaches, including Variational Inference (i.e. online retraining of the BGNN) and Langevin MCMC, would probably lead to prohibitive online cost, with potential difficulties related to the lack of robustness of these algorithms.

    \subsection{Bayesian GNN} \label{BGNN subsection}
    
    Several techniques have been proposed to quantify the uncertainty in NNs resulting in Bayesian NNs \citep{vanCamp_1993, Graves_2011, kingma2014autoencoding, blundell2015weight}. In this work we use the Bayes by Backprop method (BBB) to convert the deterministic GNN to a BGNN \citep{blundell2015weight}.
    
    In order to define the training objective, we consider a probabilistic NN where we replace the constant weights of a deterministic NN with a distribution over each weight. Before observing the data, $D$, the weight distribution is called prior, and denoted as $P(w)$. After observing the data, the weight distribution is updated, so that the distribution predicted by the BNN fits the data. The corresponding weight distribution is called posterior $P(w|D)$. For an input $x \in \mathbb{R}^n$ the output of the probabilistic NN will be a probability distribution over all possible outputs $y \in \mathbb{R}^k$. 

    The posterior weight distribution, which is formally obtained through the Bayes' rule, $P(w|D) = \frac{P(w)P(D|w)}{P(D)}$, cannot be computed analytically because it involves intractable integrals. A common approach to circumvent this problem is to perform Variational Inference (VI), which consists in approximating the posterior density by a variational distribution $q(w|\theta)$ \citep{vanCamp_1993, Graves_2011} parameterised by a set of learnable coefficients $\theta$. The new objective is to find the parameters $\theta^{opt}$ [Eq. \ref{eq:ELBO}] that minimise the Kullback-Leibler (KL) divergence between the approximate posterior and the true Bayesian posterior, which reads as

    \begin{equation} \label{eq:ELBO}
        \begin{split}
            \theta^{opt} &= \text{arg } \displaystyle{\min_{\theta} \text{KL}[ q(w|\theta) || P(w|D) ] } \\
                         &= \text{arg } \displaystyle{\min_{\theta} \int_{}^{} q(w|\theta) \text{log} \frac{ q(w|\theta) }{ P(w)P(D|w) } \,dw } \\
                         &= \text{arg } \displaystyle{\min_{\theta} [\text{KL}[ q(w|\theta) || P(w) ] } - \mathbb{E}_{q(w|\theta)}[ \text{log} P(D|w)]]
        \end{split}
    \end{equation}
    
    We consider the approximate posterior to be a fully factorised Gaussian \citep{Graves_2011}, corresponding to a squared loss \citep{blundell2015weight}. The prior is also a Gaussian distribution corresponding to L2 regularization \citep{blundell2015weight}. All weights within one layer share the same prior mean, $\mu$, and the same prior standard deviation, $\sigma$. In a forward pass of the BGNN the weights are sampled from the posterior distribution, and gradients are evaluated 
    using the usual reparametrisation trick \citep{kingma2014autoencoding}, as described in \citep{blundell2015weight}. The sampling from the variational posterior distribution is performed by sampling from a unit Gaussian scaled by the posterior standard deviation, $\sigma^*$, and shifted by the posterior mean, $\mu^{*}$. To ensure its positivity, the standard deviation is parameterised as $\sigma^{*} = \text{log} (1+ \text{exp}(\rho^{*}))$. Consequently, the variational parameters to be optimised are $\theta = (\mathbf{\hat{\mu}}, \mathbb{\hat{\rho}})$, where $\mathbf{\hat{\mu}} = (\mu, \mu^*)$ and $\mathbf{\hat{\rho}} = (\rho, \rho^*)$.

    %--------------------------------------------------------------------------------------------
    \subsection{Ensemble Kalman method for online stress correction} \label{EKF section} 
        
        The ensemble Kalman method is used to update the prior probability density of the state of a system, represented by vector $\mathbf{x} \in \mathbb{R}^n$, taking into account noisy and partial observations of that state. In the present context, the state vector corresponds to the microscale stress components predicted by the BGNN at every node, $j$, of the surface mesh. Those are concatenated in a vector of $6J$ components, where $J$ is the number of surface nodes of the inference mesh where Neumann boundary conditions are to be enforced. The state vector reads as
        
        \noindent
        \begin{equation}
            \mathbf{x} = 
            \begin{pmatrix}
            \mathbf{\hat{x}}^\text{{1}} & ... &  \mathbf{\hat{x}}^\text{{J}}
            \end{pmatrix}^T
        \end{equation}
        
        \noindent
        where
        
        \begin{equation}
            \mathbf{\hat{x}}^\text{j} = 
            \begin{pmatrix}
            S_\text{xx}^\text{j} & \:S_\text{yy}^\text{j} & \:S_\text{zz}^\text{j} & \:S_\text{yz}^\text{j} & \:S_\text{xz}^\text{j} & \:S_\text{xy}^\text{j}    
            \end{pmatrix}
        \end{equation}
        
        \noindent
        
        The Neumann boundary conditions are treated as linear observations of the state vector, through the formal definition of observation operator $\mathbf{H}$ as
        
        \begin{equation}
            \mathbf{Hx} =              
            \begin{pmatrix}
                \mathbf{S}^\text{1} \cdot \mathbf{n}^\text{1}  & ... &  
                \mathbf{S}^\text{J} \cdot \mathbf{n}^\text{J}
            \end{pmatrix}
        \end{equation}
        
        \noindent
        where $\mathbf{S^j}$ is the $[3 \times 3]$ symmetric microscale stress tensor at node $j$ and $\mathbf{n^j}$ is the $[3 \times 1]$ unit normal vector
        \newline
        
        \begin{equation}
            \mathbf{S^j} = \begin{pmatrix}
                                S_{xx}^{j} & S_{xy}^{j} & S_{xz}^{j}\\
                                S_{xy}^{j} & S_{yy}^{j} & S_{yz}^{j}\\
                                S_{xz}^{j} & S_{yz}^{j} & S_{zz}^{j}
                           \end{pmatrix}
            , \quad
            \mathbf{n^j} = \begin{pmatrix}
                                n_{x}^{j}\\
                                n_{y}^{j}\\
                                n_{z}^{j}
                           \end{pmatrix}
        \end{equation}
        
        \noindent
        Following the standard Bayesian approach to data assimilation, random noise is added to the observations so that a Gaussian likelihood is implicitly defined. The data reads as
        
        \begin{equation}
            % d = h(\mathbf{x}) + \epsilon
            \mathbf{d} = \mathbf{Hx} + \epsilon
             \qquad
            \epsilon \sim \mathcal{N}(0,\mathbf{\Sigma}_{\epsilon})
        \end{equation}
        
        \noindent
        If $\mathbf{x}$ were normally distributed then we would have an explicit formula, through Bayes' rule, for the posterior state vector, $\mathbf{x}^{\star}$. In that case, the posterior state would be given by
        
        \begin{equation}
            \mathbf{x}^{\star} = \mathbf{x} + \mathbf{G}(\mathbf{d} - \textbf{H}\mathbf{x})
        \end{equation}
        where $\mathbf{G}$ is the so-called Kalman gain defined as $\mathbf{G} = \mathbf{\Sigma} \mathbf{H}^{T}(\mathbf{H} \mathbf{\Sigma} \mathbf{H}^{T} + \mathbf{\Sigma}_{\epsilon})^{-1}$,  $\mathbf{\Sigma}$ being the prior covariance matrix.
        
        \noindent
        Our prior state density, which is the output of the BGNN, is of course non Gaussian. However, the idea of the Ensemble Kalman method, which is at the heart of the Ensemble Kalman filter, is to approximately sample the posterior density using the above formula, i.e. claiming that the prior (and therefore the posterior density as well as the observations are Gaussian-perturbed linear observations of the state) is Gaussian. This is achieved by taking $N$ samples from the BGNN posterior and creating an ensemble of state vectors
        
        \begin{equation}
            \mathbf{X} = 
            \begin{pmatrix}
            \mathbf{x}^\text{{1}} & ... &  \mathbf{x}^\text{{N}}
            \end{pmatrix}
        \end{equation}
        
        \noindent
        The formula for the posterior update can be rewritten in terms of the $N$ samples as
        \begin{equation}
            \mathbf{X}^{\star} = \mathbf{X} + \mathbf{\tilde{G}}(\mathbf{D} - \textbf{H}\mathbf{X})
        \end{equation}
        
        \noindent
        where the perturbed data matrix reads as
        
        \begin{equation}
            \mathbf{D} = 
            \begin{pmatrix}
            \mathbf{d}^\text{{1}} & ... &  \mathbf{d}^\text{{N}}
            \end{pmatrix}
        \end{equation}
        where $\mathbf{d}^\text{{i}} \sim \mathcal{N}(\mathbf{d},\mathbf{\Sigma}_{\epsilon})$, and the approximated Kalman gain matrix (see further below), $\mathbf{\tilde{G}}$, is defined by
        \begin{equation}
            \mathbf{\tilde{G}} = \mathbf{\tilde{\Sigma}} \mathbf{H}^{T}(\mathbf{H} \mathbf{\tilde{\Sigma}} \mathbf{H}^{T} + \mathbf{\Sigma}_{\epsilon})^{-1}
        \end{equation}
 
        \noindent
        In the equations above, the prior covariance matrix used in the standard Gaussian case has been substituted by the ensemble covariance matrix defined as
        % \begin{equation}
        %     \mathbf{\tilde{\Sigma}} = \frac{1}{N-1} \sum_{i=1}^{N}  \mathbf{X}^{(i)} \mathbf{X}^{{(i)}^T}
        % \end{equation}
        \begin{equation}
            \mathbf{\tilde{\Sigma}} = \frac{\mathbf{X} \mathbf{X}^T}{N-1}
        \end{equation}
                
        \noindent
        We note that that the observation matrix $\mathbf{H}$ does not need explicit defining. Only its action needs to be evaluated for every prior sample. For additional (and classical) algorithmic details, the reader may refer to [Appendix \ref{appendix:EnkF}]. The posterior update is calculated efficiently using an SVD, taking advantage of the fact that the empirical covariance matrix is of rank $N$ at most.
        
        To summarize, the online stress correction procedure that we employ here can be described using the following steps. A prior state ensemble is created by drawing $N$ samples from the BGNN posterior. Then, the Kalman update scheme described above transforms the prior empirical distribution into an approximation of the posterior empirical distribution that satisfies the homogeneous Neumann conditions up to a certain tolerance encoded by the spherical covariance matrix $\mathbf{\Sigma}_{\epsilon}$. 
        
        This tolerance may be interpreted as a discretisation error, i.e. the Neumann conditions should not be exactly enforced as (A) the BGNN is trained on the output of finite element code, which delivers stress fields that are element-wise constant and do not satisfy the balance of forces nor the Neumann conditions exactly and (B) the normal vector at a vertex is not uniquely defined (we choose here to calculate them for each triangle and then average them at shared points). These are qualitative observations, which did not help us choose the level of noise. In practice, we adjusted the level of noise (i.e. a scalar parameter) manually so that the posterior update is of appropriate quality on a validation set.
    
%#######################################################################################################################################
\section{Numerical Examples} \label{Numerical Examples}
    
    \subsection{Cubical Heterogeneous material} \label{Ellipses dataset}
    
    In this section we want to test our methodology using a dataset containing samples from a heterogeneous material. We will demonstrate how we can use our methodology to 
    
    \begin{itemize}
        \item predict the microscale stress distribution
        \item predict the maximum microscale Von Mises stress over the patches
        \item quantify the uncertainty of the prediction
        \item improve the BGNN prediction using the online correction technique introduced in section [\ref{EKF section}]
    \end{itemize}
    
        \subsubsection{FEA set-up and generation of the training dataset} \label{train_dataset_cube}
        
\begin{figure}[htb]
            \centering
            \begin{subfigure}{.5\textwidth}
              \centering
              \includegraphics[width=\linewidth]{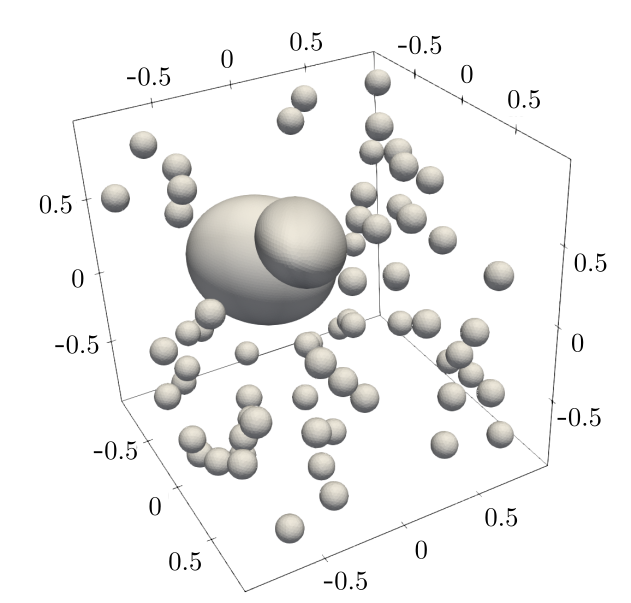}
            \end{subfigure}%
            \begin{subfigure}{.5\textwidth}
              \centering
              \includegraphics[width=\linewidth]{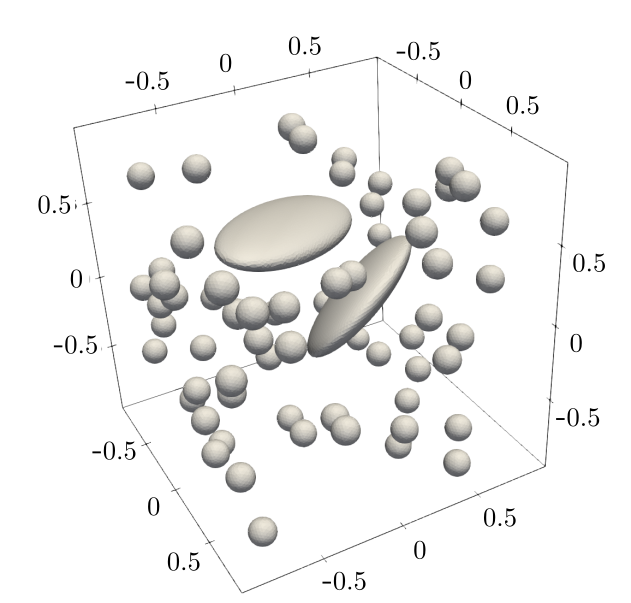}
            \end{subfigure}
            \caption{Two realisations of a cubical specimen from a heterogeneous material with a bimodal distribution of random pores: large elliptical pores and smaller equally sized spherical pores.}
            \label{fig:cube_dataset_realisations}
        \end{figure}

        For the purpose of training our model we will work with a synthetic heterogeneous material from which we extract and examine cubical specimens. The size of each specimen is 2 units along each spatial dimension. The specimen has two elliptical pores, with random position, size and orientation and a distribution of 50 to 100 spherical pores with the same radius, $R=0.08$ units. The spherical and elliptical pores are generated without intersecting. For the macroscale simulations, input of the BGNN, only the elliptical pores are taken into account and the spheres are ignored. The Young’s modulus and the Poisson ratio of the structure are 1 and 0.3 respectively. Two realisations of the cubical specimen are shown in [Fig  \ref{fig:cube_dataset_realisations}].
        
\begin{figure}[ht]
            \begin{center}
                \includegraphics[width=.5\linewidth]{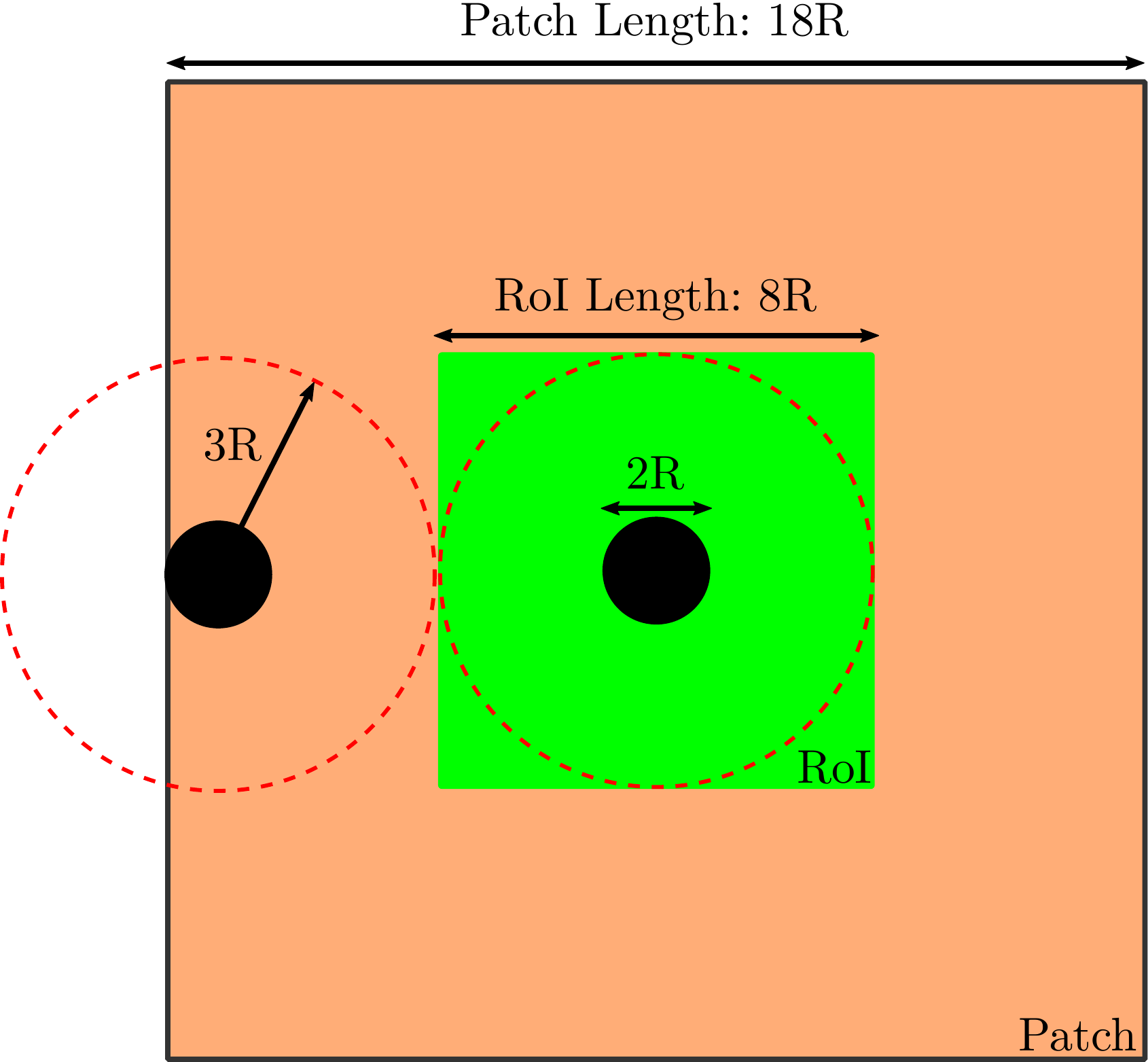}
                \caption{Sketch of a cross section of a patch. In light brown we see the patch, in green the region of interest (ROI) and with the black the spherical voids.}
                \centering
                \label{fig:patch}
            \end{center}
        \end{figure}

        As already explained in [\ref{Global-local framework}] the input to the GNN is a patch of the geometry, not the entire structure, and the prediction of the GNN happens in a sub region of the patch that we called ROI. The patch and ROI size depend on the radius of interaction of the spherical pores. Based on the work of \citep{SIFs} on stress intensity factors, we assume that the effect of spherical voids on the global stress field fades out for a distance larger than 4 radii from the centre of the spherical voids. We combine this information with our previous work \citep{krokos2021bayesian} and we choose a patch size of $[18R \; \times \; 18R]$ and a ROI size of $[8R \; \times \; 8R]$, where $R$ is the radius of the spherical voids, as can be seen in [Fig  \ref{fig:patch}].
        
\begin{figure}[htb]
            \centering
            \begin{subfigure}{.45\textwidth}
              \centering
              \includegraphics[width=\linewidth]{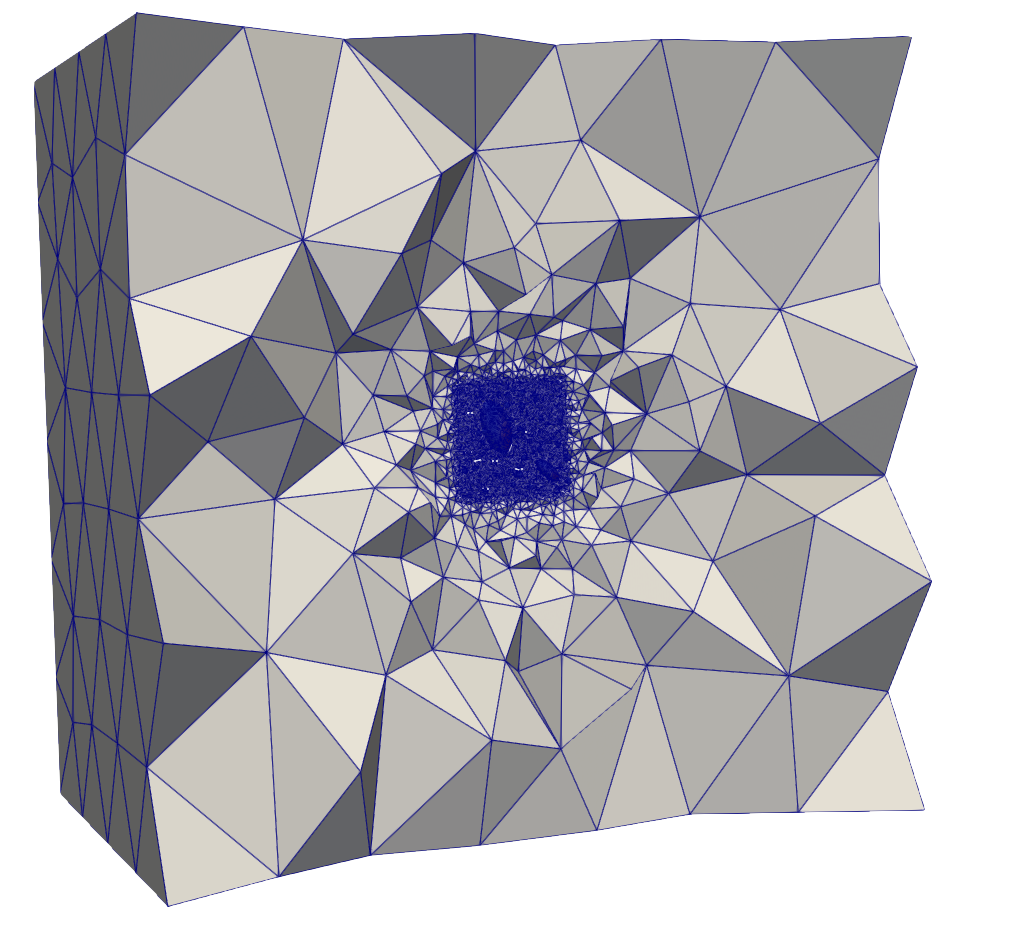}
              \caption{}
              \label{fig:buffer:a}
            \end{subfigure}%
            \begin{subfigure}{.45\textwidth}
              \centering
            \includegraphics[width=\linewidth]{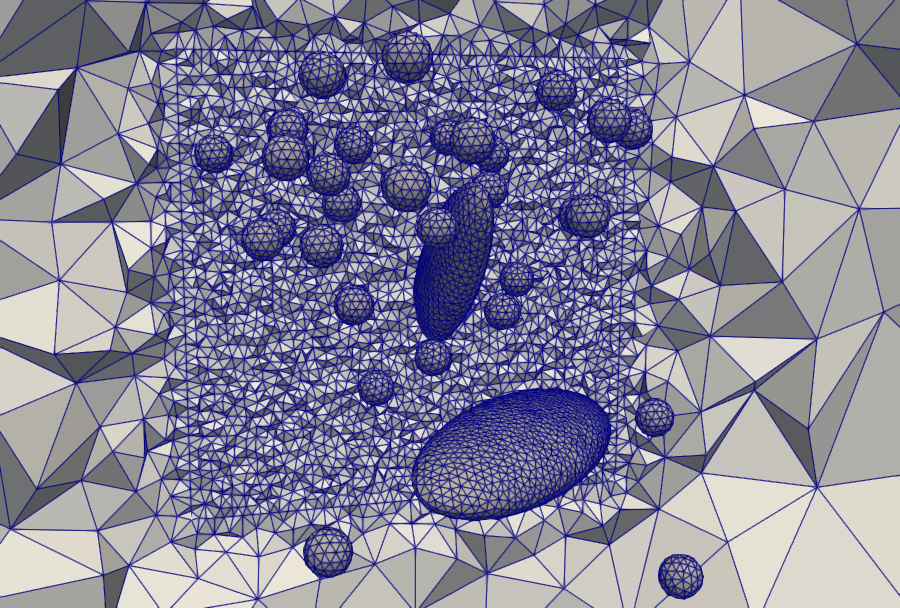}
              \caption{}
              \label{fig:buffer:b}
            \end{subfigure}
            \caption{On the left (a) we can see a volume mesh of the multiscale structure. The buffer area where the mesh is coarse is visible. In the centre we can see the dense mesh area from where we extract the patches. The shape of the dense mesh area is a cube, but here only one face of this cube is visible. On the right (b) we can see a zoom in the dense mesh region. We want to emphasise that the fine scale features visible are inside the dense mesh area.}
            \label{fig:buffer}
        \end{figure}

        The patches that are extracted never intersect with the exterior boundaries of the FE mesh. Additionally, we apply homogeneous Dirichlet boundary conditions away from the material specimen, as represented in [Fig  \ref{fig:buffer}], [Eq. \ref{eq:BCs}].
        
        \begin{align}
            u &= 
            \begin{bmatrix}
                E_{ \text{xx} } & E_{ \text{xy} } & E_{ \text{xz} } \\
                E_{ \text{xy} } & E_{ \text{yy} } & E_{ \text{yz} } \\
                E_{ \text{xz} } & E_{ \text{yz} } & E_{ \text{zz} }
            \end{bmatrix}
            (X - X_0)^\top
        \label{eq:BCs}
        \end{align}
        
        where $E_{xx}$, $E_{xy}$, $E_{xz}$, $E_{yy}$, $E_{yz}$, $E_{zz}$ are far-field load parameters, $X$ is the position of a point in $\mathbb{R}^3$ and $X_0$ is the initial position of the centre of the body in $\mathbb{R}^3$.

        A linear rescaling technique allows us to extrapolate results from a given load amplitude to another. This technique is fully detailed in our previous paper \citep{krokos2021bayesian}, and recalled in [Appendix \ref{appendix:preprocessing}]. The principle is to normalise the input (macro) stress tensor by the maximum over the patch of a chosen measure of this stress tensor, yielding stress fields whose pointwise measure ranges from -1 to 1. After inference, the output (micro) stress is multiplied by the normalisation factor. As our mechanical model is linear, this procedure yields equivariant predictions with respect to changes in load amplitude. The scaling measure chosen in this work is the maximum absolute value of all components of the stress tensor.

        The dataset we use consists of 200 FE simulations completed in 43.5 CPU hours on an Intel\textsuperscript{\tiny\textregistered} Xeon\textsuperscript{\tiny\textregistered} Gold 6148 CPU @ 2.40GHz CPU. From every FE simulation we extract 5 patches, at random locations in the cubic specimen, resulting in 1,000 data points, 900 for training and 100 for testing.
   
        %-----------------------------------------------------------------------------------------------------------------------------------
        \subsubsection{Training parameters} \label{GNN_parameters}
        
        For the training of the network we used a batch size of 2 and the Adam optimiser with an initial learning rate of $1 \cdot 10^{-4}$ decreasing by 5\% every 50 epochs. Additionally, we identified some key parameters for the training of the GNN and performed several tests to identify their optimum values. We concluded that for achieving optimum behaviour we will use a GNN with 5 GN blocks, residual connection, independent encoder, 128 filters and a maximum of 10 neighbours per node. We stress that the neighbours are calculated as a preprocessing step and thus remain the same throughout the GNN training. The total number of trainable parameters for the network is 842,654. More details can be found in [Appendix \ref{appendix:GNN_parameters}].

        In the GNN architecture examined in this work, there are two types of layers that have learnable parameters, namely the linear and layer normalisation layer. For both layers PyTorch Geometric’s default initialisation methods are being used. For the linear layer, the weights are initialised using the kaiming uniform method, as described by \citep{he2015delving}, with the default parameters used by PyTorch, namely $a=\sqrt{5}$. The biases are initialised using uniform initialisation, where values are drawn from a uniform distribution between 0 and 1. For the layer normalisation the weights are initialised to 1 and the biases to 0. The same methods are applied to all the GNNs trained in this work.
        
        Training on 900 patches using the parameters identified above requires 6 hours on an NVIDIA V100 GPU with 16GB of RAM with a maximum memory requirement of 4.3GB.

        %-----------------------------------------------------------------------------------------------------------------------------------
        \subsubsection{GNN graph} \label{GNN graph}

\begin{figure}[htb]
            \centering
            \begin{subfigure}{.5\textwidth}
              \centering
              \includegraphics[trim={5cm 2cm 2cm 5cm}, width=\linewidth]{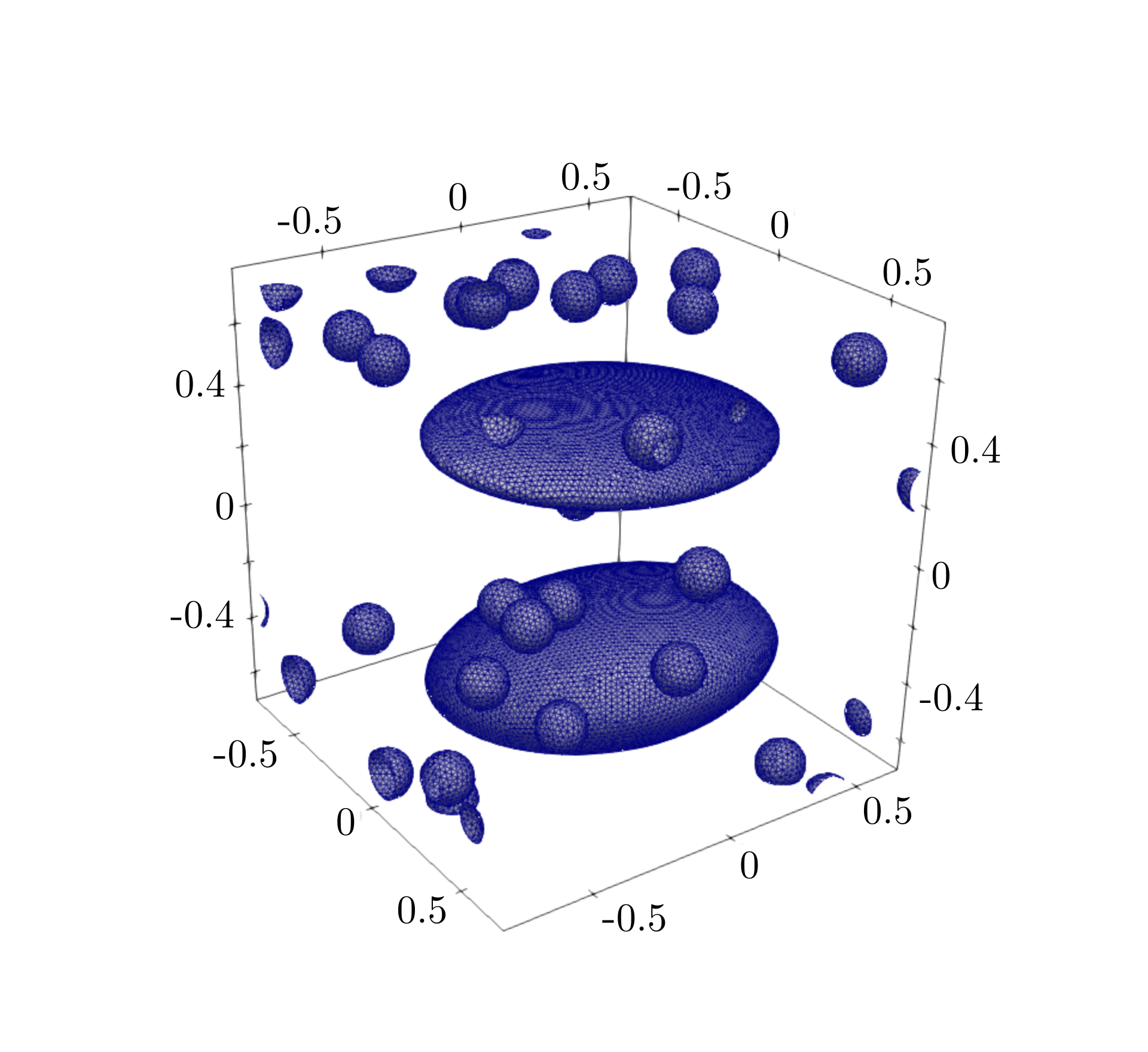}
              \caption{FE mesh of the patch}
            \end{subfigure}%
            \begin{subfigure}{.5\textwidth}
              \centering
              \includegraphics[width=1\linewidth]{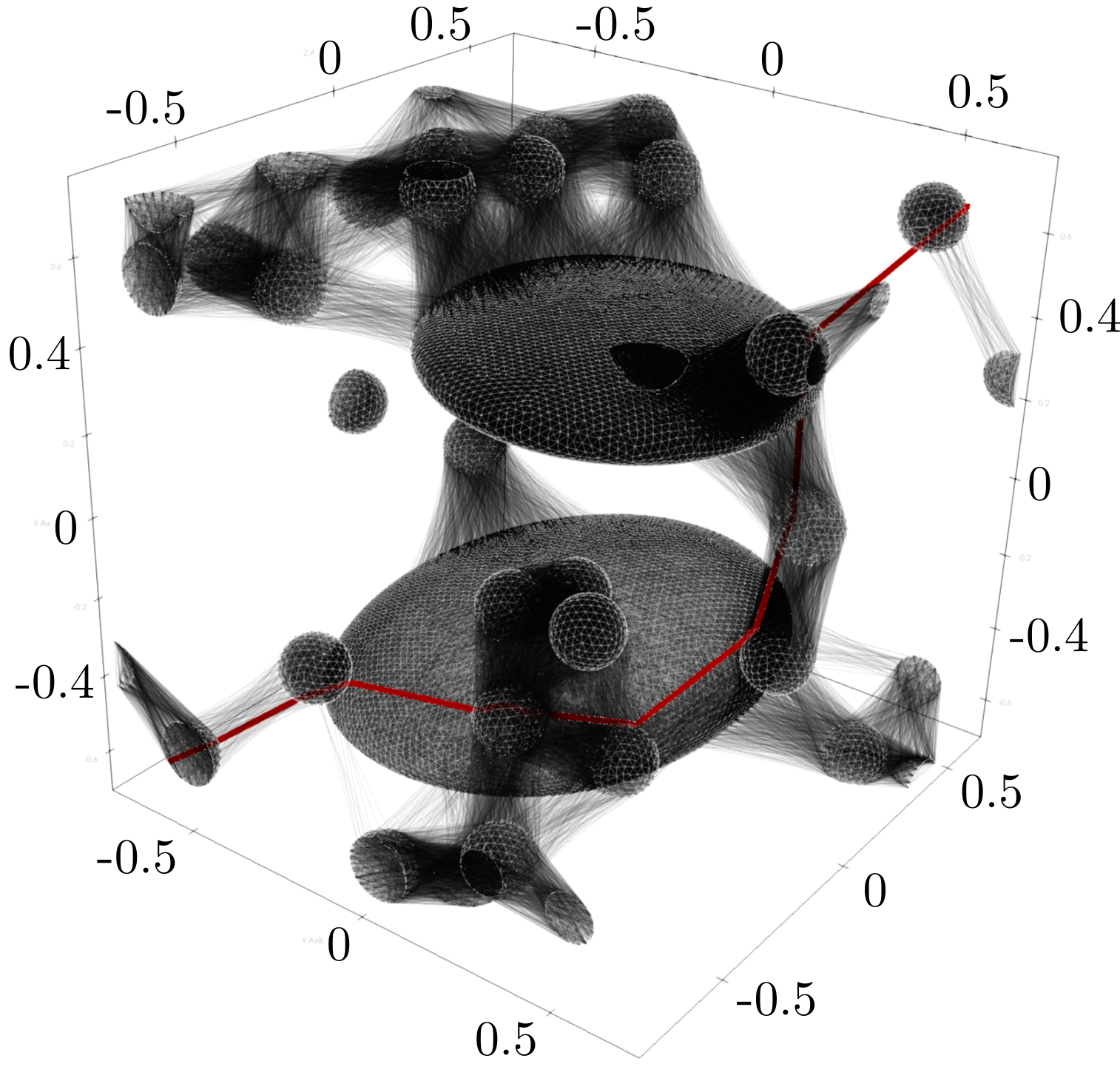}
              \caption{Graph of the patch used for the GNN training}
            \end{subfigure}
            \caption{
            On the left (a) we can see the surface mesh of a patch. On the right (b) we can see the graph of the same patch that will be used for the GNN training. The graph connects the nodes of the surface mesh using Euclidean edges. With red we can see the shortest path that connects two points of the graph located on opposite corners.}
            \label{fig:GNN_graph}
        \end{figure}

        In this section we provide information about the graph used for the GNN training. The graph is constructed using Euclidean edges as described in [Section \ref{neighbourhood}]. An example can be found in [Fig \ref{fig:GNN_graph}], where both the surface mesh and the graph can be seen. A threshold of 10 neighbours per node is being used and a radius of 0.32 for the radius neighbours. The patch has 20,113 points. The shortest path to connect 2 points of the graph located on opposite corners is 24 steps.

        %-----------------------------------------------------------------------------------------------------------------------------------
        \subsubsection{BGNN MAP prediction}

        In this section we will examine the mean prediction provided by the BGNN, without referring to the uncertainty estimation. All the results refer to the validation set.
        
\begin{figure}[p]
            \centering
            \begin{subfigure}[b]{0.9\textwidth}
              \centering
              \includegraphics[ trim={5cm 0cm 5cm 0cm},, width=\linewidth]{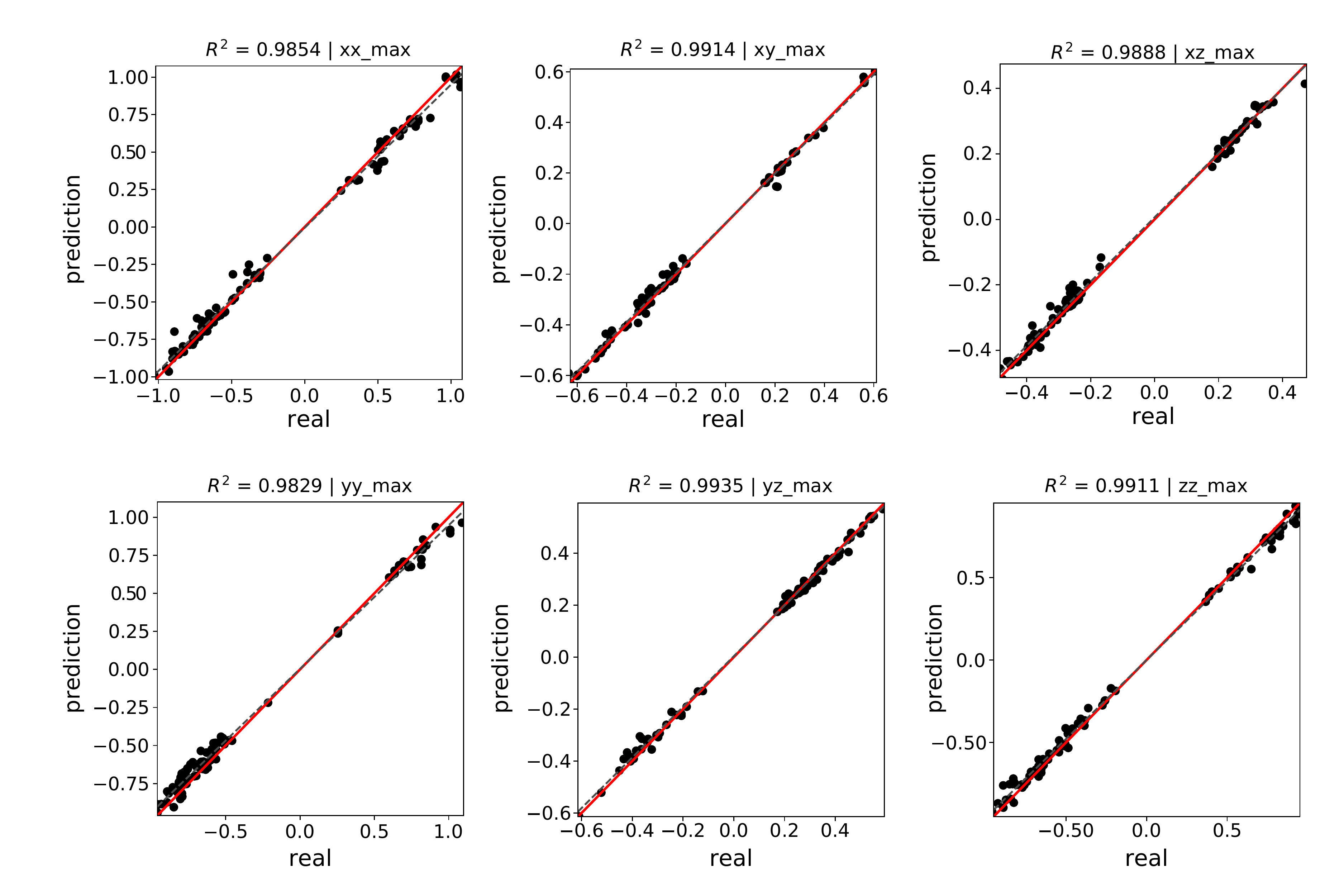}
              \caption{maximum microscale components}
              \label{fig:max_predictions:a}
            \end{subfigure}%
            \vspace{20pt}
            \begin{subfigure}[b]{0.5\textwidth}
              \centering
            \includegraphics[trim={0cm 0cm 0cm 1cm}, width=\linewidth]{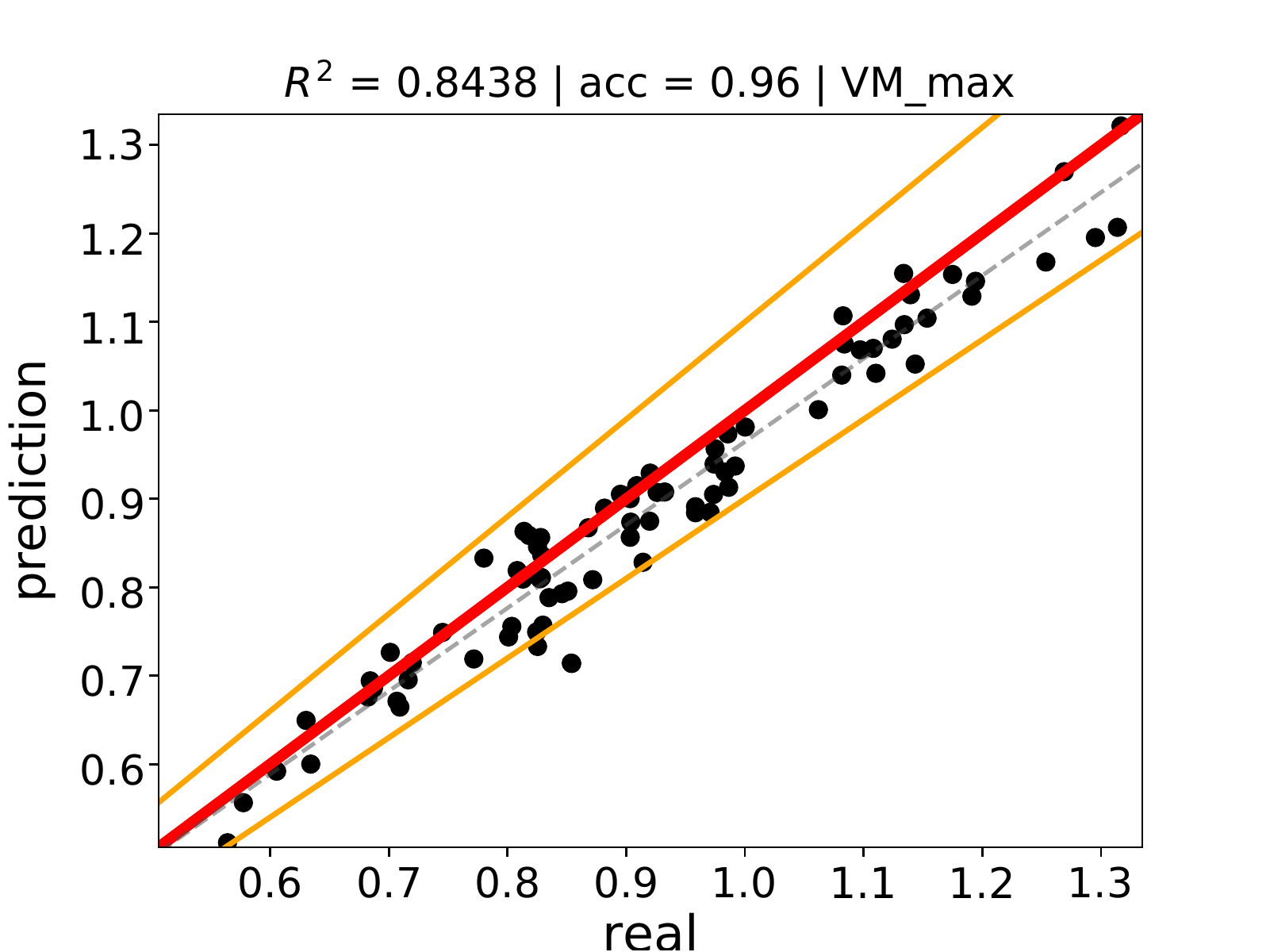}
              \caption{maximum microscale Von Mises stress}
              \label{fig:max_predictions:b}
            \end{subfigure}
            \caption{In both subfigures the x-axis corresponds to the FE prediction in the ROI of each patch in the validation set and the y-axis to the BGNN prediction in the ROI of each patch in the validation set. In the top subfigure, (a), we observe 6 diagrams corresponding to each of the maximum absolute microscale stress components. In the bottom subfigure, (b), we observe the maximum Von Mises stress.}
            \label{fig:max_predictions}
        \end{figure}

        In [Fig \ref{fig:max_predictions:a}] we observe that the maximum stress components predicted by the BGNN are in a good agreement with the ones calculated by the FEA. Specifically, the coefficient of determination, $R^2$, for all the 6 stress components is larger than 0.98. In [Fig \ref{fig:max_predictions:b}] we observe similar results for the maximum Von Mises stress. Although the points are more scattered and the coefficient of determination dropped to 0.84, the accuracy is 96\%.
        
\begin{figure}[htb!]
            \begin{center}
                \includegraphics[width=\linewidth]{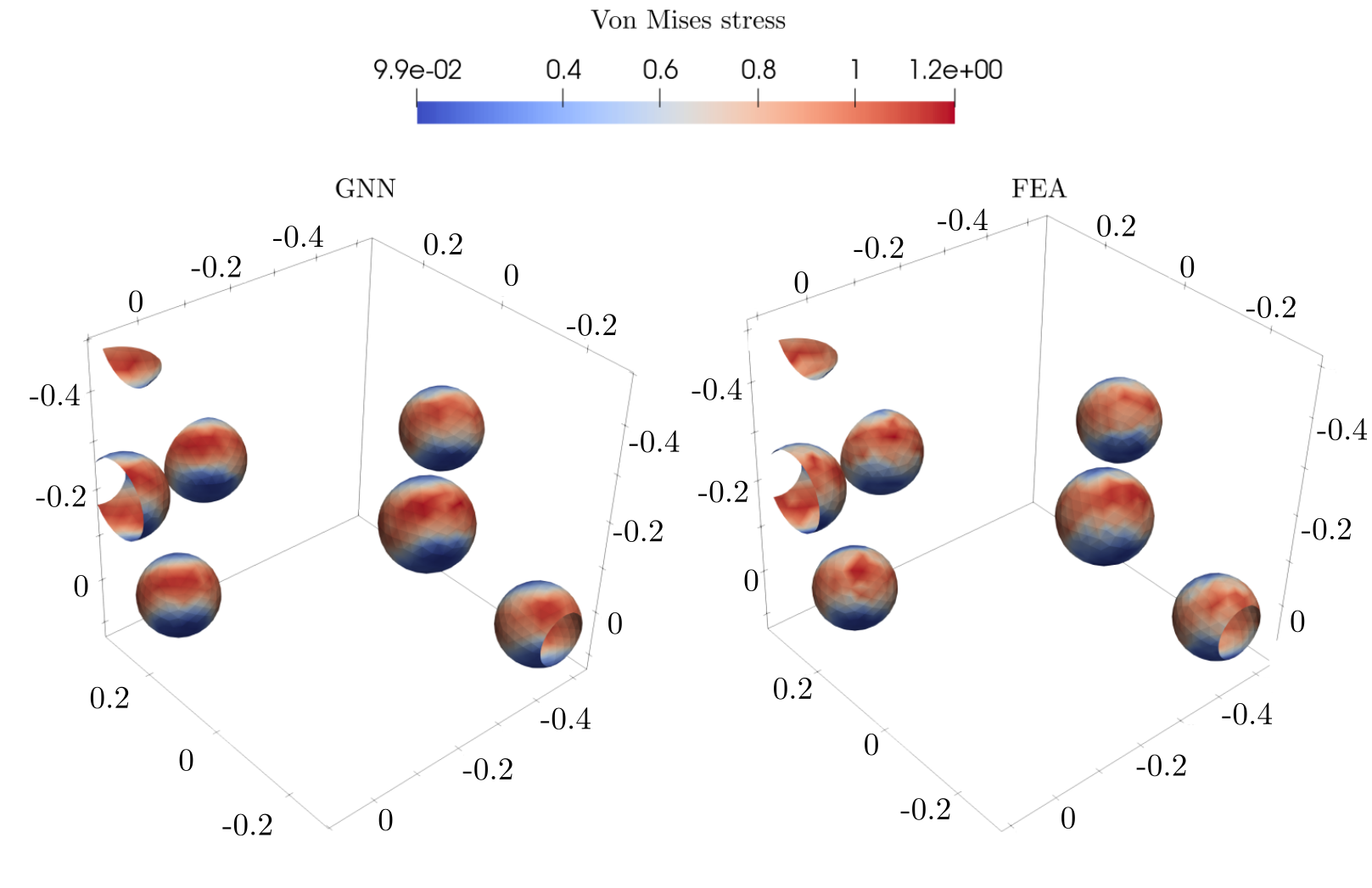}
                \caption{Comparison between the Von Mises stress distribution as calculated by FEA (right) and the Von Mises stress distribution as predicted by the GNN (left) on a patch from the validation set.}
                \centering
                \label{fig:example_0}
            \end{center}
        \end{figure}

        Additionally, we examine the stress distribution on the FE mesh. In [Fig \ref{fig:example_0}] we compare the microscale Von Mises stress in the ROI of a patch as predicted by the GNN with the one calculated by FE simulations. We observe that the predicted stress distribution is very close to the ground-truth.
        
        % \clearpage
        %-----------------------------------------------------------------------------------------------------------------------------------
        \subsubsection{BGNN uncertainty estimation}
        
\begin{figure}[htb]
              \centering
              \includegraphics[width=.7\linewidth]{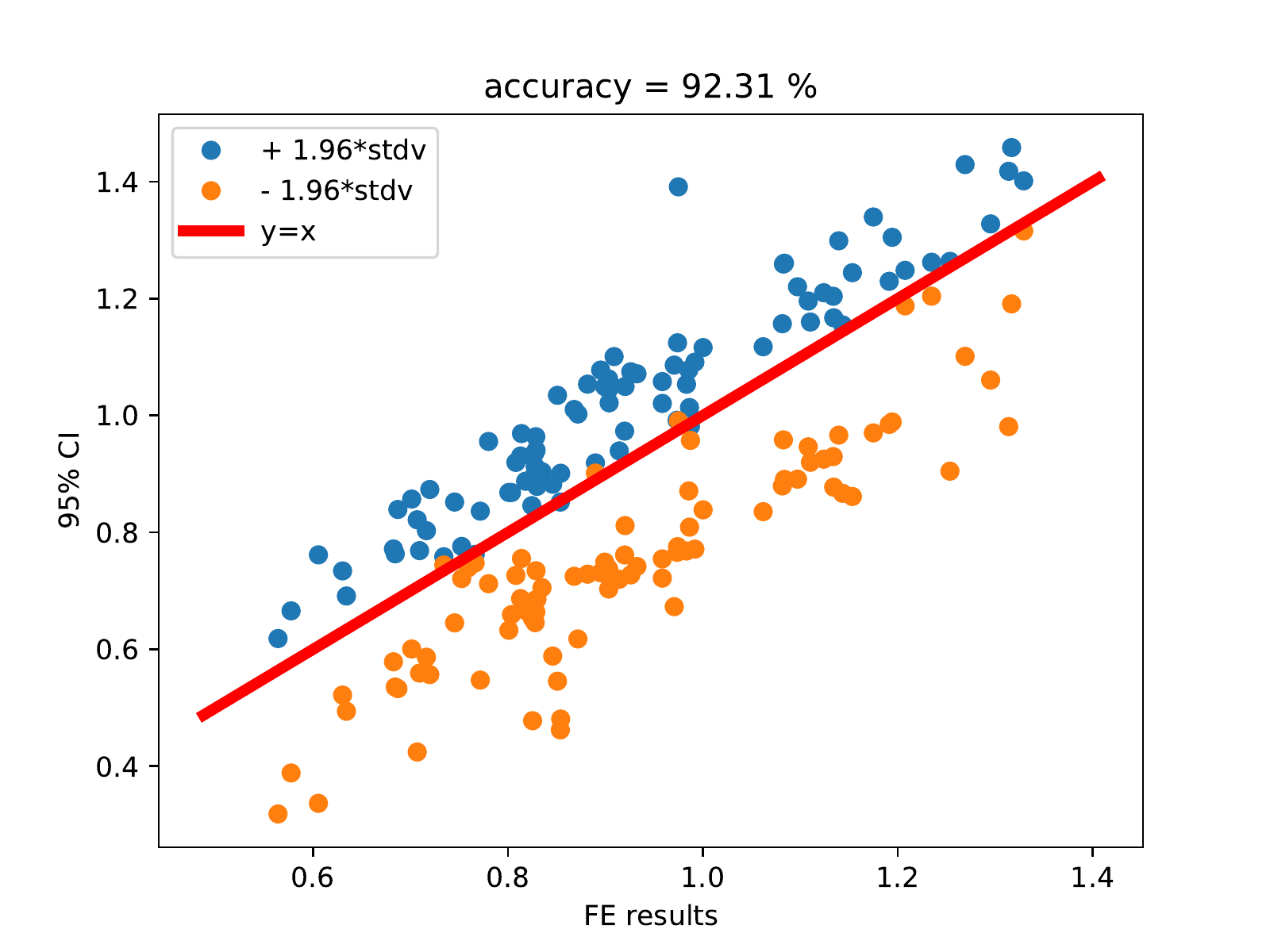}
        \caption{In this figure we see the upper and lower 95\% CIs for the maximum Von Mises stress prediction, blue and orange points respectively. The BGNN is able to include the real value in its 95\% CIs in 92\% of the cases.}
        \label{fig:BBB_CI}
        \end{figure}

        After studying the quality of the predictions provided by the mode of the BGNN, we propose to evaluate the ability of the BGNN to quantify the uncertainty of the prediction and provide credible intervals (CIs) that reflect the error in the prediction. From [Fig \ref{fig:BBB_CI}] we can see that for the validation set the percentage of points inside the 95\% CIs is 92\%, which is satisfactory.
        
        \clearpage
    %-----------------------------------------------------------------------------------------------------------------------------------
    \subsection{Online bias correction}
    
    In this section we will demonstrate the use of the Ensemble Kalman method, section [\ref{EKF section}], to improve the BGNN prediction. We will examine two realistic cases, first case is to improve the prediction of an under trained network and second case is to improve the prediction of a network in an extrapolation regime.
        
        %------------------------------------------------------------------------------------------------------------------------------------------
        \subsubsection{Under trained GNN}
        
\begin{figure}[htb]
            \centering
            \begin{subfigure}{.5\textwidth}
              \centering
              \includegraphics[width=\linewidth]{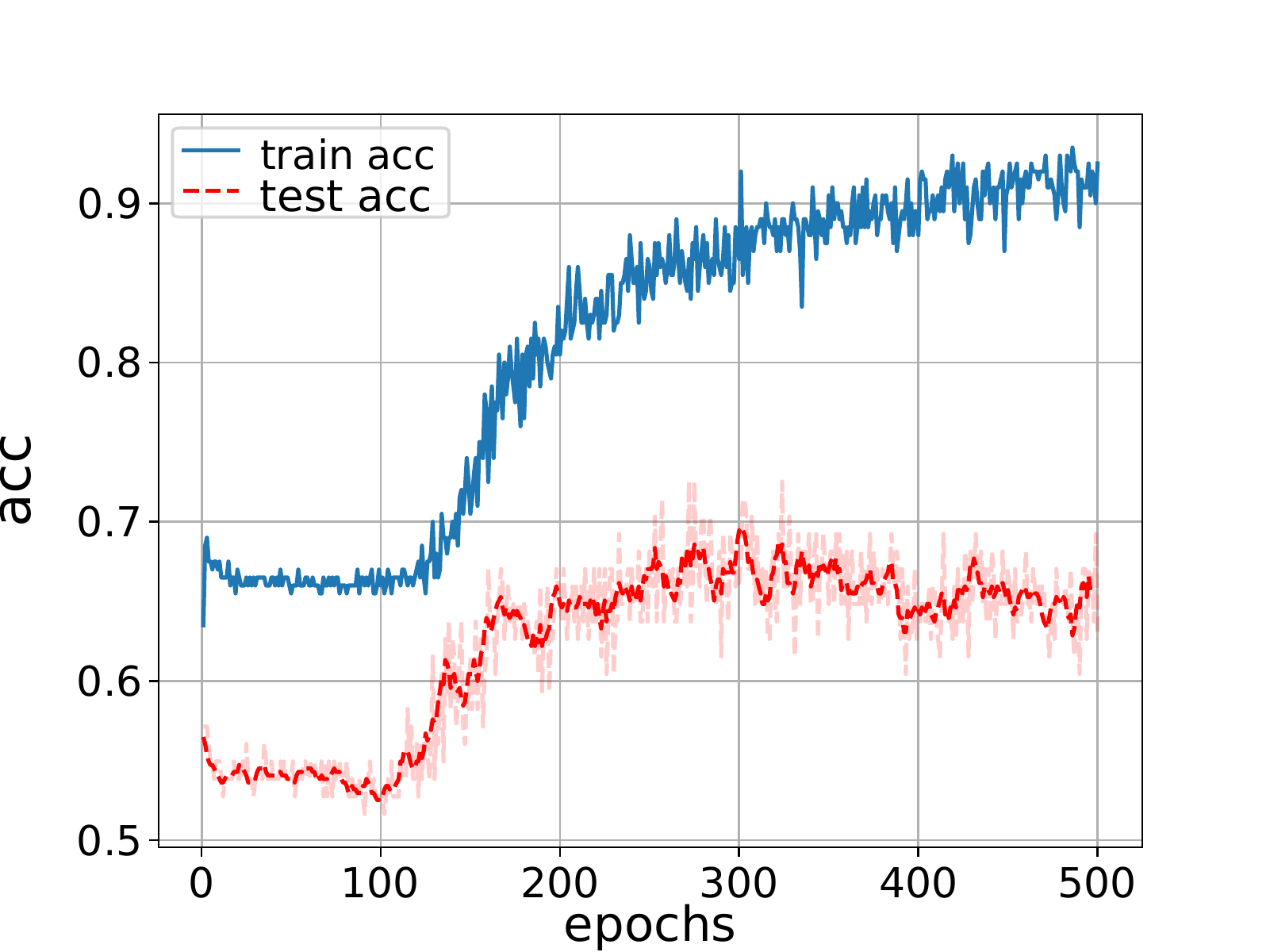}
              \caption{32 FE simulations}
              \label{fig:underTrained_VS_Trained:a}
            \end{subfigure}%
            \begin{subfigure}{.5\textwidth}
              \centering
              \includegraphics[width=\linewidth]{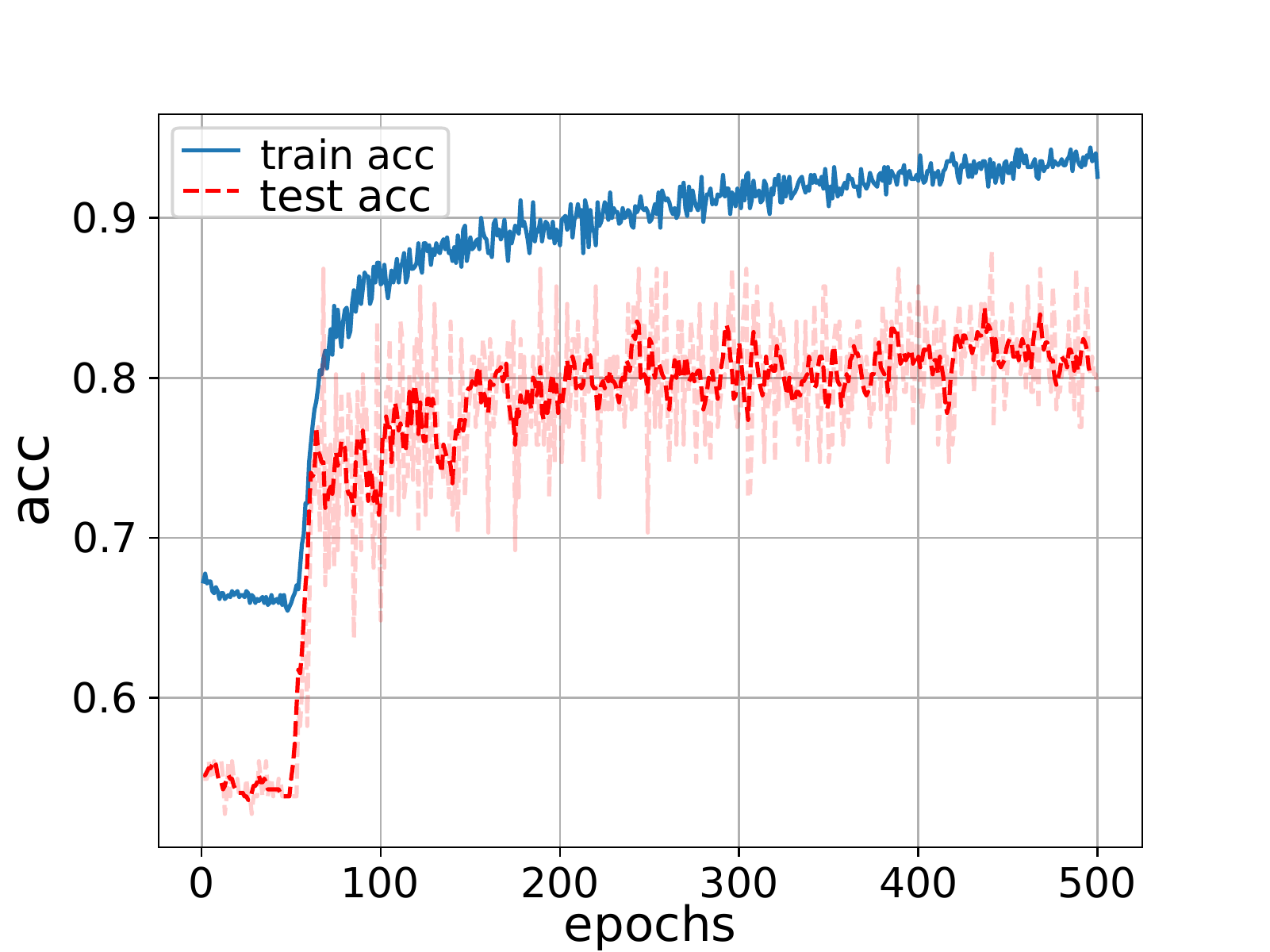}
              \caption{100 FE simulations}
              \label{fig:underTrained_VS_Trained:b}
            \end{subfigure}
            \caption{In both diagrams we see accuracy curves defined using the mean BGNN prediction for the maximum Von Mises stress. Specifically, we see the accuracy as function of the training epochs for the training set (blue) and the test set (red). In the diagram on the left (a) the GNN is trained with 32 FE simulations, 160 patches, and in the diagram on the right (b) the GNN is trained with 100 FE simulations, 500 patches.}
            \label{fig:underTrained_VS_Trained}
        \end{figure}

        With the phrase under trained we refer to a network for which we use a small dataset for training. In [Fig \ref{fig:underTrained_VS_Trained}] we show the accuracy plots for the mean BGNN prediction as a function of the epochs for 2 GNNs trained with the same parameters but the one on the left uses data from 32 FE simulations for training and the one on the right uses data from 100 FE simulations for training. The BGNN trained with more data performs better than the undertrained network. 
        
\begin{figure}[htb!]
            \begin{center}
                \includegraphics[width=.75\linewidth]{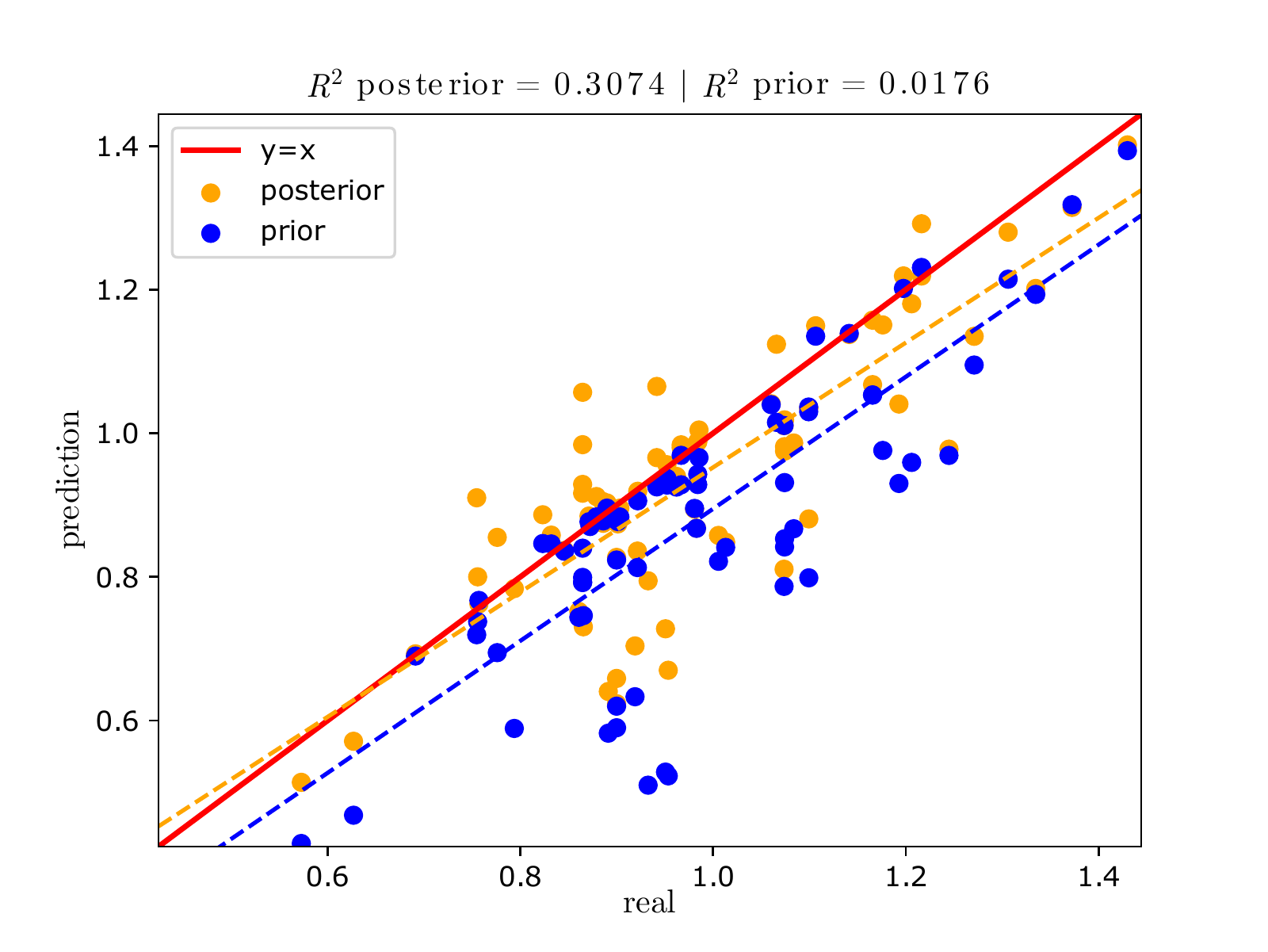}
                \caption{A diagram showing the posterior (yellow points) maximum Von Mises prediction and the prior (blue points) maximum Von Mises prediction. We observe that the posterior, after the stress update, has a larger coefficient of determination compared to the prior.}
                \centering
                \label{fig:prior_vs_post_mean_undertrained}
            \end{center}
        \end{figure}

        Using the Ensemble Kalman method we want to update the output distribution of the BGNN, prior, and obtain the posterior. In [Fig \ref{fig:prior_vs_post_mean_undertrained}] we can see the posterior and the prior for the prediction of the maximum Von Mises stress on the validation set. The coefficient of determination between the posterior and the FE results has a value of 0.3074 which is larger than the  coefficient of determination between the prior and the FE results that has a value of 0.0176. Also, we can see that best line that fits the posterior data is closer to the $y=x$ line (ideal result) compared to the best line that fits the prior data, i.e. the output of the BGNN, without the online correction. Therefore, the posterior mean better fits the data compared to the prior mean and thus the online stress correction resulted in an improvement of the mean BGNN prediction.
        
\begin{figure}[htb]
            \begin{center}
                \includegraphics[width=\linewidth]{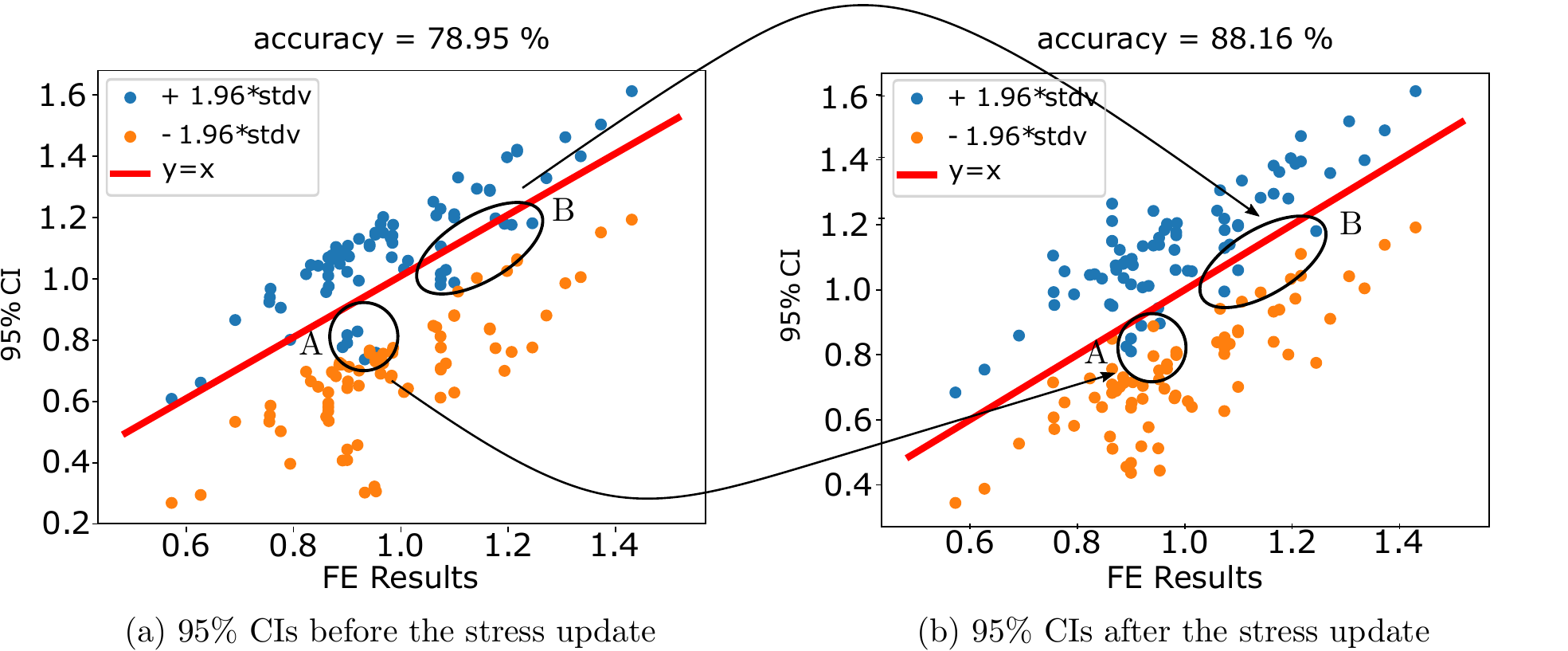}
                \caption{Comparison between the prior and posterior 95\% CIs for the maximum Von Mises stress in the ROI of the patches for a BGNN trained with only 32 FE simulation data. The diagram on the left, (a), is showing the upper and lower 95\% CIs, blue and orange points respectively, before the stress update. While the diagram on the right, (b), is after the stress update. Two clusters, A and B, are marked before and after the stress correction. This highlights that blue points, +95\% CIs, which were below the $y=x$ line for the prior case, have moved towards the $y=x$ line and in a lot of cases surpassed it, in the posterior case.}
                \centering
                \label{fig:EKF_UQ_smallData}
            \end{center}
        \end{figure}

        Furthermore, we want to examine the posterior distribution and not only the posterior mean value. In [Fig \ref{fig:EKF_UQ_smallData}] we see the comparison between the prior 95\% CIs (left) and the posterior 95\% CIs (right) for the validation set. Firstly, we can observe that even in the prior case, where we have only used 32 FE simulations, the 95\% CIs give a good estimation of where the real value is. We can see that the percentage of points that are inside the 95\% CIs is 79\%. Additionally, we see that this value climbs to 88\% in the posterior case. This clearly means that by using the online stress update we were able to improve the BGNN prediction without adding more data to the training dataset. We note that we only update the mean value, and the posterior standard deviation is taken equal to the prior one.
        
\begin{figure}[htb]
            \begin{center}
                \includegraphics[width=\linewidth]{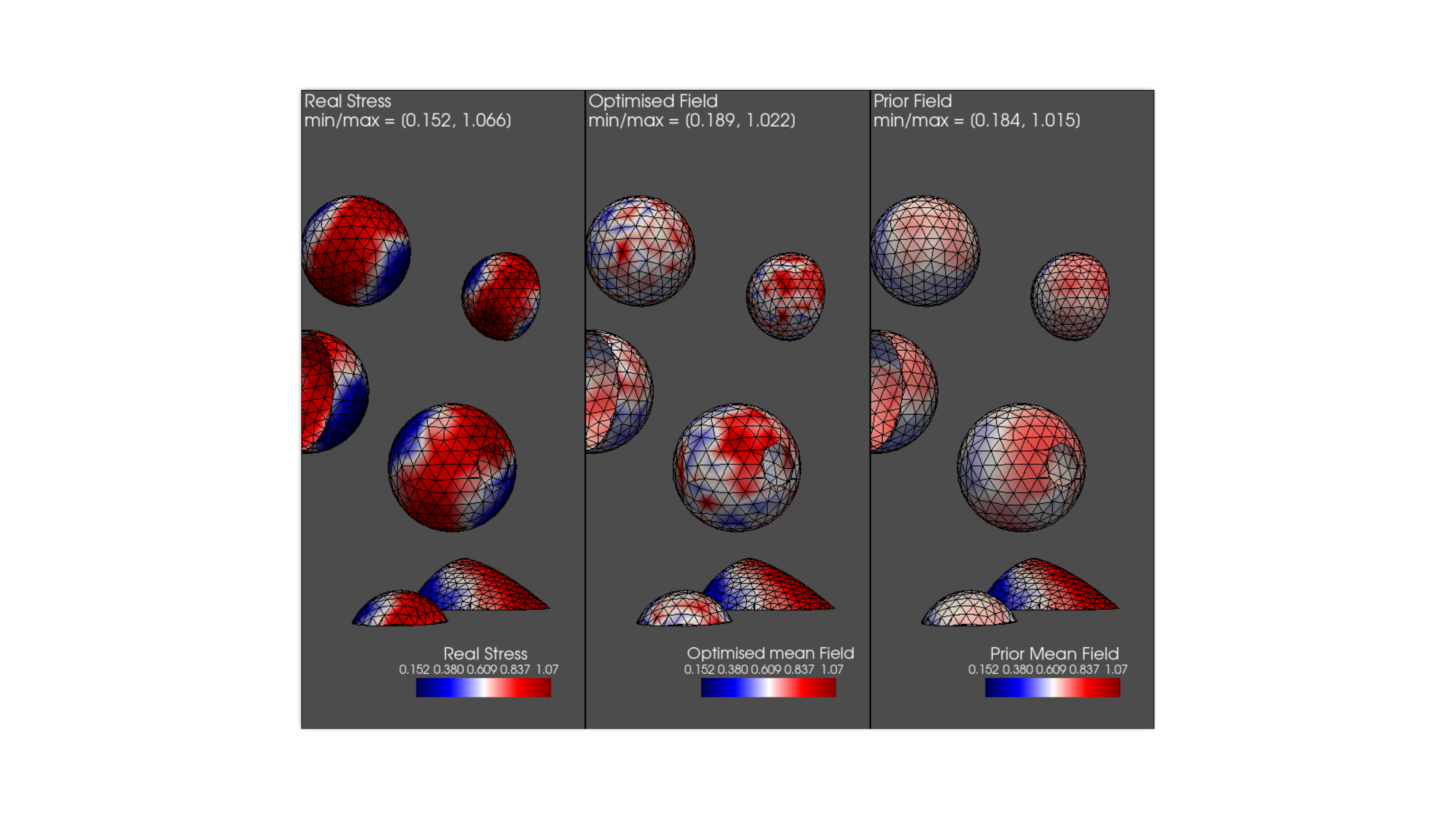}
                \caption{Comparison between the Von Mises stress before and after the online stress update with the one calculated through FE simulations. On the left we see the FE results, in the middle the posterior and on the right the prior. For the posterior and prior we show the mean prediction.}
                \centering
                \label{fig:EKF_UQ_smallData_distribution}
            \end{center}
        \end{figure}

        Lastly, in [Fig \ref{fig:EKF_UQ_smallData_distribution}] using a patch from the validation set we compare the mean Von Mises stress distribution on the mesh before and after the stress update with the one calculated through FE simulations. We can see that the posterior, middle image, is closer to the FE results, image on the left, compared to the prior, image on the right.
        
        %\clearpage
        %------------------------------------------------------------------------------------------------------------------------------------------
        \subsubsection{Prediction for out-of-training microstructural inputs}
        
\begin{figure}[htb]
            \centering
            \begin{subfigure}{.5\textwidth}
              \centering
              \includegraphics[width=\linewidth]{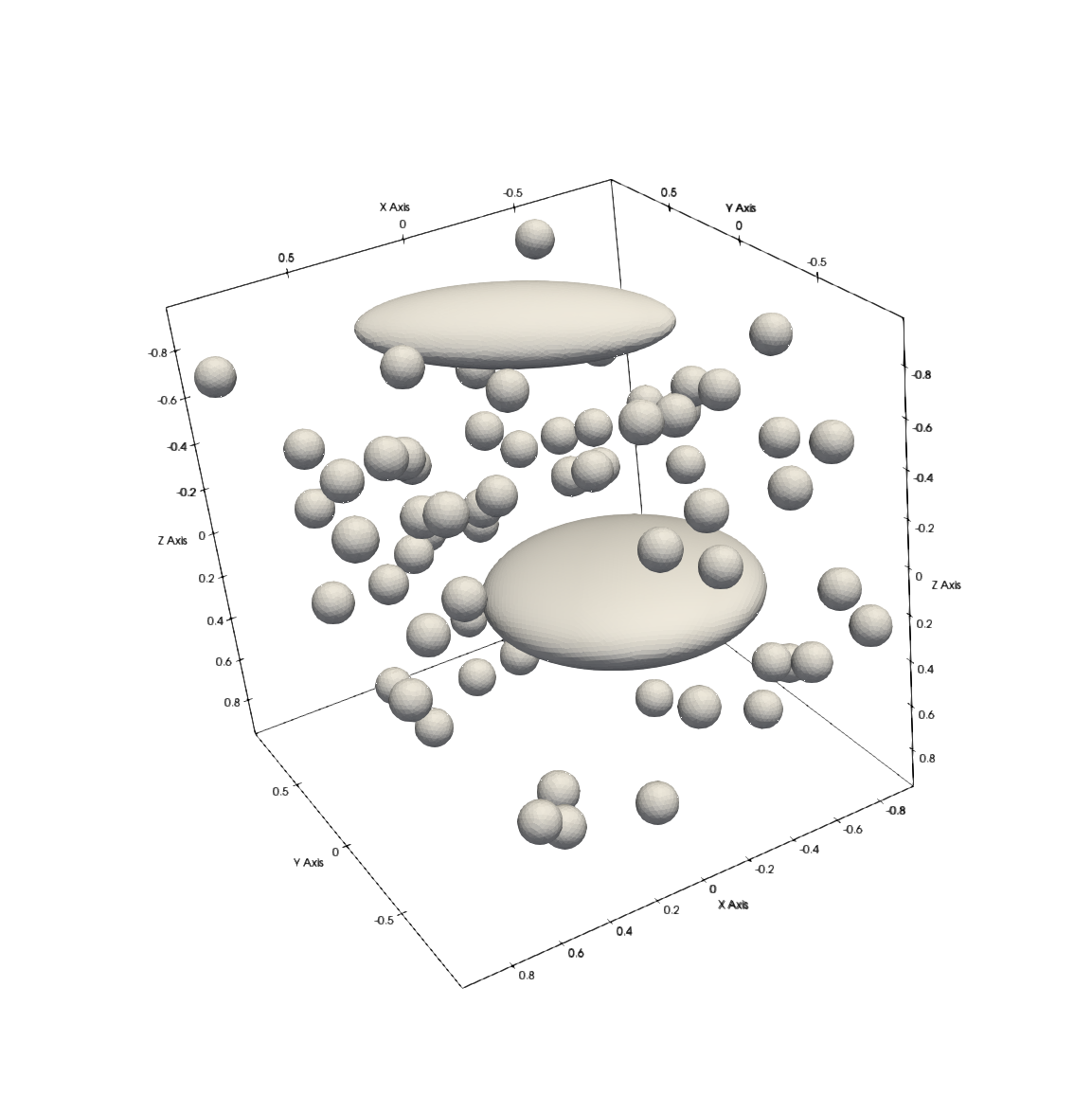}
              \caption{realisation of the training dataset}
              \label{fig:cube_spheres_ellipses:a}
            \end{subfigure}%
            \begin{subfigure}{.5\textwidth}
              \centering
              \includegraphics[width=.75\linewidth]{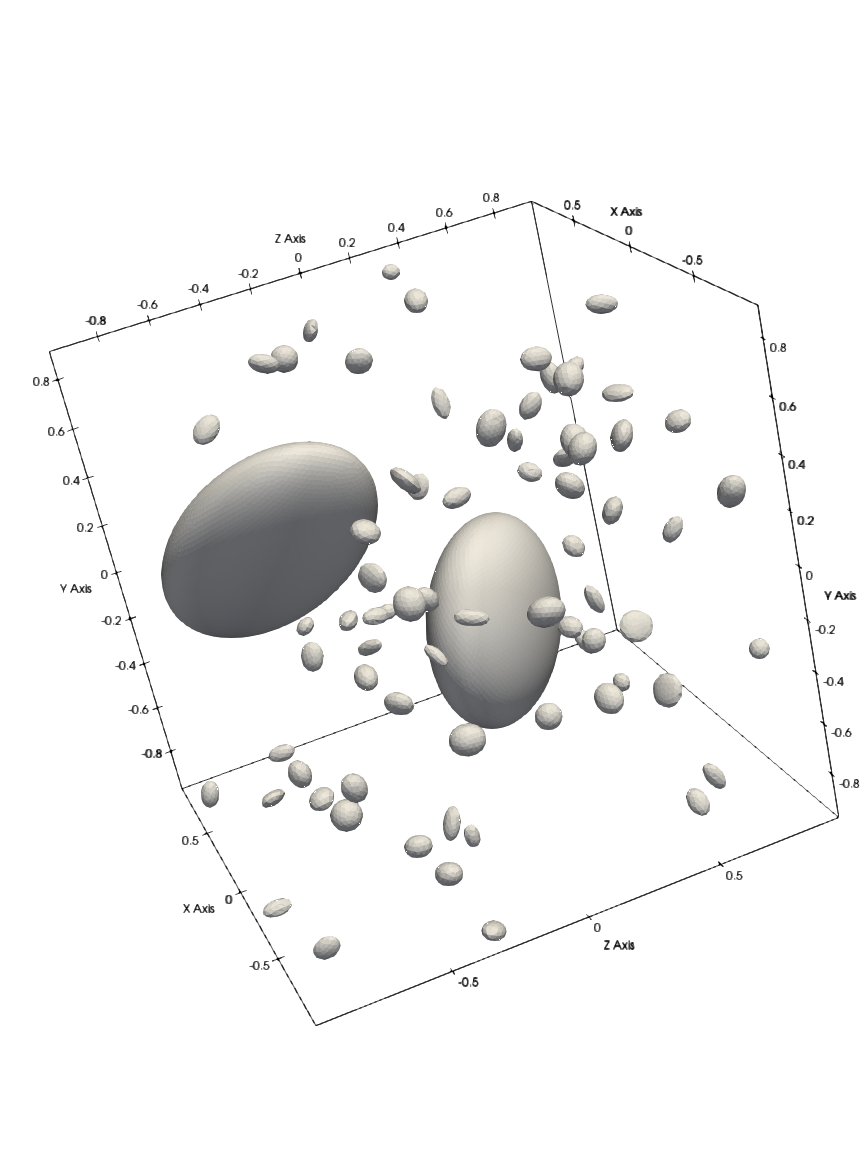}
              \caption{realisation of the test dataset}
              \label{fig:cube_spheres_ellipses:b}
            \end{subfigure}
            \caption{On the left (a) we see a realisation of the training dataset and on the right (b) of the test dataset. For the training dataset the porous phase is composed of spheres while for the test set of ellipsoids.}
            \label{fig:cube_spheres_ellipses}
        \end{figure}

        In this section, with the phrase \say{out-of-training} we refer to a network trained with spherical defects but evaluated in cases that have elliptical defects. The BGNN is extrapolating which is a particularly undesirable but common scenario for real applications. In [Fig \ref{fig:cube_spheres_ellipses}] we show a realisation of the training dataset on the left and a realisation of the test dataset on the right.
        
\begin{figure}[htb]
            \begin{center}
                \includegraphics[width=\linewidth]{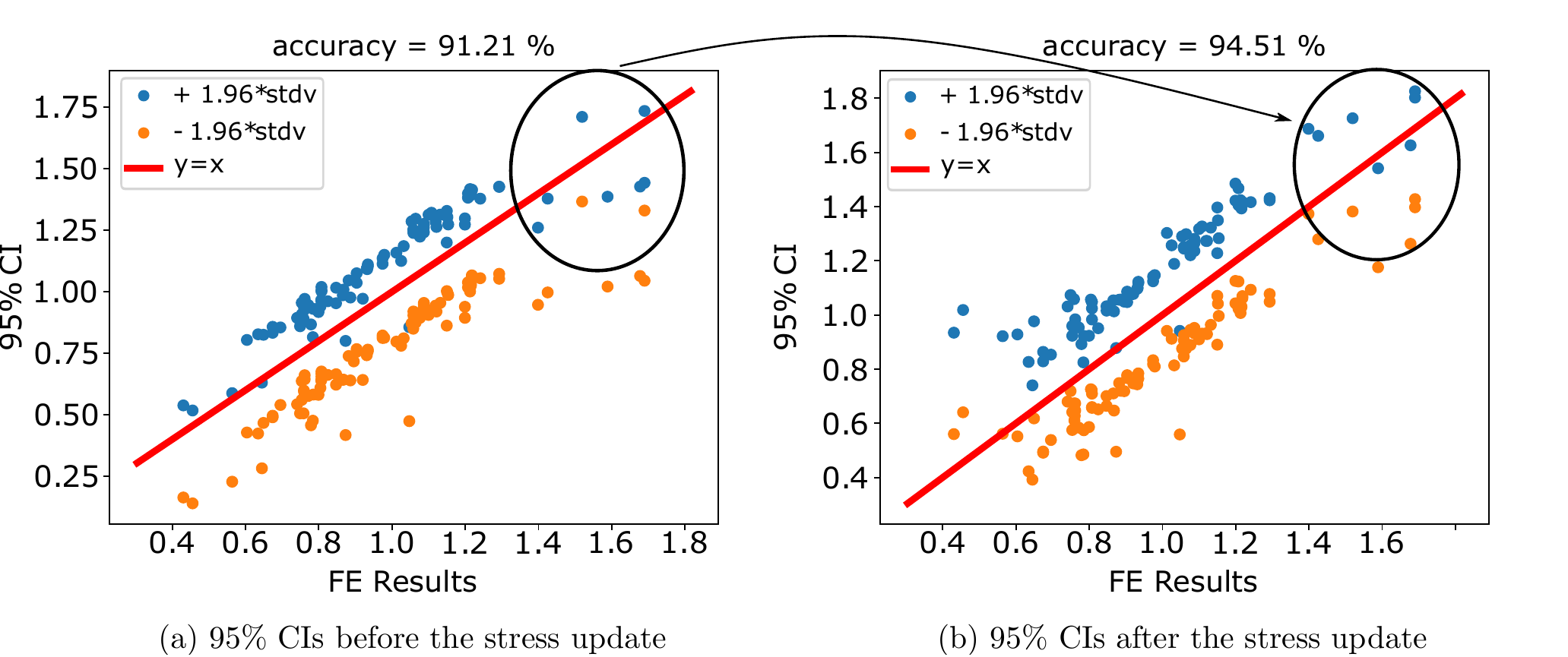}
                \caption{Comparison between the prior and posterior 95\% CIs for the maximum Von Mises stress in the ROI of the patches for a BGNN trained on a dataset with spheres as microscale features and tested on a dataset with ellipses as microscale features. The diagram on the left, (a), is showing the upper and lower 95\% CIs, blue and orange points respectively, before the stress update. While the diagram on the right, (b), is after the stress update. A cluster of points is marked before and after the stress correction. This  highlights that blue points, +95\% CIs, which were below the $y=x$ line for the prior case, have moved towards the $y=x$ line and in a lot of cases surpassed it, for the posterior case.}
                \centering
                \label{fig:EKF_UQ_OOD}
            \end{center}
        \end{figure}

        In [Fig \ref{fig:EKF_UQ_OOD}] we see the comparison between the prior 95\% CIs (left) and the posterior 95\% CIs (right). Firstly, we can observe that even in the prior case, where the BGNN extrapolates, the 95\% CIs give a good estimation of where the real value is. We can see that the percentage of points that are inside the 95\% CIs is 91\%. Additionally, we see that this value climbs to 95\% in the posterior case. Most importantly we see that using the ensemble Kalman method we managed to avoid underpredicting the maximum values. This clearly means that by using the Ensemble Kalman method we were able to improve the BGNN prediction without adding more data to the training dataset. We note that we only update the mean value, and the posterior standard deviation is taken equal to the prior one. 
        
        We stress that the results of this section do not suggest that the GNN framework could be used for extrapolation away from the dataset. Instead, they indicate that given a reasonably accurate prediction where a downward bias is observed for the highest stress values, the proposed online stress correction technique may be deployed to correct this bias upward.
        
    \clearpage
    
    %------------------------------------------------------------------------------------------------------------------------------------------
    \subsection{Dogbone specimen}\label{const dogbone dataset}

    After demonstrating the applicability of our GNN and studying the online correction technique, in this subsection we apply the proposed deep learning methodology to a more challenging case. We also provide multiple examples of the Von Mises stress distribution on the surface of the mesh using data from the validation set. 
    
        \subsubsection{Training dataset}
        
\begin{figure}[htb]
            \begin{center}
                \includegraphics[width=0.95 \linewidth]{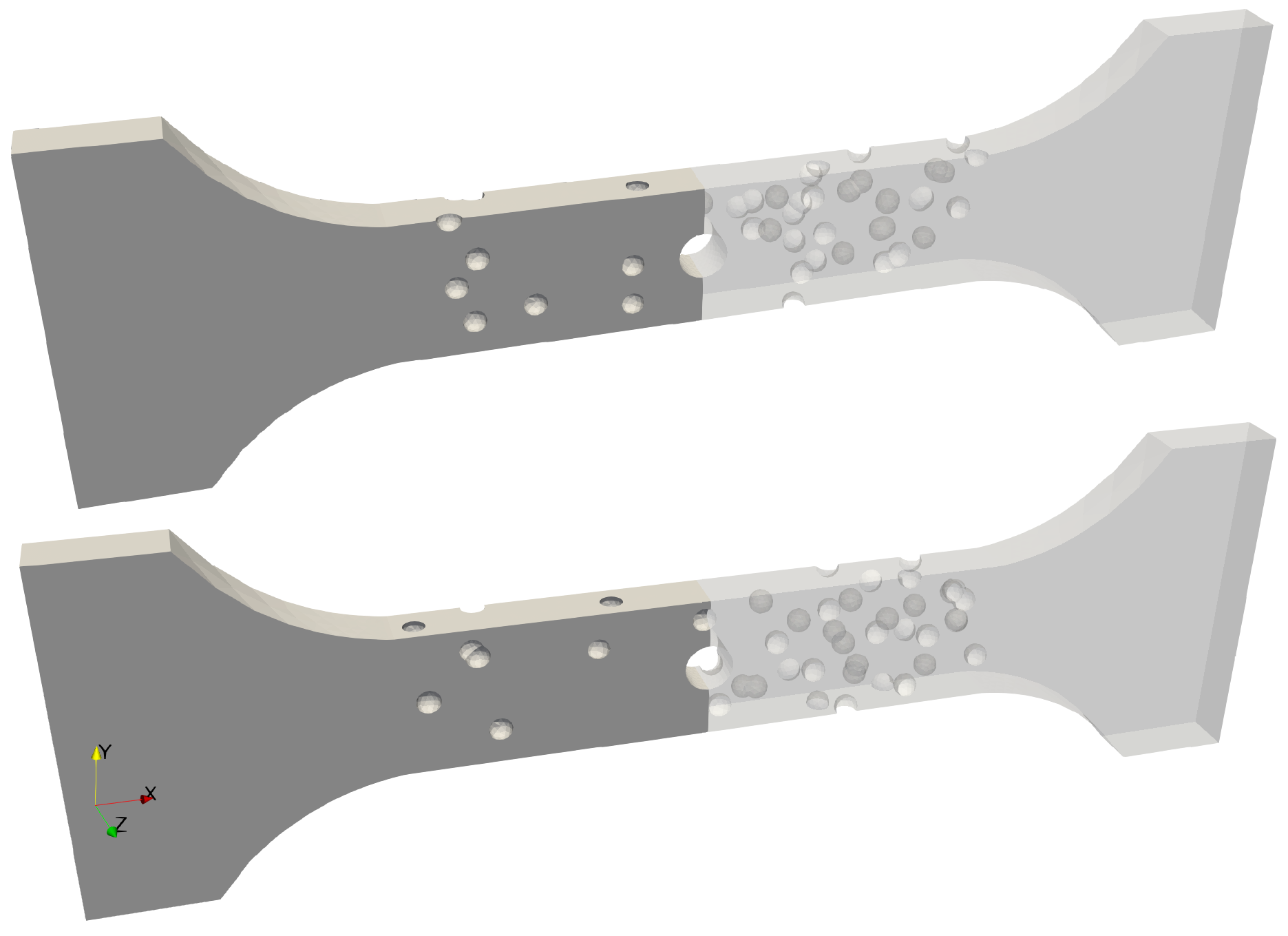}
                \caption{Two realisations of the dogbone used to train the GNN. Each dogbone has a cylinder-shaped hole and the porous material is defined via a random distribution of spherical voids.}
                \centering
                \label{fig:ConstDogEBoneGeoms}
            \end{center}
        \end{figure}

        The geometry that we choose to model is a dogbone with a single hole in the middle that does not change from one realisation to another, and with a random distribution of 50 to 100 spherical pores. The size of the dogbone is 2 units along the $x$ axis (length), 0.6 units along the $y$ axis (height) and 0.08 units along the $z$ axis (width). The radius of the hole is 0.032 units, and the radius of the spherical pores is 0.016 units. The spherical pores may intersect with each other, with the boundaries of the geometry and the hole. For the macroscale simulations, input of the BGNN, only the hole is taken into account and the spherical pores are ignored. In [Fig \ref{fig:ConstDogEBoneGeoms}] we show two examples of the geometry.
        For the boundary conditions we apply displacements on those faces of the dogbone specimen that are perpendicular to the $x$ axis. The prescribed displacement has the same magnitude for both sides, and is applied along the positive outer normal direction. Zero displacement is applied along the other two spatial directions. Lastly, The Young’s modulus and the Poisson ratio of the structure are 1 and 0.3 respectively.
        
        We performed 500 FE simulations completed in 57 CPU hours on an Intel\textsuperscript{\tiny\textregistered} Xeon\textsuperscript{\tiny\textregistered} Gold 6148 CPU @ 2.40GHz CPU. From every FE simulation we extract 10 patches resulting in 5,000 data points, 4,000 for training and 1,000 for testing. As explained in section [\ref{train_dataset_cube}] the patch is of size $[18R \; \times \; 18R]$ and the ROI is of size $[8R \; \times \; 8R]$, where $R$ is the radius of the spherical pores.
        
        \subsubsection{GNN parameters} \label{dogbone GNN parameters}
        
        For the GNN training we used the Adam optimiser with an initial learning rate of $1 \cdot 10^{-4}$ decreasing by 5\% every 50 epochs. Additionally, we started with the parameters that we identified in [\ref{GNN_parameters}], namely 5 GN blocks, residual connection, independent encoder, 128 filters and a maximum of 10 neighbours per node. After experiments we concluded that in this case the GNN can benefit from a maximum of 20 neighbours per node but all the other hyperparameters remained the unchanged. We stress that the neighbours are calculated as a preprocessing step and thus remain the same throughout the GNN training. The maximum GPU memory required for training the GNN is 8.2GB using a batch size of 4. Training of the GNN was performed on an NVIDIA V100 GPU with 16GB of RAM.
        
        In this more challenging problem, we investigate the idea of using dual convolutions. Instead of only considering the Euclidean neighbourhood we also consider the Geodesic neighbourhood of every node in order to perform joint Geodesic and Euclidean convolutions as explained in [\ref{neighbourhood}]. After experiments, we concluded that in our case the optimum value for the ratio between Geodesic and total convolutions is 75\%. The total number of trainable parameters for the network is 691,196. For more details the reader can refer to [Appendix \ref{appendix:Geod_and_Eucl_conv}].

        Additionally, throughout this work we made the choice to predict the full stress tensor instead of the Von Mises stress only and to work on patches of the structure instead of the entire structure. Using data from the dogbone dataset we performed experiments to support these choices in [Appendix \ref{appendix:stress_vs_max}] and [Appendix \ref{appendix:full_vs_patches}] respectively.

        \subsubsection{Numerical examples with deterministic GNN}
        
\begin{figure}[htb]
            \centering
            \begin{subfigure}{.5\textwidth}
              \centering
              \includegraphics[width=\linewidth]{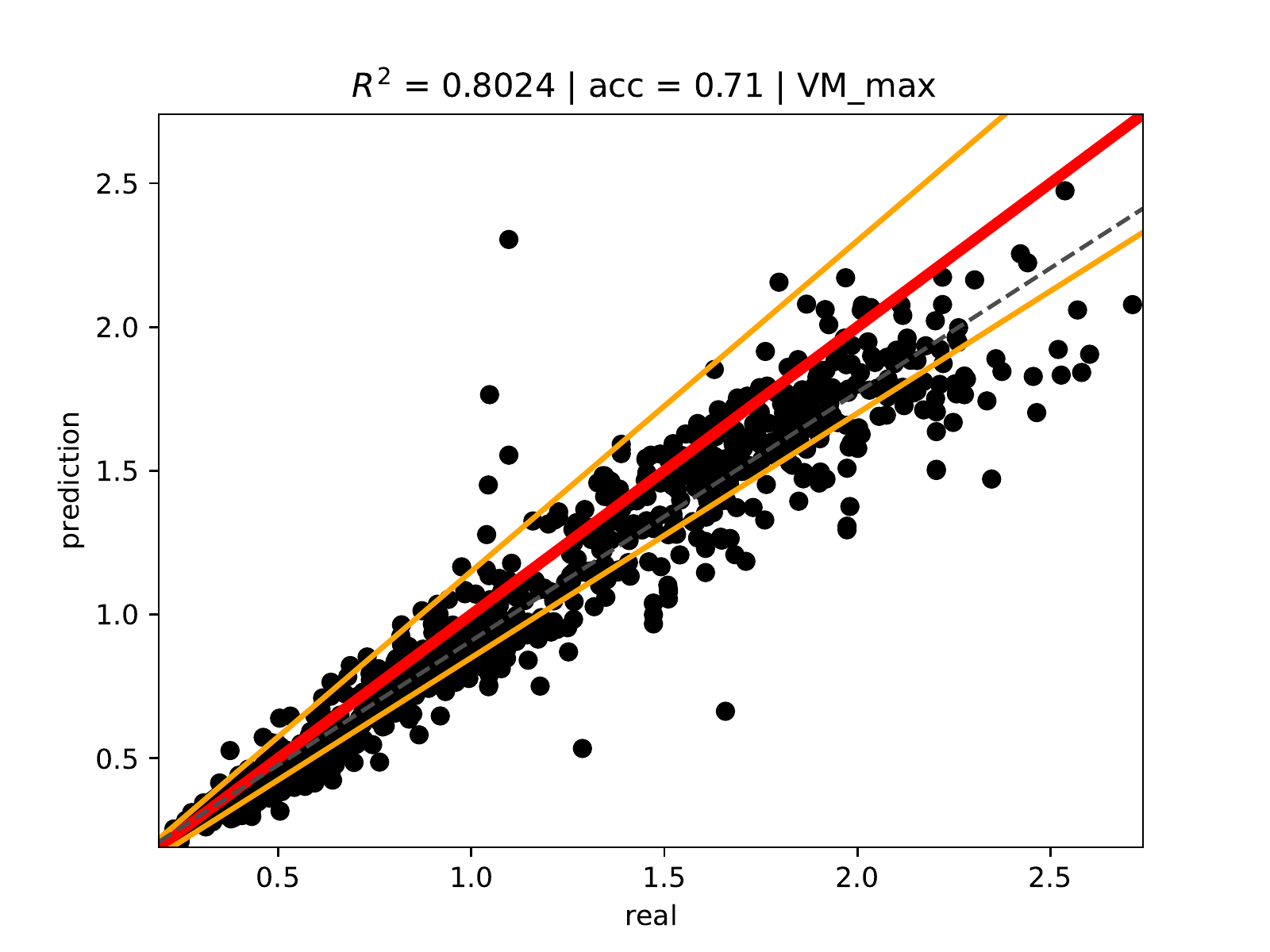}
              \caption{evaluate on patches}
              \label{fig:VM_max_singleHole_data4000:a}
            \end{subfigure}%
            \begin{subfigure}{.5\textwidth}
              \centering
              \includegraphics[width=\linewidth]{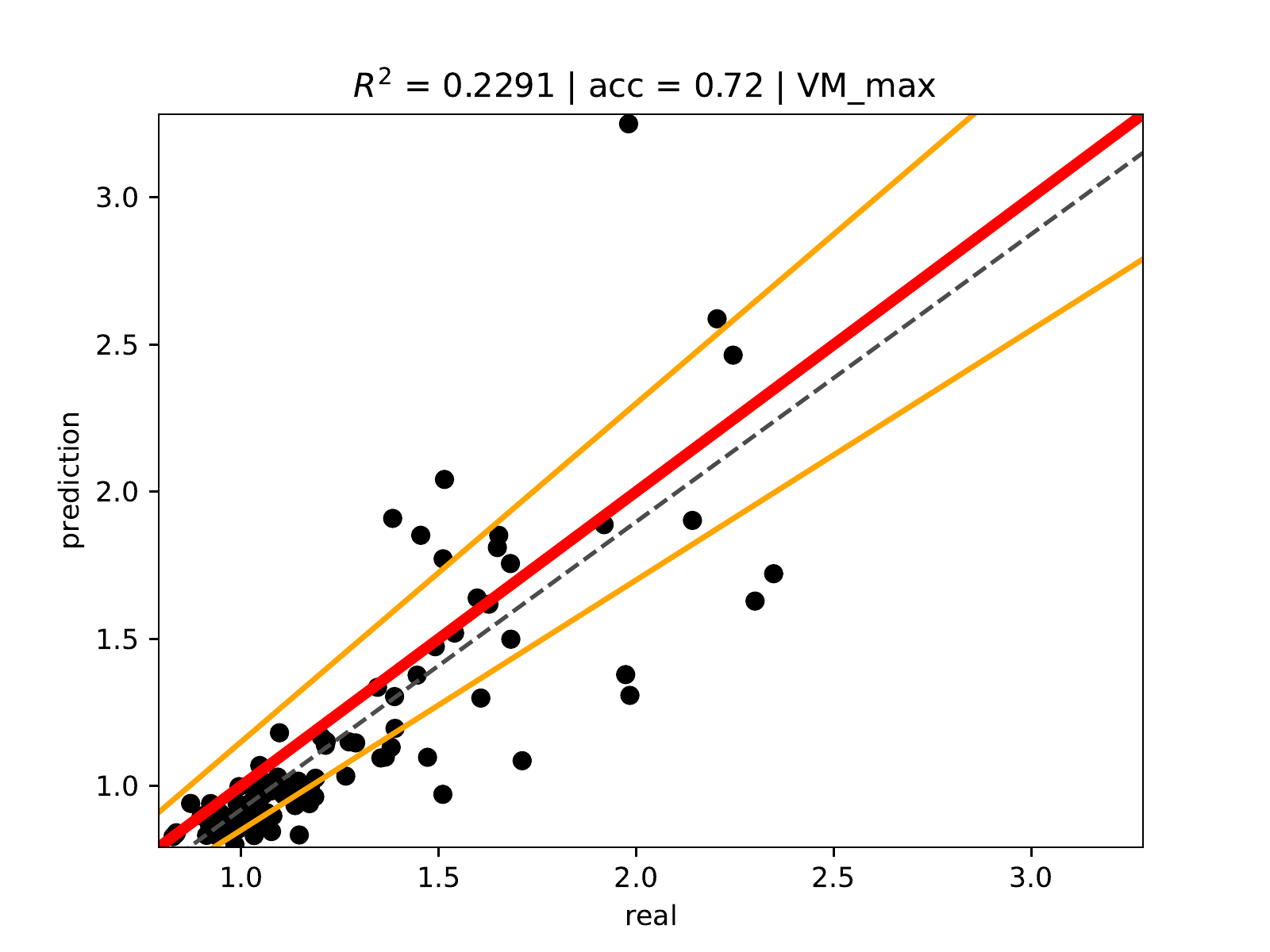}
              \caption{evaluate on the entire structure}
              \label{fig:VM_max_singleHole_data4000:b}
            \end{subfigure}
            \caption{In both subfigures the x-axis corresponds to the maximum Von Mises stress as calculated by FEA and the y-axis as calculated by the GNN. For the left subfigure, (a), we see results for the ROI of the patches while for the right subfigure, (b), for the entire dogbone.}
            \label{fig:VM_max_singleHole_data4000}
        \end{figure}

        All the numerical examples, plots and figures presented in this section refer to the validation set. As can be seen in [Fig \ref{fig:VM_max_singleHole_data4000:a}] when training with 4000 patches the accuracy for the maximum Von Mises stress is 71\% which is considerably lower than the accuracy reported for the cubical heterogeneous material case, 96\%. Visually the accuracy can be interpreted as the percentage of points inside the 2 yellow lines. In [Fig \ref{fig:VM_max_singleHole_data4000:b}] we can also evaluate the performance of the GNN on the entire dogbone specimen, and not only at patch level. The accuracy does not change considerably in this case, but the data are more spread around the $y=x$ line which is indicated by the lower coefficient of determination $R^2$. 
        
\begin{figure}[h]
            \centering
            \begin{subfigure}{\textwidth}
              \centering
              \includegraphics[width=\linewidth]{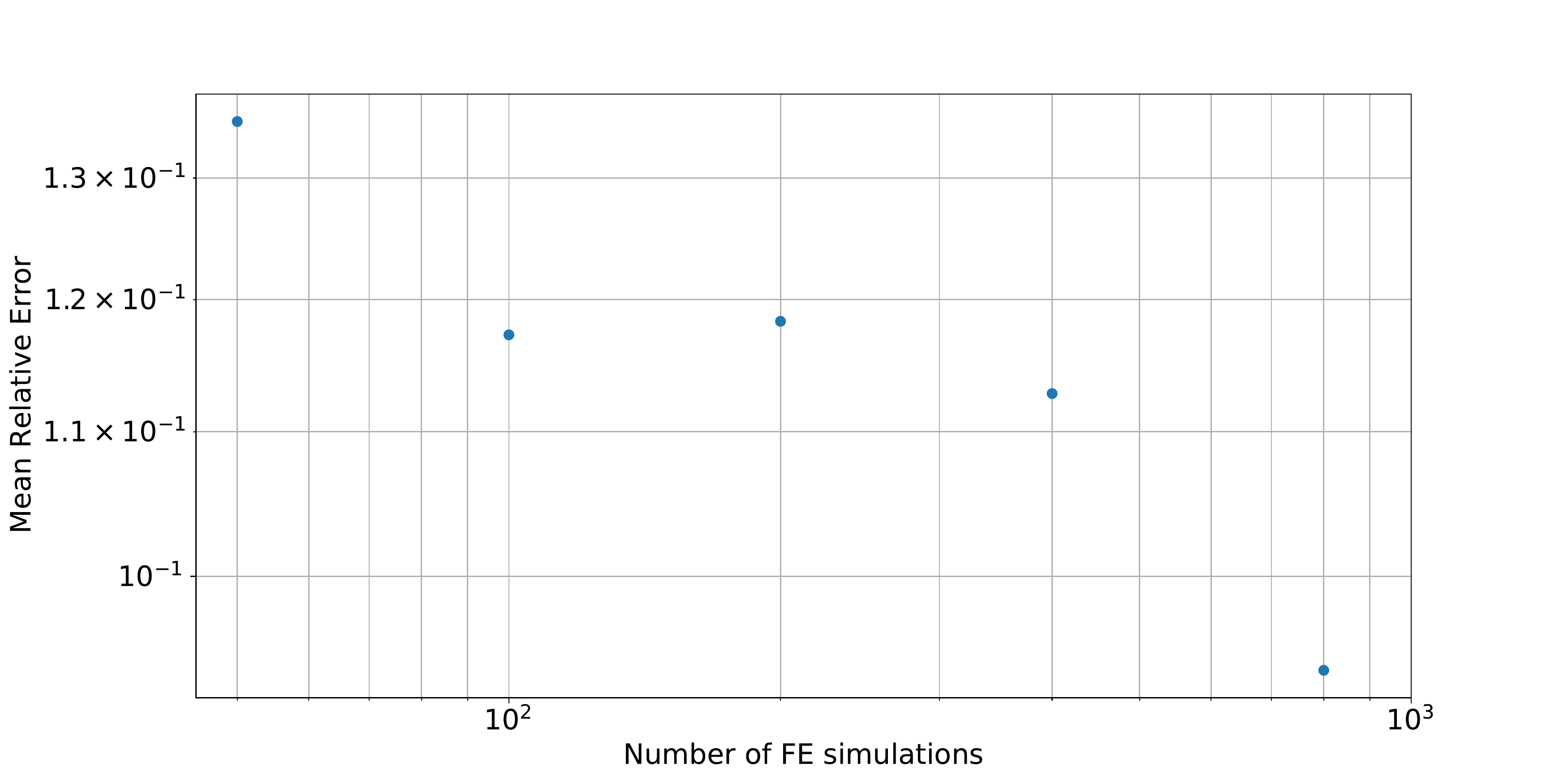}
              \caption{Mean relative error}
              \label{fig:SingleCylinder_convergennce:a}
            \end{subfigure}%
            \vfill
            \begin{subfigure}{\textwidth}
              \centering
              \includegraphics[width=\linewidth]{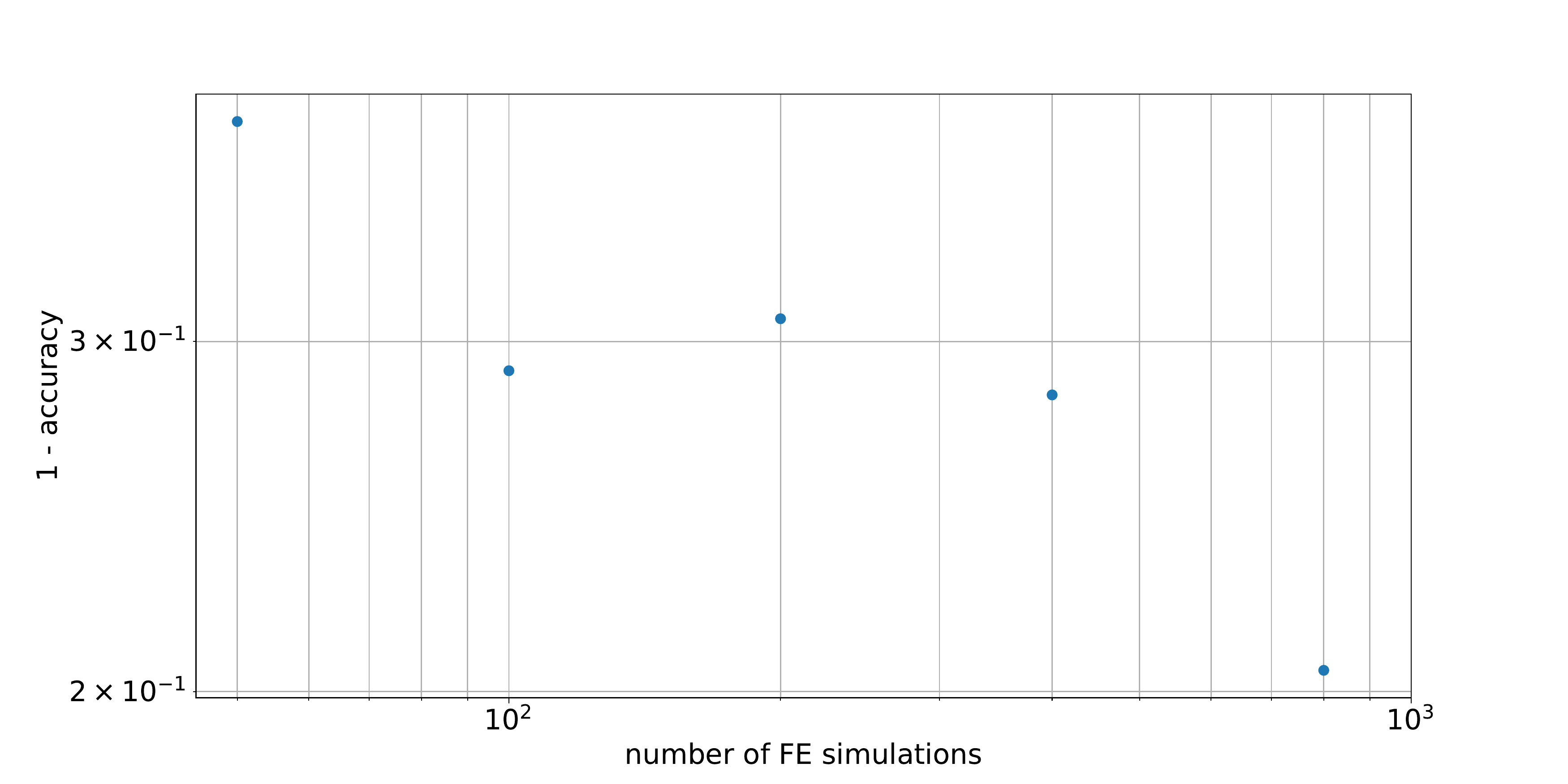}
              \caption{\protect\say{1-accuracy} for the 15\% error threshold}
              \label{fig:SingleCylinder_convergennce:b}
            \end{subfigure}
            \caption{In the diagram on top (a) we see the mean relative error between the maximum Von Mises stress in the ROI of the patches as calculated by FEA and as predicted from the GNN. In the diagram on the right (b) we have plotted \protect\say{1-accuracy} for the 15\% error threshold. In both figures the x axis corresponds to the number of data used to train the GNNs. Both figures are plotted in a log-log scale.}
            \label{fig:SingleCylinder_convergennce}
        \end{figure}

        Additionally, we investigate the performance of the network with respect to the size of the training dataset. We train 5 GNNs with patches generated from 50, 100, 200, 400 and 800 FE simulations. From each FE simulation we extract 10 patches. First, we investigate the convergence properties of the GNN with respect to the number of  FE simulations used for their training. In [Fig \ref{fig:SingleCylinder_convergennce:a}] we see the convergence plot for the mean relative error. Here we define the relative error as the relative error between the maximum Von Mises stress in the ROI of the patches as predicted by the GNN and as calculated by FEA. In [Fig \ref{fig:SingleCylinder_convergennce:b}] we see the convergence plot for the accuracy metric. From these figures we can conclude that the accuracy increases and the mean relative error decreases as more data is added to the training set. Convergence seems to be linear in log-log scale, indicating that a fixed fraction of the error is eliminated by multiplying the dataset size by a given ratio. Such a relationship would allow us, in principle, to determine how much data is needed to achieve a given accuracy for the GNN predictions. However, this linear relationship cannot continue to infinity, as the true macro to micro FEA map is not an injective function. Sources of non-injectivity are, amongst others, the non-deterministic and/or non-smooth meshing procedure (meaning that a small perturbation of the microscale morphology could result in different mesh, which would then generate a larger difference in terms of FEA stress predictions) and the fact that finite size buffer regions lead to an indeterminacy in the description of the surroundings of the ROI. Finally, our GNN architecture is not a universal approximator (the number of neurons is finite), which could trigger a stagnation of the convergence plot for larger datasets. The study of such inflexion points in the convergence behaviour of the GNN training is left to future studies.

\begin{figure}[h]
            \centering
            \begin{subfigure}{.5\textwidth}
              \centering
              \includegraphics[width=\linewidth]{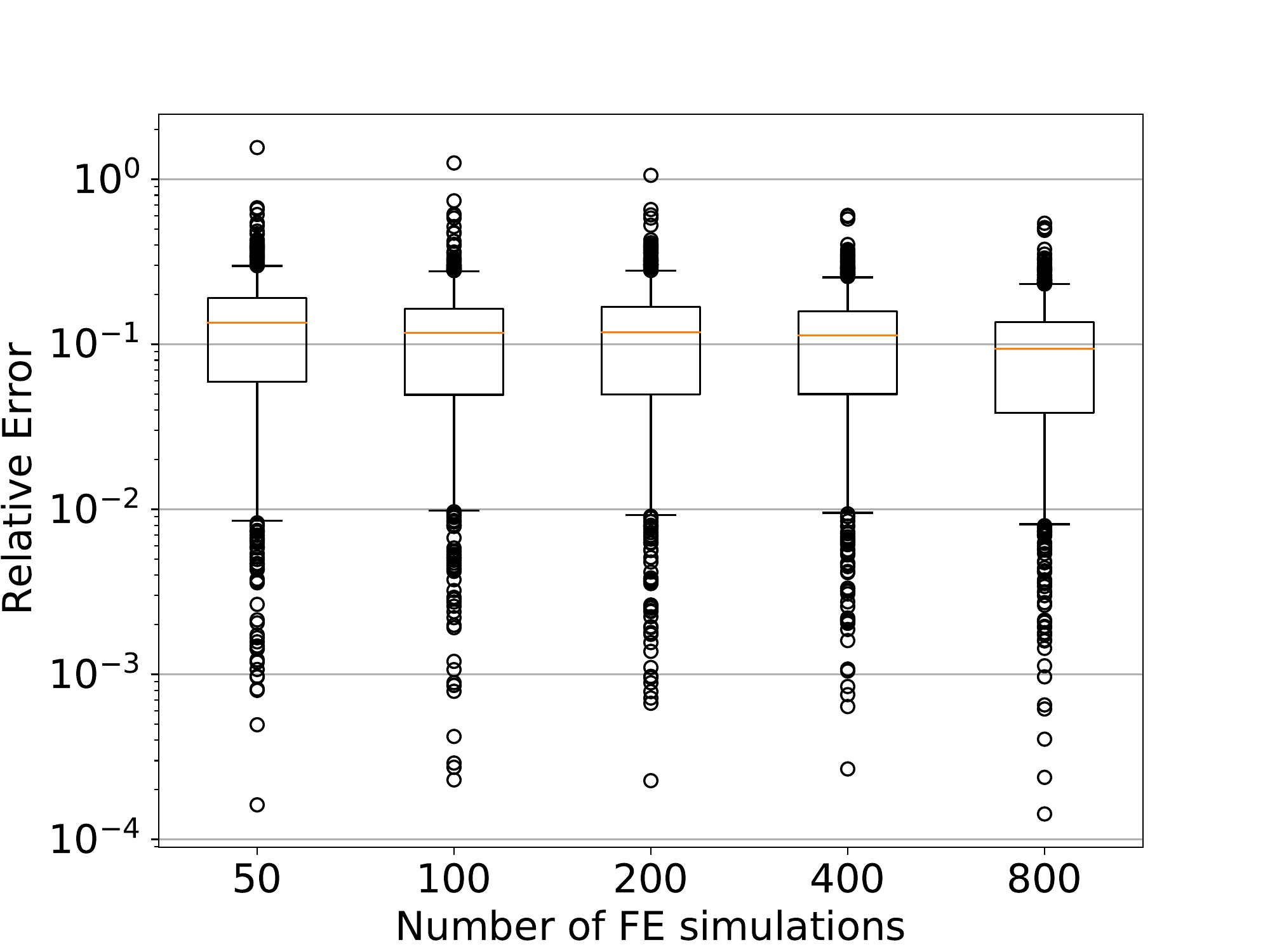}
              \caption{Box plot with outliers}
              \label{fig:SingleCylinder_relative_error_boxplots:a}
            \end{subfigure}%
            \begin{subfigure}{.5\textwidth}
              \centering
              \includegraphics[width=\linewidth]{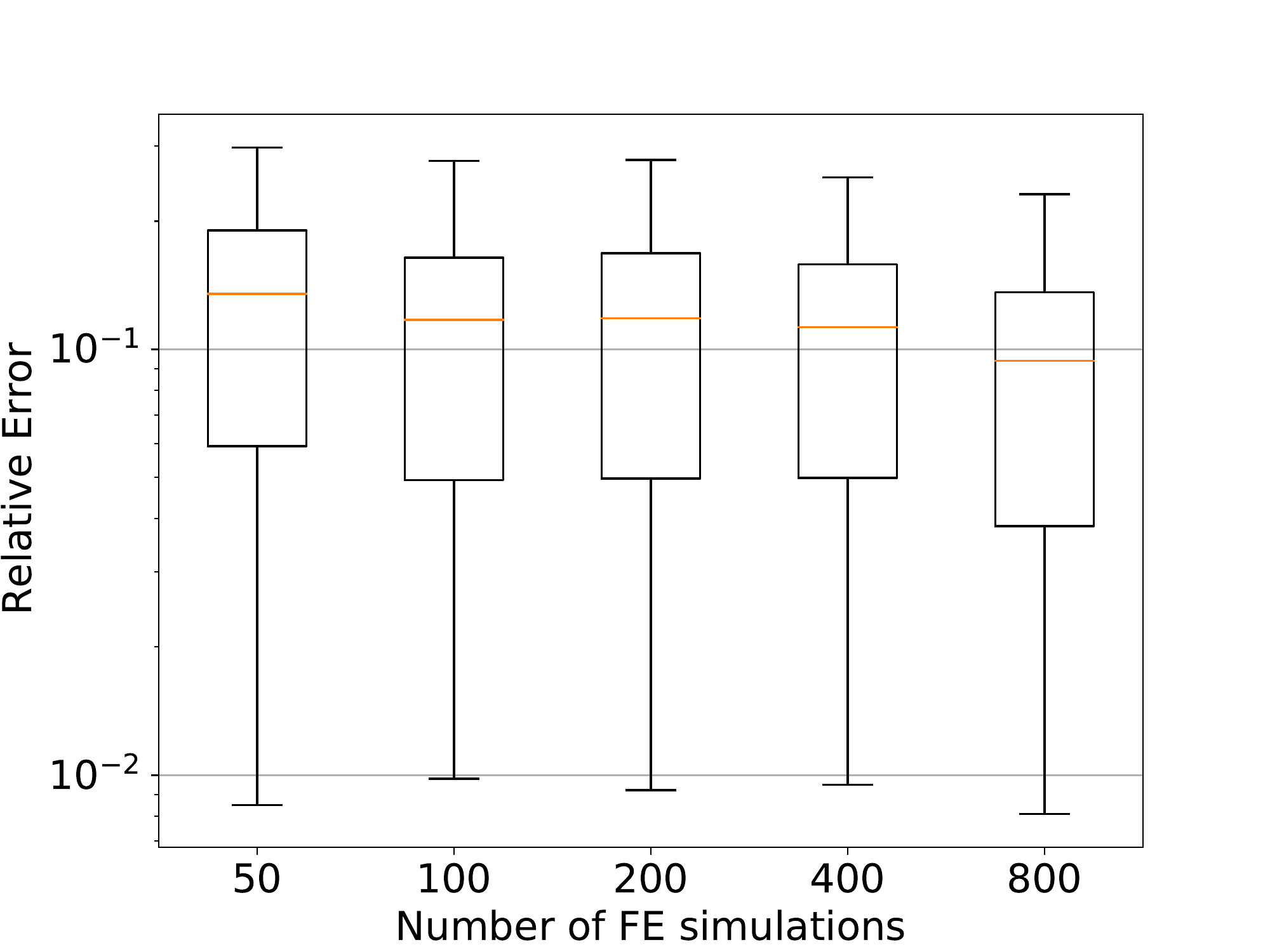}
              \caption{Box plot without outliers}
              \label{fig:SingleCylinder_relative_error_boxplots:b}
            \end{subfigure}
            \caption{In the diagram on the left (a) we see 5 box plots for the mean relative error in the ROI of the patches in the test set. Each box plot is defined by 5 numbers. The 2 whiskers at the top and bottom correspond to the 95th and 5th percentiles respectively. The top and bottom limits of the box corresponds to the 75th and 25th percentiles respectively. Finally, the yellow line corresponds to the mean of the data. Each box plot corresponds to a GNN trained with different number of data. In the diagram on the right (b) we can see the same box plots without the outliers so that the trend is clearly visible. Each box plot corresponds to 1,000 points.}
            \label{fig:SingleCylinder_relative_error_boxplots}
        \end{figure}

        Furthermore, to give some insights into the distribution of relative error across the patches, we provide box plots for the relative error [Fig \ref{fig:SingleCylinder_relative_error_boxplots}]. From [Fig \ref{fig:SingleCylinder_relative_error_boxplots:a}] we can see that the predictions with the highest relative error values (above the 95th percentile) keep getting more accurate as more data is added to the training set. In  [Fig \ref{fig:SingleCylinder_relative_error_boxplots:b}] we can see that the 95 percentile (denoted by the upper whisker) and the mean relative error (denoted by the yellow horizontal line) follow the same trend.

\begin{figure}[h]
            \centering
            \begin{subfigure}{.5\textwidth}
              \centering
              \includegraphics[width=\linewidth]{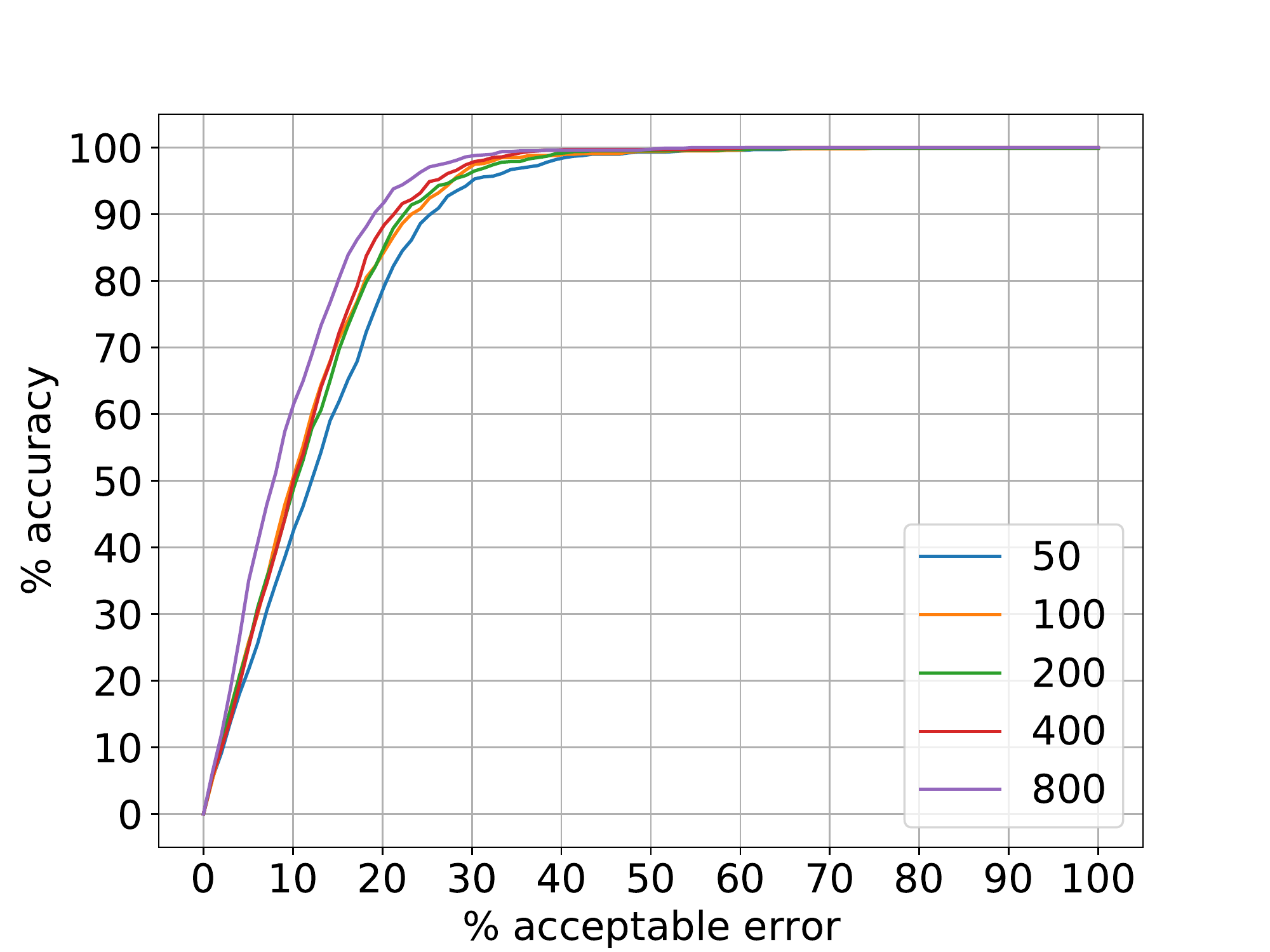}
              \caption{Accuracy VS error threshold}
              \label{fig:SingleCylinder_size_study:a}
            \end{subfigure}%
            \begin{subfigure}{.5\textwidth}
              \centering
              \includegraphics[width=\linewidth]{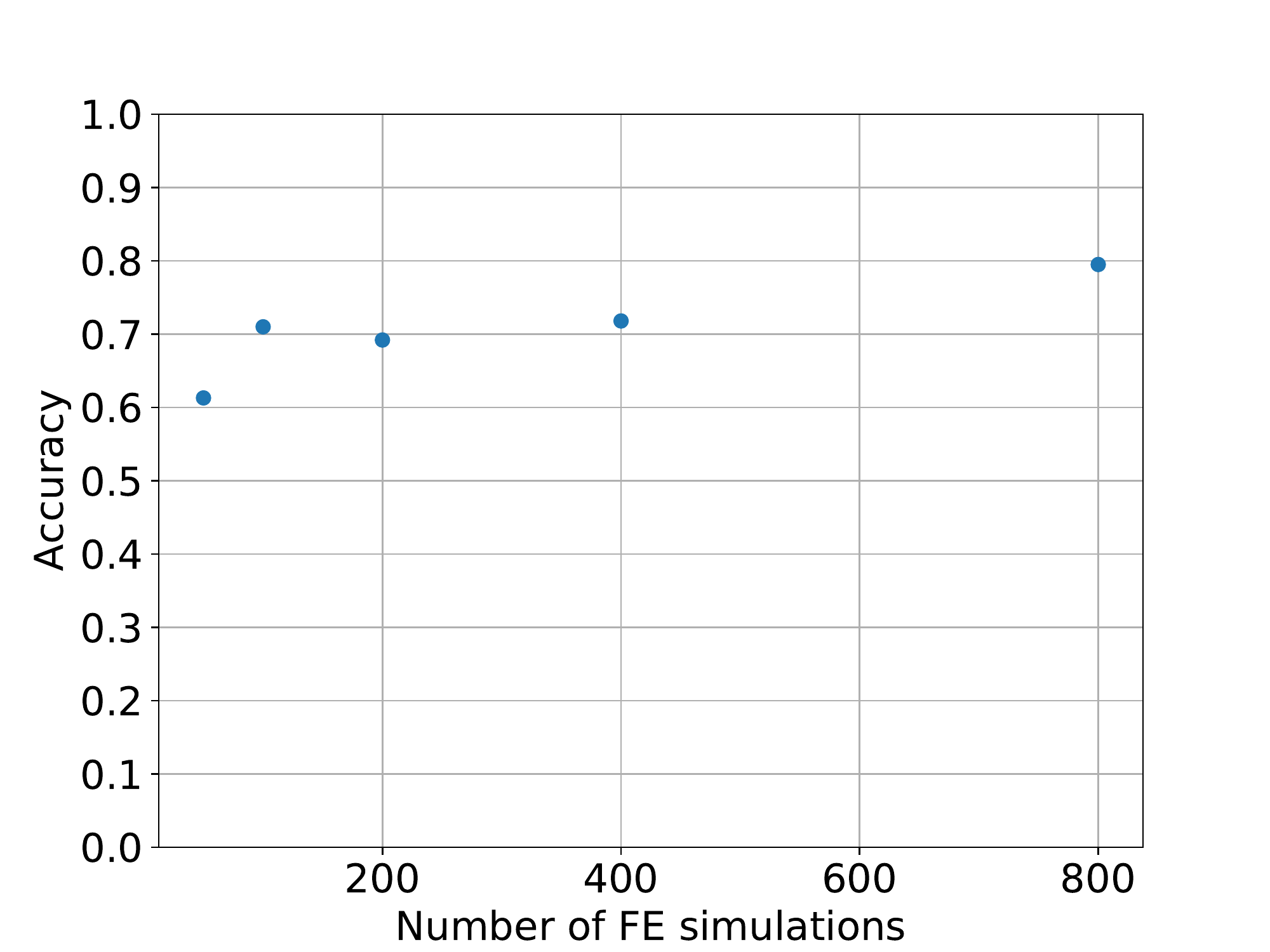}
              \caption{Accuracy for the 15\% error threshold}
              \label{fig:SingleCylinder_size_study:b}
            \end{subfigure}
            \caption{In the diagram on the left (a) we see accuracy curves with respect to the threshold used for the relative error. Coloured lines correspond to GNNs trained with datasets of different size, namely patches extracted from 50, 100, 200, 400 and 800 FE simulations. In the diagram on the right (b) we see the accuracy for the 15\% error threshold with respect to number of data used to train the GNNs.}
            \label{fig:SingleCylinder_size_study}
        \end{figure}

        Moreover, we examine the accuracy that can be achieved with the given training datasets. In [Fig \ref{fig:SingleCylinder_size_study:a}] we can see the accuracy values that can be achieved with respect to the relative error threshold for GNNs trained with different number of FE simulations. We observe a sharp increase of the accuracy with respect to the threshold values, as long as the threshold remains under 20\%. For even looser threshold values, the accuracy slowly converges to 100\%. In [Fig \ref{fig:SingleCylinder_size_study:b}] we can see the accuracy values that can be achieved for the 15\% relative error threshold. We can observe that the accuracy increases as we increase the size of the training dataset, from 60\% for the GNN trained with 50 FE simulations to 80\% for the GNN trained with 800 FE simulations.
        
        We note that with only 50 direct FE simulations, and for the 20\% threshold value, the accuracy reaches 80\%. This shows that obtaining general trends in the microscale field does not require an overwhelming number of direct numerical simulation. This is due to the nature of the random field that the GNN is being trained to reproduce. A single direct FE simulation provides more and more micromechanical information as the size of the computational domain increases, for fixed statistics of the pores' spatial distribution. We will see that this is confirmed by the fact that the prediction error is usually at its largest in regions where the geometrical specificity of the interaction between the specimen's boundary and the microstructure is sparsely represented in the dataset (\textit{i.e.} around the cylindrical hole in our case)

        A similar study has been conducted in [Appendix \ref{appendix:triaxiality}] but the triaxiality has been used as quantity of interest instead of the Von Mises stress. Similar trends have been observed, which is why they are not included in the core of the paper.
        
\begin{figure}[htb]
            \begin{center}
                \includegraphics[width=.80\linewidth]{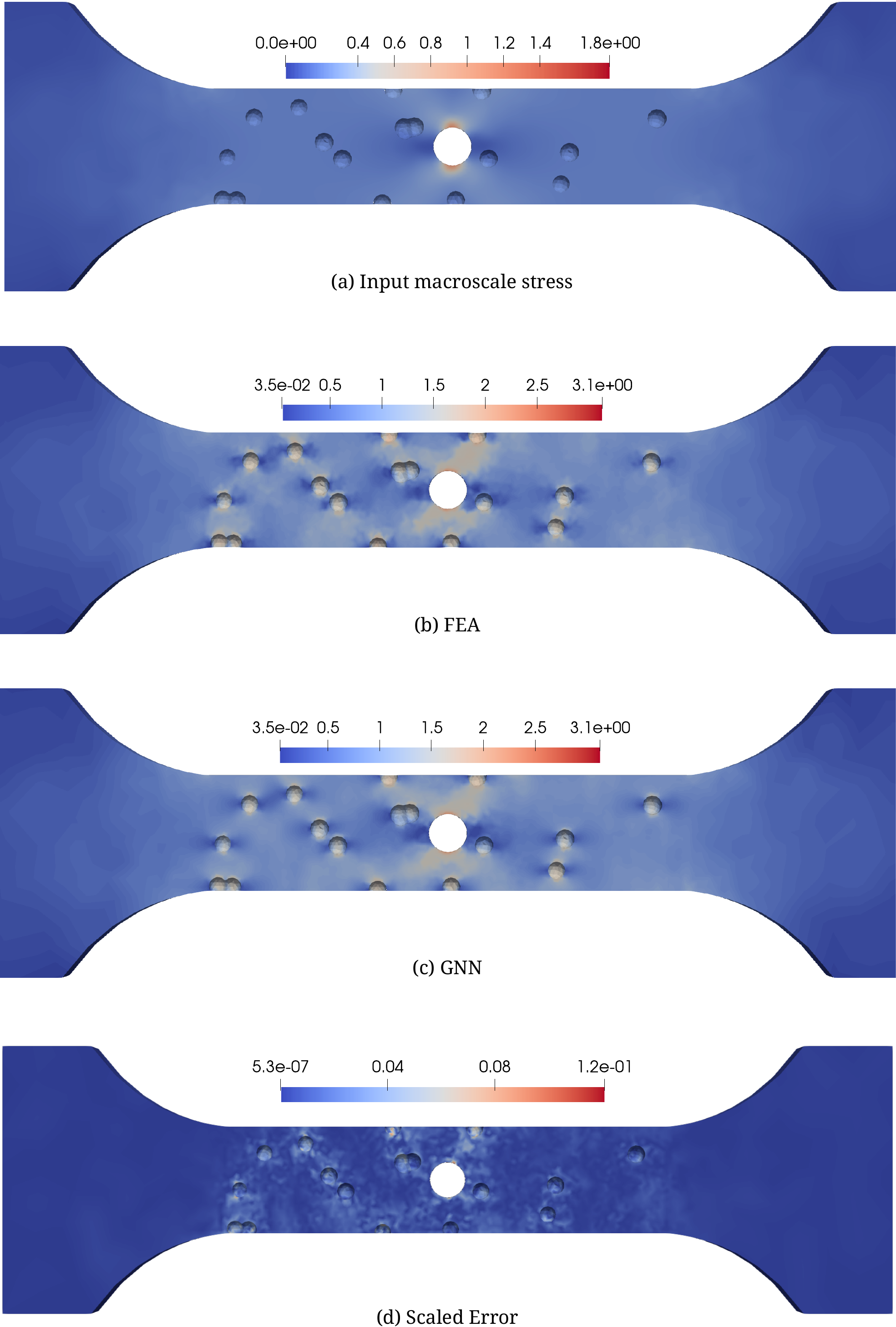}
                \caption{Comparison between the microscale Von Mises stress distribution as calculated by FEA (b) and the one predicted by the GNN (c). We can also see the input macroscale Von Mises stress interpolated on the microscale mesh (a) and the scaled error distribution (d) which is defined here as $|VM_\text{FEA} - VM_\text{NN}|/\text{max}(VM_\text{FEA})$, where $VM_\text{NN}$ is the microscale Von Mises stress predicted by the GNN and  $VM_\text{FEA}$ is the microscale Von Mises stress calculated by FEA. The mean relative error in the dense mesh area of the specimen is 0.1225 and the mean scaled error is 0.0164. The GNN result is reconstructed from the union of multiple patch predictions where only the ROI is extracted.}
                \centering
                \label{fig:variable_dogbone_file75}
            \end{center}
        \end{figure}

\begin{figure}[htb]
            \begin{center}
                \includegraphics[trim={0 0 0 4cm}, width=\linewidth]{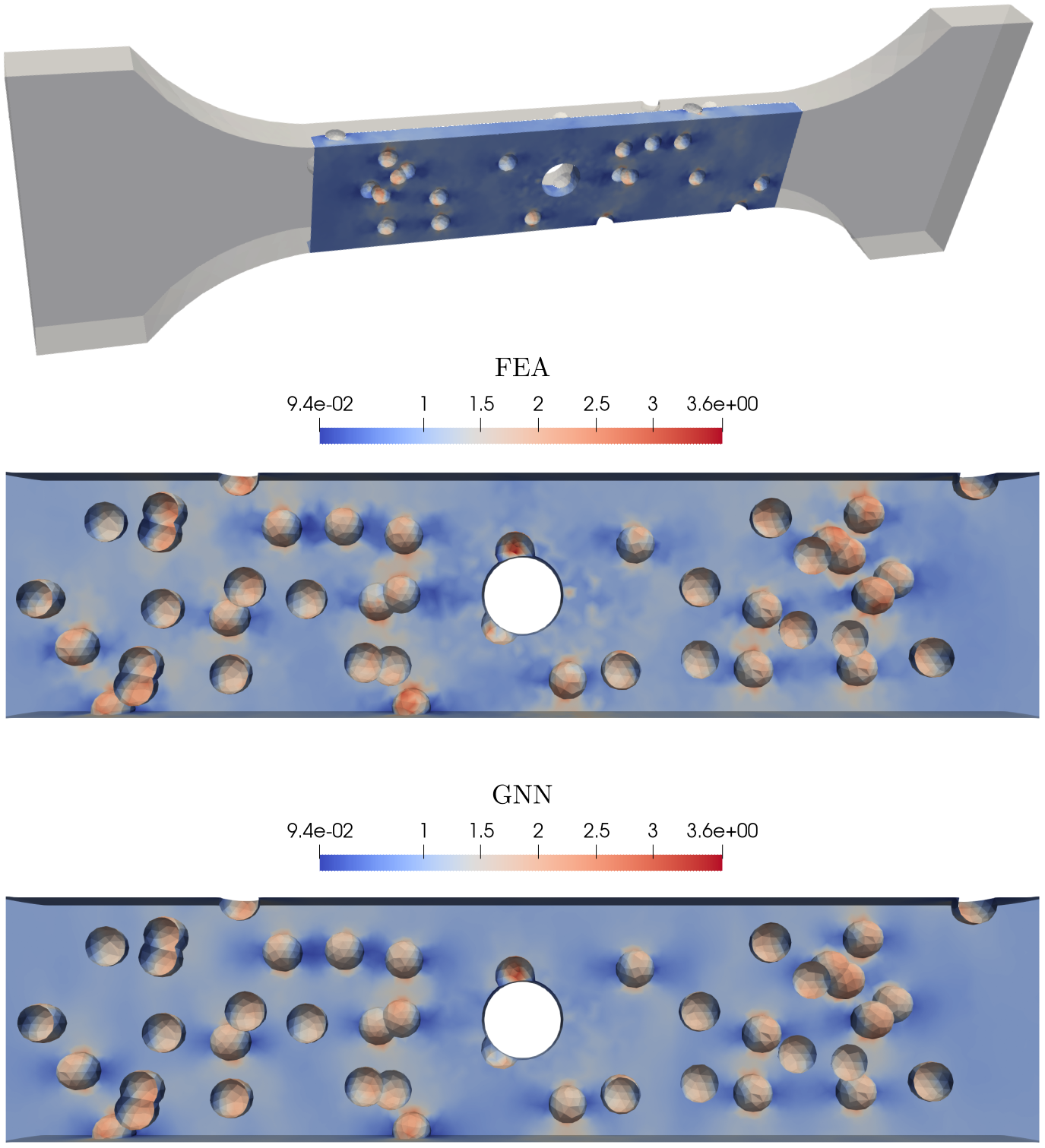}
                \caption{Comparison between the Von Mises stress distribution, extracted from a cross section of the structure (top), as calculated by FEA (middle), and as predicted by the GNN (bottom). We can observe that the two distributions are qualitatively very similar. The GNN result is reconstructed from the union of multiple patch predictions where only the ROI is extracted. The mean relative error in the dense mesh area of the specimen is 0.1072 and the mean scaled error is 0.01162. The scaled error is defined as $|VM_\text{FEA} - VM_\text{NN}|/\text{max}(VM_\text{FEA})$, where $VM_\text{NN}$ is the microscale Von Mises stress predicted by the GNN and  $VM_\text{FEA}$ is the microscale Von Mises stress calculated by FEA.}
                \centering
                \label{fig:file_12_crossection}
            \end{center}
        \end{figure}

        After examining the relative error and the accuracy of the prediction we now focus on the stress distribution within the structure.
        Firstly, in [Fig \ref{fig:variable_dogbone_file75}] we compare the microscale Von Mises stress predicted by the GNN with the one calculated by FE simulations. We observe that the predicted stress distribution is close to the ground truth. The GNN result is reconstructed from the union of multiple patch predictions where only the ROI is extracted. This is possible if we align one corner of the ROI with a corner of the image and use a sliding window equal to the size of the ROI along each direction. This procedure is explained in detail in  \citep{krokos2021bayesian}. Also, we want to examine the stress distribution in the inner structure of a dogbone. In [Fig \ref{fig:file_12_crossection}] we compared the predicted Von Mises stress with the ground truth on a cross section of a dogbone structure. In this case we see more complex interactions but the GNN predictions remain close to the FE results.
        
\begin{figure}[p]
            \centering
            \begin{subfigure}[b]{\textwidth}
              \centering
              \includegraphics[ width=\linewidth]{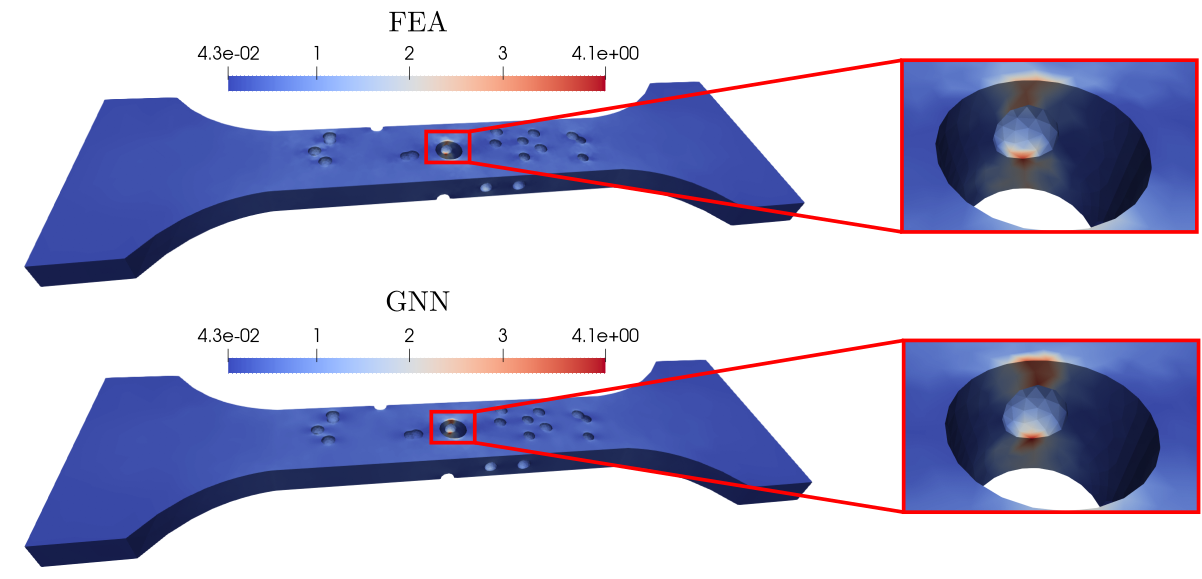}
              \caption{spherical void - cylinder-shaped hole interaction}
              \label{fig:max_zoom:a}
            \end{subfigure}%
            \vspace{20pt}
            \begin{subfigure}[b]{\textwidth}
              \centering
            \includegraphics[width=\linewidth]{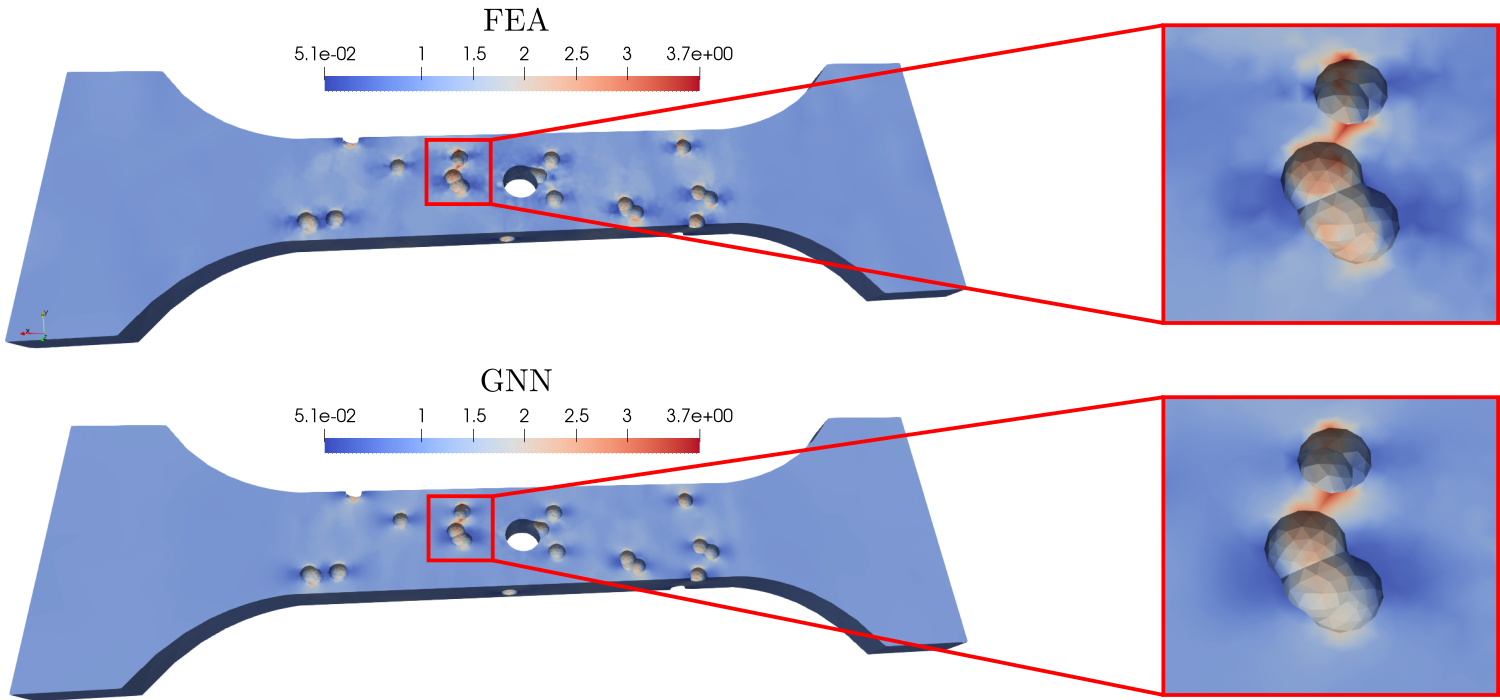}
              \caption{spherical void - spherical void interaction}
              \label{fig:max_zoom:b}
            \end{subfigure}
            \caption{In both subfigures we compare the maximum Von Mises stress as calculated by FEA and as predicted by the GNN. In the top subfigure (a), the maximum is due to an interaction between spherical voids and the cylinder-shaped hole while in the bottom, (b), due to an interaction between 2 spherical voids. The GNN result is reconstructed from the union of multiple patch predictions where only the ROI is extracted. The mean relative error in the dense mesh area of the specimens is 0.1156 for (a) and 0.1093 for (b) while the mean scaled error is 0.0114 for (a) and 0.0176 for (b). The scaled error is defined as $|VM_\text{FEA} - VM_\text{NN}|/\text{max}(VM_\text{FEA})$, where $VM_\text{NN}$ is the microscale Von Mises stress predicted by the GNN and  $VM_\text{FEA}$ is the microscale Von Mises stress calculated by FEA.}
            \label{fig:max_zoom}
        \end{figure}

        Moreover, we are interested in studying the predicted stress distribution in the highly stressed regions of the specimen. In [Fig \ref{fig:max_zoom:a}] we have zoomed in a region where a spherical void interacts strongly with the cylinder-shaped hole, and in [Fig \ref{fig:max_zoom:b}] we have zoomed in at a strong interaction between 2 spherical voids. We observe that in both cases the GNN correctly locates the area where the maximum Von Mises stress occurs, and predicts its value and the stress distribution around it with a qualitatively satisfying level of accuracy. Quantitative measures of accuracy are provided in the caption of the figure.

        Lastly, we focus on the location of the maximum error between the GNN prediction and the FEA. In 3 of the 5 examples that we present in [Appendix \ref{appendix:dogbone_examples}], the maximum error is located on the cylindrical hole. In the rest of them, the second largest error is associated with the cylindrical hole. This is unsurprising, since in order to create the training dataset we used a random sampling strategy to extract patches from the specimen so the cylindrical hole is underrepresented in the training dataset. A better sampling strategy can be devised but this is left to a future study.
        
    \subsubsection{GNN prediction dependency on the input mesh }

\begin{figure}[htb] 
            % [0,0]
              \begin{subfigure}[b]{0.5\linewidth}
                \centering
                \includegraphics[width=\linewidth]{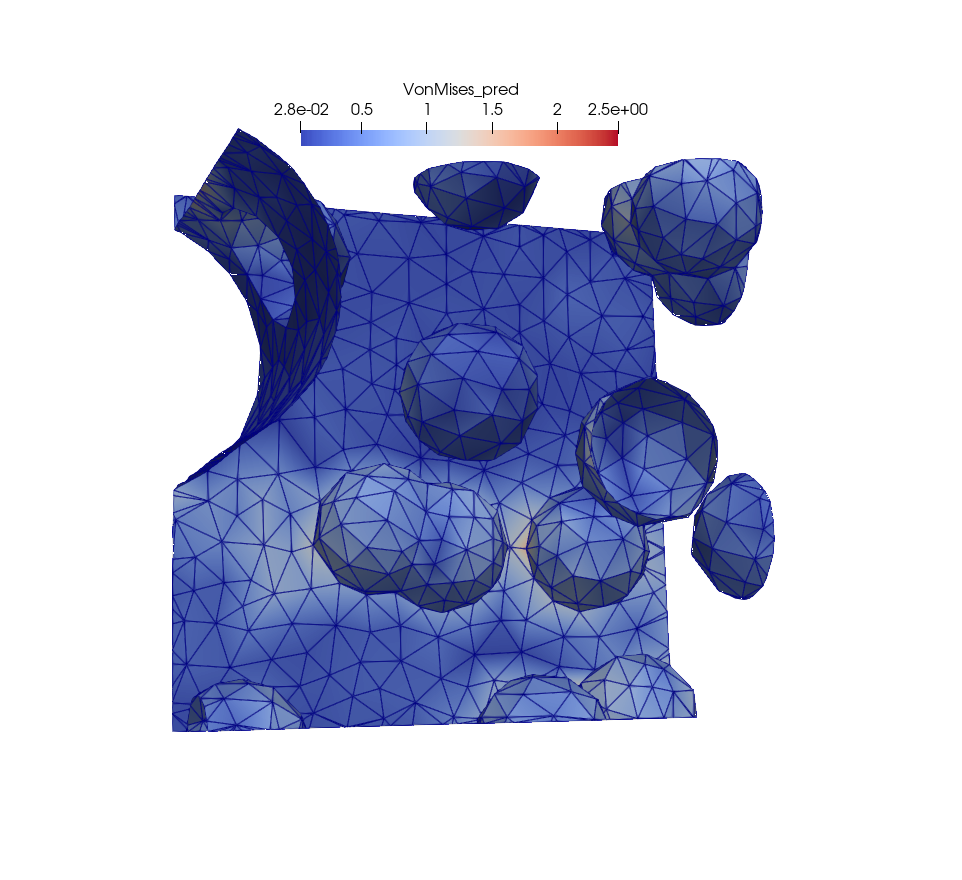} 
                \caption{edge length = 0.008} 
                \label{fig:mesh_dependency:a} 
                \vspace{4ex}
              \end{subfigure}%% 
            % [0,1]
              \begin{subfigure}[b]{0.5\linewidth}
                \centering
                \includegraphics[width=\linewidth]{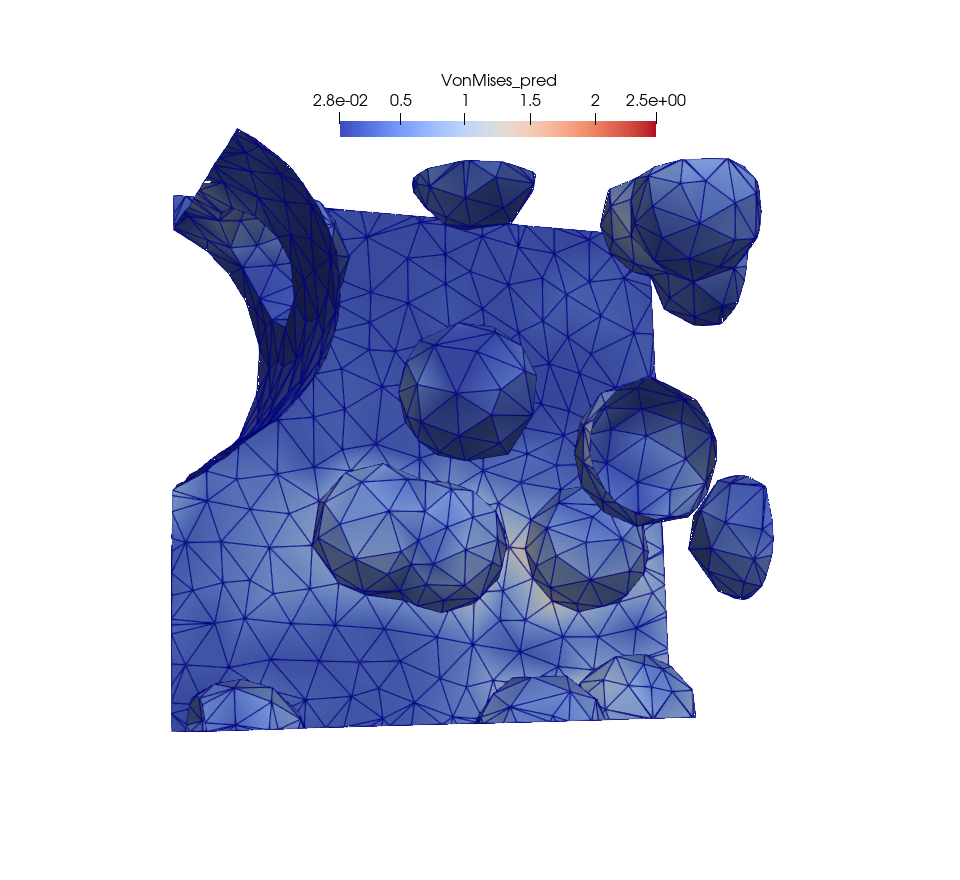} 
                \caption{edge length = 0.009} 
                \label{fig:mesh_dependency:b} 
                \vspace{4ex}
              \end{subfigure} 
            %  [1,0]
              \begin{subfigure}[b]{0.5\linewidth}
                \centering
                \includegraphics[width=\linewidth]{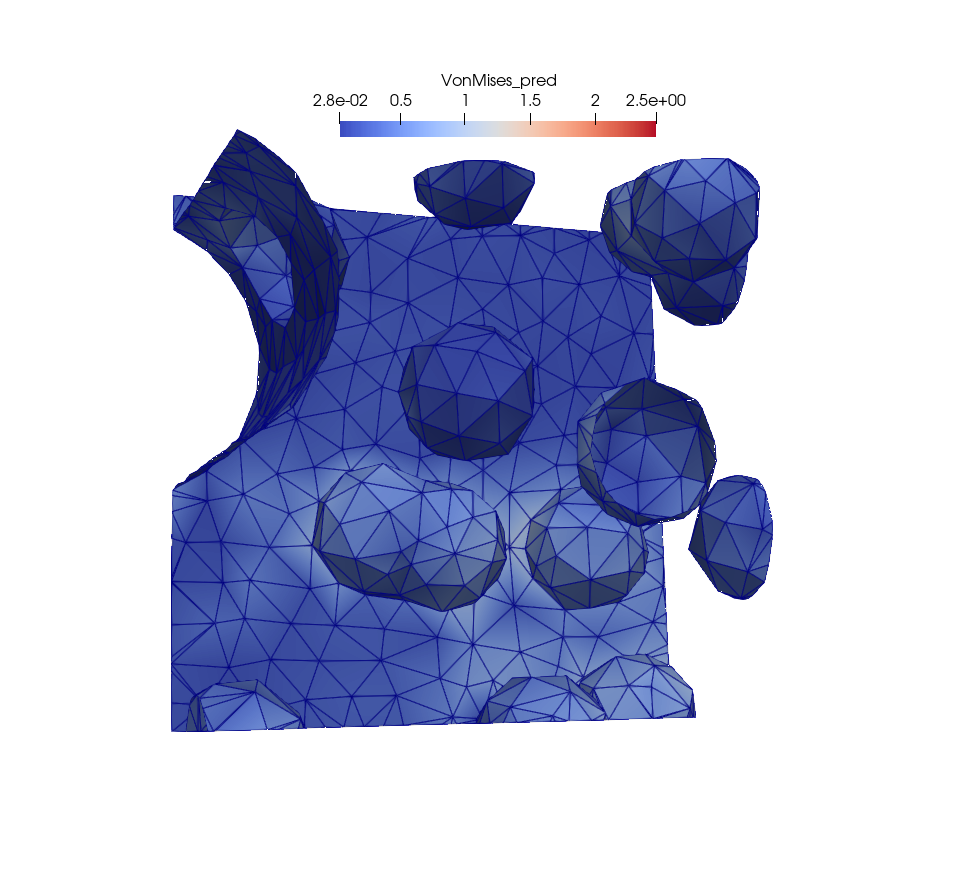}
                \caption{edge length = 0.010} 
                \label{fig:mesh_dependency:c} 
                % \vspace{4ex}
              \end{subfigure}%%
            %  [1,1]
              \begin{subfigure}[b]{0.5\linewidth}
                \centering
                \includegraphics[width=\linewidth]{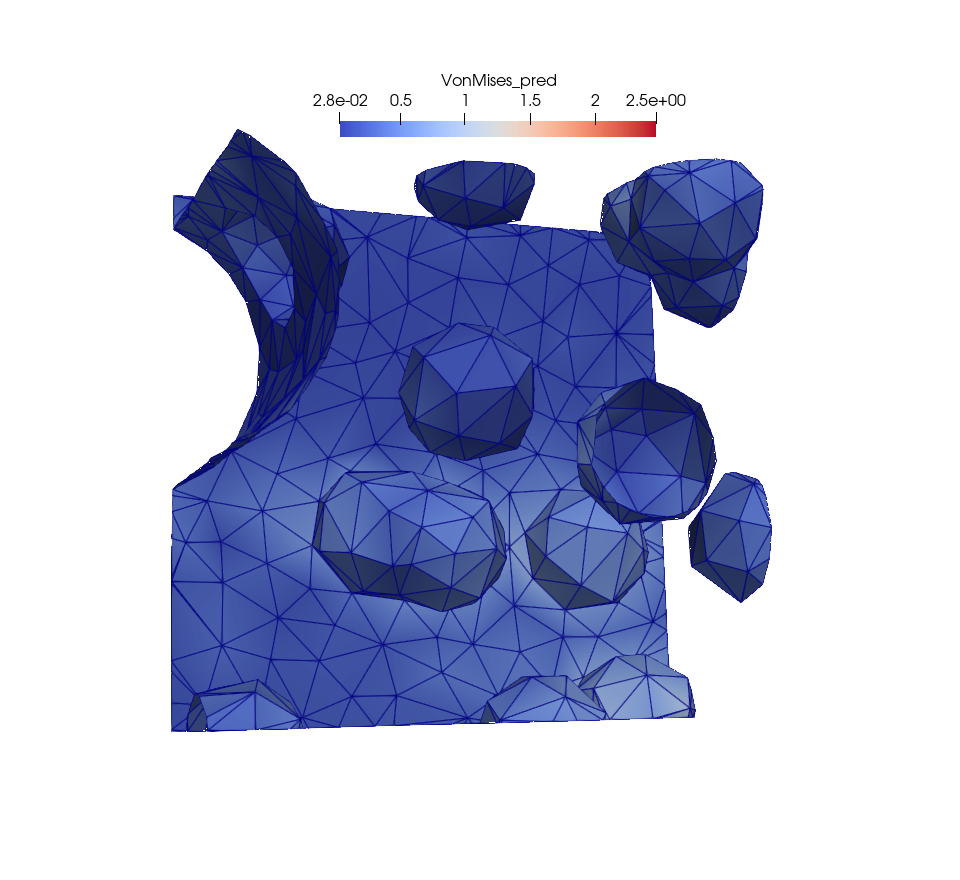}
                \caption{edge length = 0.011} 
                \label{fig:mesh_dependency:d} 
                % \vspace{4ex}
              \end{subfigure} 
              
              \caption{
              Evaluation of the performance of the GNN on input meshes of different density. In all of the figures the microscale Von Mises stress is plotted.}
              
              \label{fig:mesh_dependency} 
        \end{figure}

        In this section we aim to examine the dependency of the GNN prediction on the density of the input mesh. To this end, we propose to mesh one single porous dogbone realisation with four different mesh densities. The four different meshes have typical edge lengths 0.008, 0.009, 0.010 and 0.011. In [Fig \ref{fig:mesh_dependency}] we can see the results obtained by applying the NN to the patch. We  observe that the stress distribution does not change considerably from the 0.008 to the 0.009 case but as the mesh gets coarser, like in the 0.011 case, the GNN does not manage to capture the highly stressed area correctly. We highlight that the GNN was trained using examples meshed with a typical edge length of 0.008. 
        
    \clearpage
    \subsection{Variable dimension dogbone specimen}
        
        In this section we demonstrate the ability of our method to be used for exploration of responses in a space of shapes described by a few geometrical parameters. To this end, we perform geometric learning in a family of dogbone shapes where not only the spherical void distribution will be different from realisation to realisation but also the cylindrical-hole and the dimensions of the dogbone. We focus on the dependence of the quality of the results with respect to the size of the dataset both in terms of maximum Von Mises stress and stress distribution.
        
        \subsubsection{Training dataset}
        
\begin{figure}[htb!]
                \begin{center}
                    \includegraphics[width=.6\linewidth]{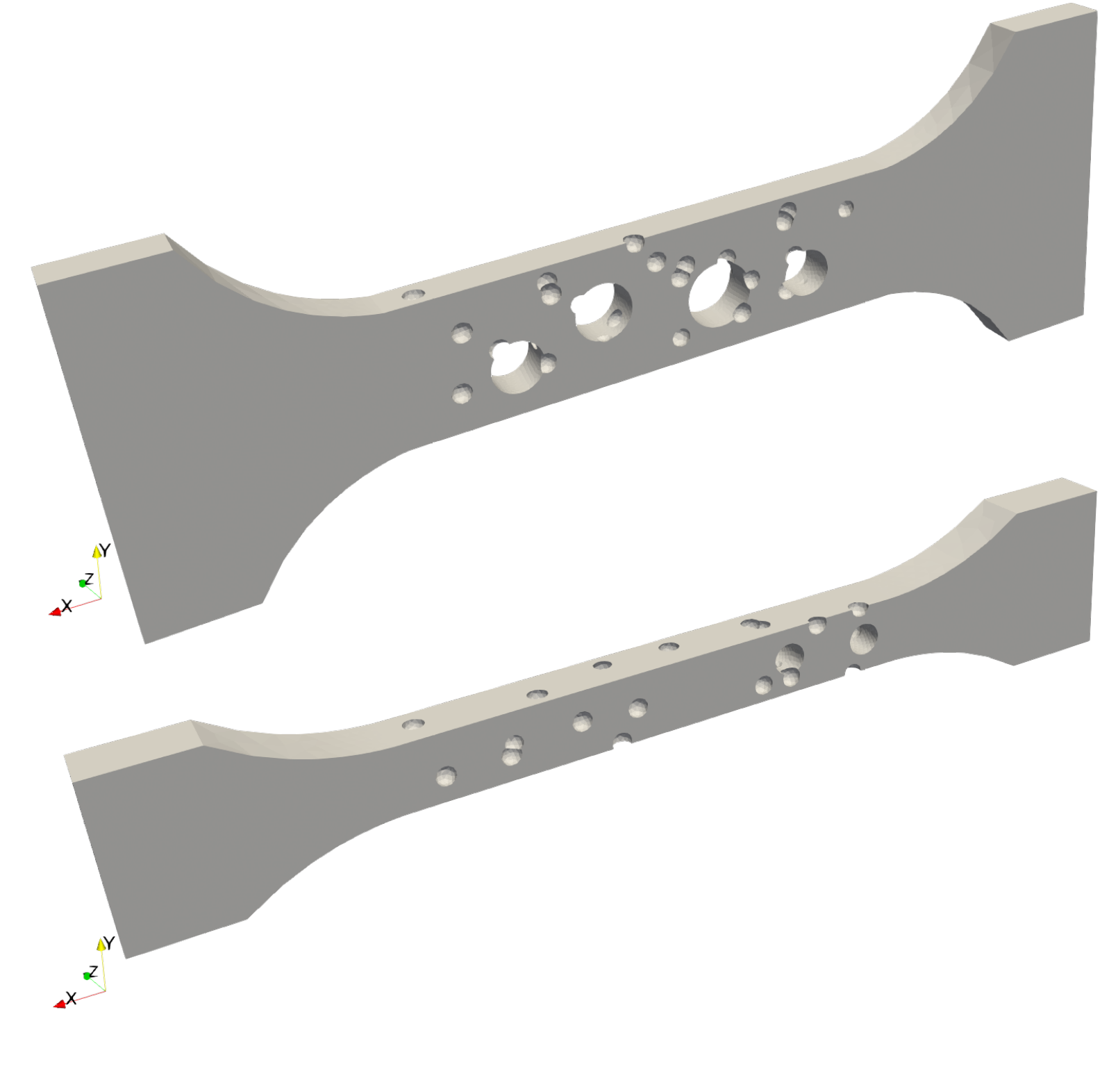}
                    \caption{Two realisations of the variable dogbone structure. In the top subfigure we observe a dogbone with large height, small width and 4 cylindrical holes. In the bottom subfigure we observe a dogbone with small height, large width and 2 cylindrical holes.}
                    \centering
                    \label{fig:variable_dogbone_examples}
                \end{center}
            \end{figure}

            To create this dataset we follow the same process as in the \say{Dogbone} section, [\ref{const dogbone dataset}]. The difference lays on the fact that here the dimension of the specimen is not constant. The length of the dogbone is equal to 2, the width varies from 0.06 to 0.12 and the height to varies from 0.24 to 0.7. Additionally, the number of cylindrical holes varies from 1 to 4 in every realisation, and their radii from 0.024 to 0.06. Two realisations of the variable dogbone geometry can be found in [Fig \ref{fig:variable_dogbone_examples}].
            
        \subsubsection{Numerical examples with variable dimension dogbone specimen}
        
\begin{figure}[htb!]
            \centering
            \begin{subfigure}{.5\textwidth}
              \centering
              \includegraphics[width=.9\linewidth]{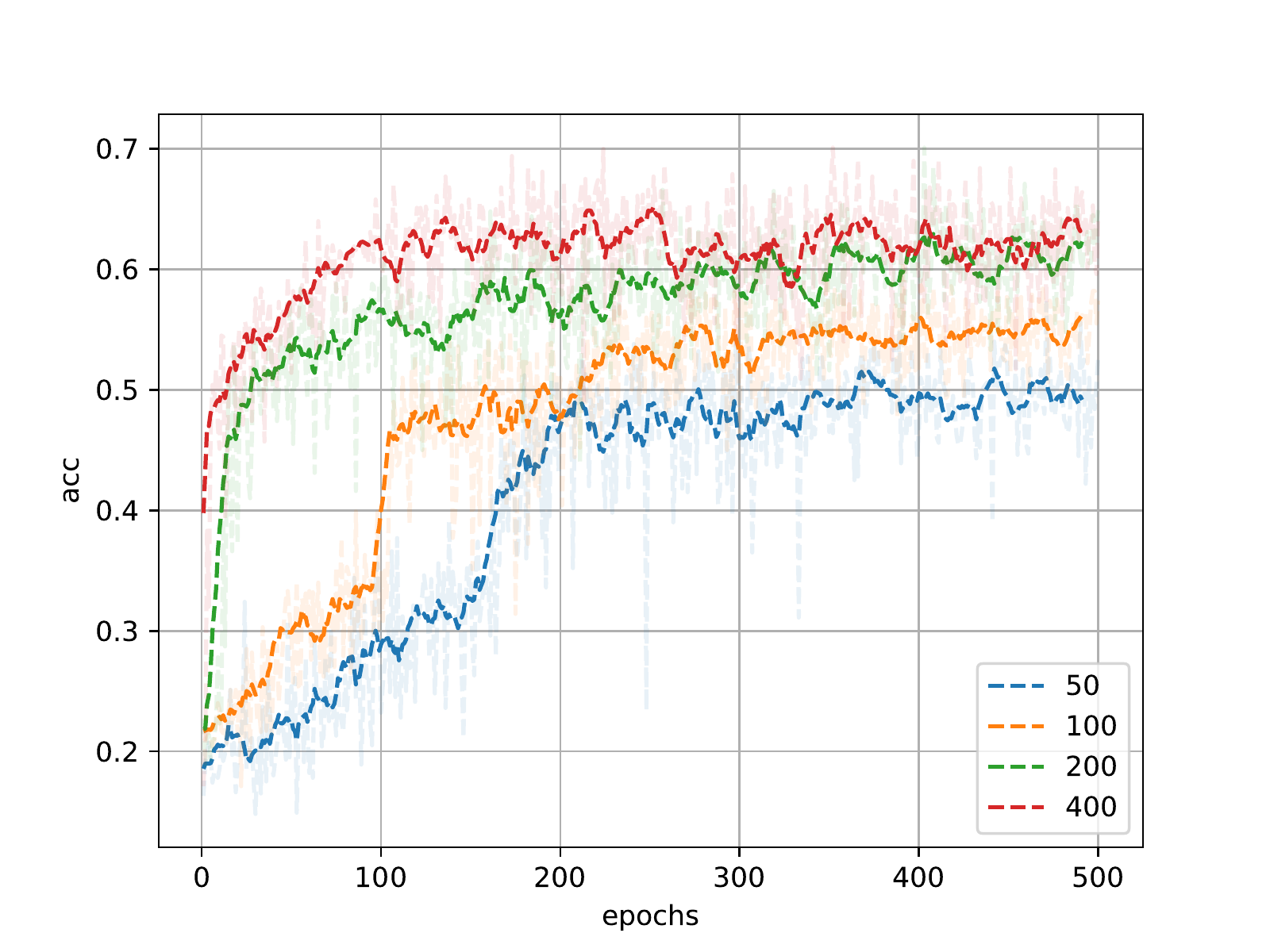}
              \caption{Accuracy}
              \label{fig:variable_dogbone_accuracy_curves:a}
            \end{subfigure}%
            \begin{subfigure}{.5\textwidth}
              \centering
              \includegraphics[width=\linewidth]{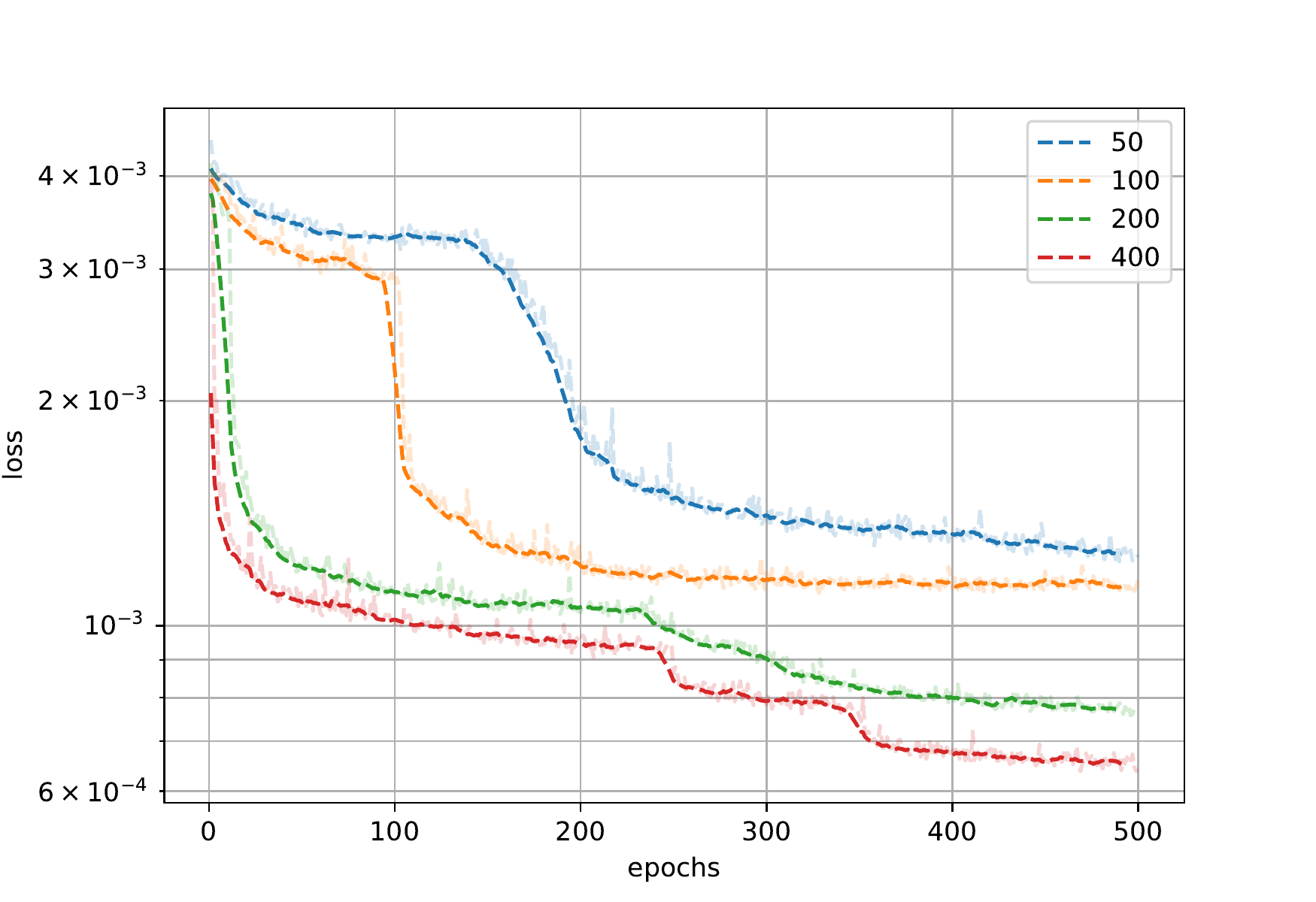}
              \caption{Loss}
              \label{fig:variable_dogbone_accuracy_curves:b}
            \end{subfigure}
            \caption{In the diagram on the left (a) we see accuracy curves defined using the maximum Von Mises stress and a 15\% threshold for the relative error. Specifically, we see the accuracy as function of the training epochs. In the diagram on the right (b) we see the loss function as a function of the epochs. For both diagrams, coloured lines correspond to GNNs trained with datasets of different size, namely patches extracted from 50, 100, 200 and 400 FE simulations.}
            \label{fig:variable_dogbone_accuracy_curves}
        \end{figure}

        All the numerical examples, plots and figures presented in this section refer to the validation set. We perform a similar experiment as in the section [\ref{const dogbone dataset}], where we investigate the dependency of the accuracy on the size of the dataset [Fig \ref{fig:variable_dogbone_accuracy_curves}]. The accuracy increases as we increase the size of the dataset, from 50.21\% for the GNN trained with 50 FE simulations to 64.30\% for the GNN trained with 400 FE simulations. Compared to the previous family of dogbone specimen, where the dimensions of the structure and the cylindrical hole did not change from one realisation to another, we can observe that the accuracy that can be achieved using 400 FE simulations decreases from 71\% to 64\% for the variable dogbone case. This result was to be expected since the space of shapes that we attempt to learn has more parameters than before. In order to achieve similar accuracy, we create additional 600 FE simulations and we train the GNN for 100 epochs using as initial weights the ones determined by the GNN trained with 400 FE simulations. The accuracy achieved for this case is 70\%. 
        
\begin{figure}[htb!]
            \begin{center}
                \includegraphics[trim={0cm 0cm 0cm 0cm}, width=.95\linewidth]{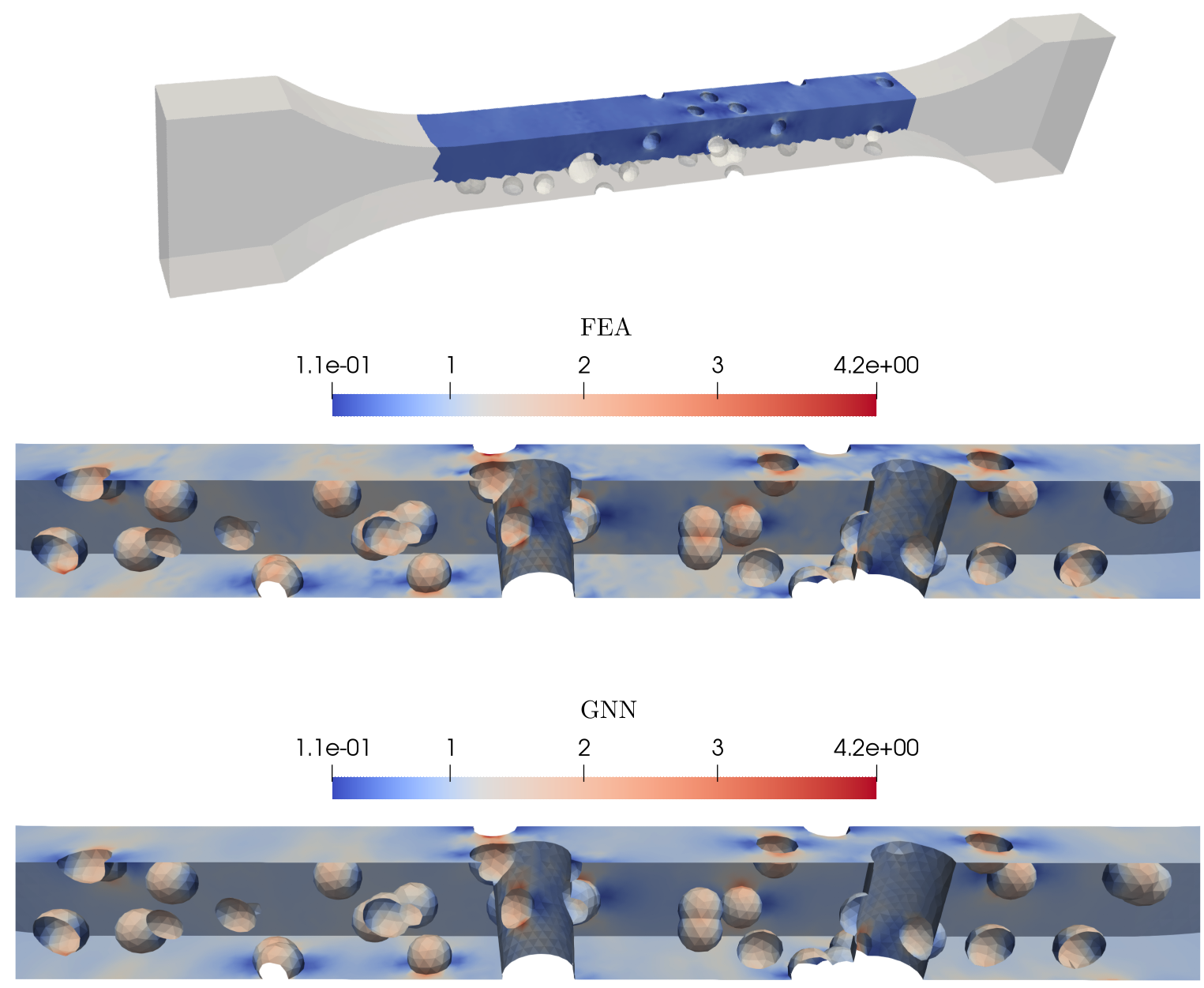}
                \caption{Comparison between the Von Mises stress distribution, extracted from the top of the structure (top), as calculated by FEA (middle), and as predicted by the GNN (bottom). We can observe that the two distributions are qualitatively very similar. The GNN result is reconstructed from the union of multiple patch predictions where only the ROI is extracted. The mean relative error in the dense mesh area of the specimen is 0.1285 while the mean scaled error is 0.0117. The scaled error is defined as $|VM_\text{FEA} - VM_\text{NN}|/\text{max}(VM_\text{FEA})$, where $VM_\text{NN}$ is the microscale Von Mises stress predicted by the GNN and  $VM_\text{FEA}$ is the microscale Von Mises stress calculated by FEA.}
                \centering
                \label{fig:variable_dogbone_file98}
            \end{center}
        \end{figure}

        Additionally, in order to examine the stress distribution, in the following figure we have extracted the upper part of the structure so that we can observe the stress distribution inside. This can be found in [Fig \ref{fig:variable_dogbone_file98}], where we can see that the GNN successfully reconstructed the Von Mises distribution.

\begin{figure}[htb]
            \begin{center}
                \includegraphics[width=\linewidth]{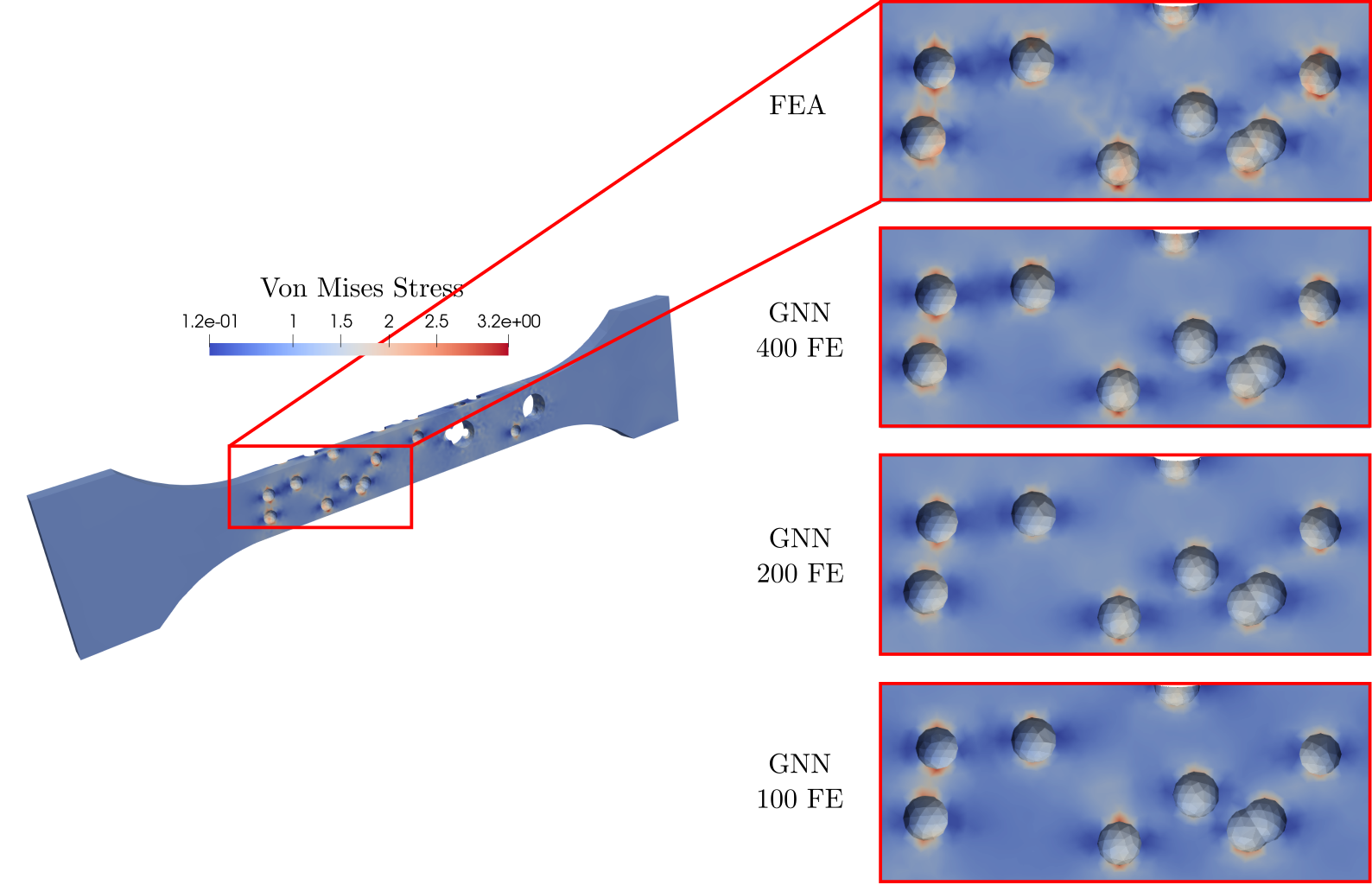}
                \caption{On the left the Von Mises distribution on the surface of a dogbone structure as calculated by FE simulations. On the right, we can see a zoom of an area of the dogbone. From top to bottom we see the FE simulation results, the GNN prediction for a GNN that was trained with patches extracted from 400, 200 and 100 FE simulations. We can observe that all the GNN predictions are very close to the FE results. The GNN result is reconstructed from the union of multiple patch predictions where only the ROI is extracted. The mean relative error in the dense mesh area of the specimens is 0.1664, 0.1263 and 0.1058 for the 100, 200 and 400 FE simulations respectively. The mean scaled error in the dense mesh area is 0.0199, 0.0169 and 0.0143 for the 100, 200 and 400 FE simulations respectively. The scaled error is defined as $|VM_\text{FEA} - VM_\text{NN}|/\text{max}(VM_\text{FEA})$, where $VM_\text{NN}$ is the microscale Von Mises stress predicted by the GNN and  $VM_\text{FEA}$ is the microscale Von Mises stress calculated by FEA.}
                \centering
                \label{fig:variable_dogbone_file92_half}
            \end{center}
        \end{figure}

\begin{figure}[htb]
            \begin{center}
                \includegraphics[width=\linewidth]{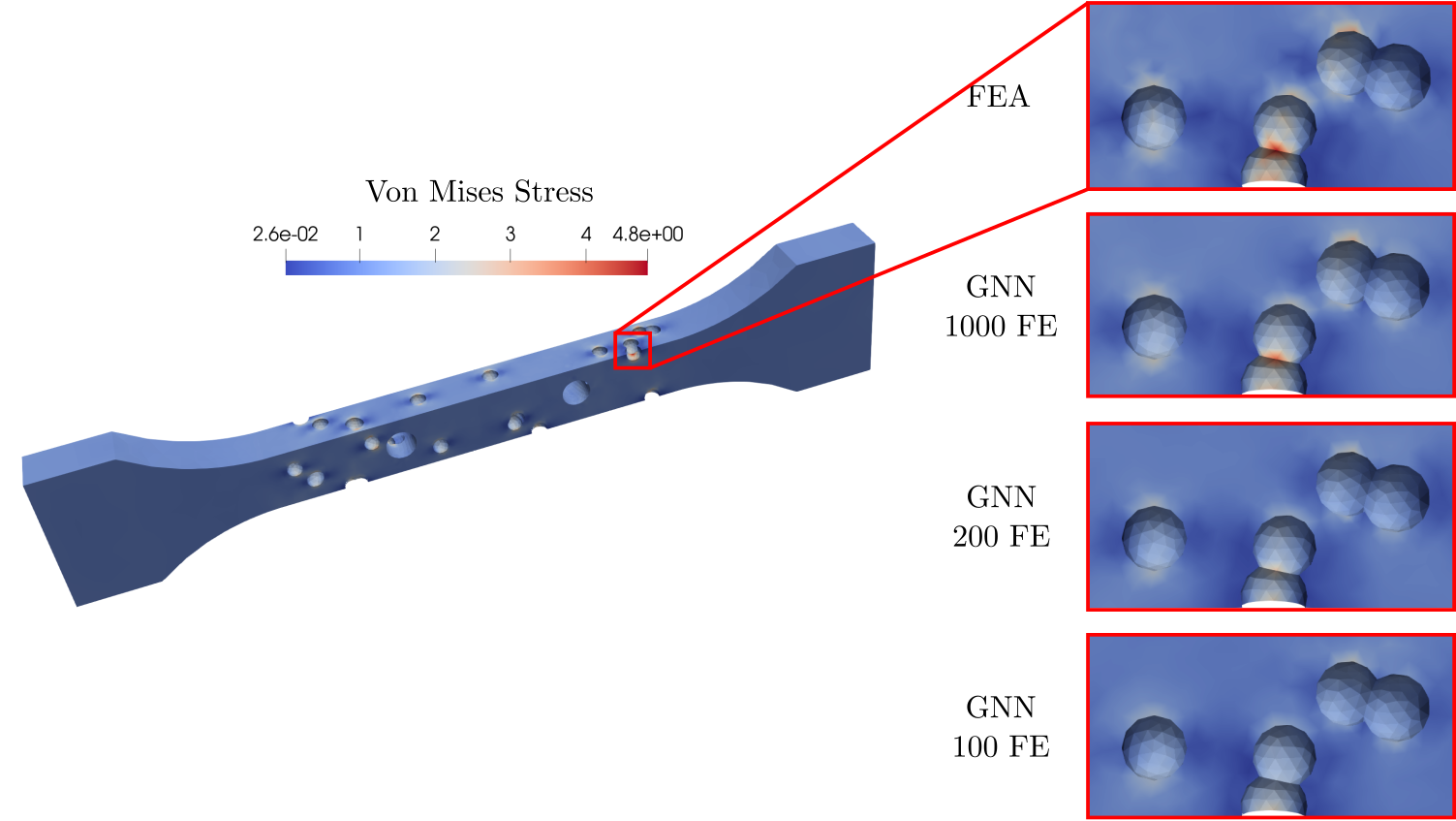}
                \caption{On the left the Von Mises distribution on the surface of a dogbone structure as calculated by FE simulations. On the right, we can see a zoom of the maximum Von Mises stress area. From top to bottom we see the FE simulation results, the GNN prediction for a GNN that was trained with patches extracted from 400, 200 and 100 FE simulations. We can observe that only the GNN that was trained using 400 FE simulations was able to correctly predict the maximum Von Mises stress that is created from the interaction between two spherical voids. The GNN result is reconstructed from the union of multiple patch predictions where only the ROI is extracted. The mean relative error in the dense mesh area of the specimens is 0.1994, 0.1384 and 0.1153 for the 100, 200 and 1000 FE simulations respectively. The mean scaled error in the dense mesh area is 0.0135, 0.0110 and 0.0090 for the 100, 200 and 1000 FE simulations respectively. The scaled error is defined as $|VM_\text{FEA} - VM_\text{NN}|/\text{max}(VM_\text{FEA})$, where $VM_\text{NN}$ is the microscale Von Mises stress predicted by the GNN and  $VM_\text{FEA}$ is the microscale Von Mises stress calculated by FEA.}
                \centering
                \label{fig:variable_dogbone_file91}
            \end{center}
        \end{figure}

        Lastly, we wish to examine the quality of the stress distribution with respect to the size of the training dataset. In [Fig \ref{fig:variable_dogbone_file92_half}] we see the Von Mises prediction for a case where not many strong interactions are present. We can observe that all the GNNs even the one trained with 100 FE simulations successfully predicts the correct Von Mises stress distribution. On the other hand, in [Fig \ref{fig:variable_dogbone_file91}] we can see a case where two intersecting defects, located at the corner of the domain, strongly interact and we can observe that the GNNs trained with less than 200 FE results fail to predict this interaction.
        
\section*{Additional remarks and insights}

\paragraph{Influence of the choice of the structural domain}

Upon closer inspection of [Fig. \ref{fig:in_out}], the choice we made to consider the cylindrical hole as a macroscopic geometrical feature was not an obvious one. Indeed, the patch exemplified in [Fig. \ref{fig:in_out}] is affected by the macroscopic stress concentration emanating from the hole, but the RoI itself is not, the buffer region being wide enough to diffuse this sharp gradient in the solution field. In that case, considering the hole as an additional element of microscopic randomness, and consequently using as input to the GNN a homogeneous stress field over the entire gage section of the specimen, would have been possible. At this stage of our investigations, we do not know the effect of such a modification on the quantity of training examples needed to achieve a given level of accuracy. This is left to future studies.

However, for a larger hole with respect to the pore and patch sizes, the picture would be very different indeed. The stress concentration created by the hole would penetrate the RoI. Without macroscopic solution representing the effect of the hole, as done for instance to produce the result displayed in [Fig. \ref{fig:variable_dogbone_file75}], the GNN would not be given any information about this stress concentration, which would result in erroneous GNN predictions. Therefore, the choice of the patch size and the choice of the elements that may be considered as elements of the microstructure cannot be made independently. At the very least, the patch should be larger than the largest microstructural element. Ultimately, the patch must be provided with all the information needed to characterise the mechanical loading applied to the RoI by the remainder of the structure. This information is encoded in the restriction of the computed macroscopic field to the patch domain.

As to why we state that our methodology works \say{without scale separation}, this may require some additional insights. Our ROIs can include arbitrary interactions between the microstructural elements (\textit{i.e.} the pores) and the boundaries, as opposed to what can be done in homogenisation frameworks, when scale separation is assumed. Moreover the macroscopic stress fields that are provided as \say{far field conditions} for the ROIs may vary arbitrarily fast in space, \textit{i.e.} macroscopic mechanical gradients are not limited to low-order polynomiality over our microstructural volume elements. Finally, the patches may be chosen as large as required, as discussed in the previous paragraph, formally encompassing direct numerical simulation when their size tends to the size of the structure. Fundamentally, our ROIs and the associated microstructural elements are explicitly positioned in space, while RVEs are associated with macroscopic material point in homogenisation theories, thereby excluding the possibility for detailed interaction between microscopic and macroscopic elements.

\paragraph{Finite element modelling errors}

In all our studies, we have chosen a relatively large element size, so that our finite element simulations can run without parallelisation. The element size is a fraction of the pores' diameters. The GNN learns to produce finite element solutions with that level of mesh refinement. As all simulations in our training set are performed with similar mesh sizes, we cannot expect the GNN to perform well with other mesh sizes. And of course, we cannot expect the GNN to produce solutions that are more accurate than the finite element approximations that are provided as training data.

Furthermore, we have chosen to use the FE mesh as GNN mesh. This is a limitation, in the sense that when using finer FEA meshes, we would need to deepen the GNN architecture to obtain the same level of information diffusion. This can be easily remedied by separating the two meshes, \textit{i.e.} choosing an FEA mesh density that is tailored to provide accurate mechanical fields, and using another GNN mesh size that is tailored to the machine learning task at hand.

% \clearpage
%#######################################################################################################################################
\section{Conclusions}

    In this work we have developed a multiscale Neural-Network-based (NN) surrogate modelling approach to circumvent the need for 3D direct numerical analysis of elastic stress simulations in porous materials and structures. 
    Given an inexpensively computed macroscopic approximation of the stress field, where the pore network is ignored, and a fine-scale representation of the geometry of the structure and of the network of microscale pores, the NN corrects the macroscopic stress field to produce fields that emulate the output of the direct numerical simulation (DNS). The NN is trained in a supervised manner using selected DNS examples.
    
    The developed framework is based on geometric learning. We have proposed to use a Graph Neural Network (GNN) to perform convolution-based learning over the surface of the porous structure, which is represented by a mesh of triangles. This surface mesh represents both the structure boundary and the surface of the pores. An advantage of using geometric learning for the surrogate modelling of porous structures is that we eliminate the need to perform unnecessary algorithmic operations associated with the analysis of bulk information, which would typically be required in a standard voxel-based CNN approach. In addition, the GNN architecture allows us to operate on a variable resolution mesh of the geometry rather than a representation with fixed spatial resolution. Our choice to deploy geometric learning is supported by three observations. Firstly, the surface representation is a comprehensive representation of the geometry of the specimen. Secondly the trace of the smooth macro stress over the surface mesh is empirically found to be a sufficiently rich structure-scale information for the GNN to produce correct microscale corrections of the stress state in the porous material. At last, we have shown numerically that the maximum microscale stress field in the porous network that we analysed in this contribution was always found to be located at the surface of the porous structure. 

    In terms of GNNs, we employ a well-established Encoder-Processor-Decoder architecture and use a state-of-the-art convolution introduced in \citep{battaglia2018relational}. We add two elements of novelty specifically tailored to the multiscale nature of our problem. Firstly, we operate on patches extracted from the porous structure, instead of the structure itself. In [Appendix \ref{appendix:full_vs_patches}] we show how this strategy leads to better generalisation. Lastly, in order to perform geometric learning on the surface of the porous structures we use dual geodesic and Euclidean convolutions. The Euclidean convolutions are necessary to connect disconnected parts of the input mesh, i.e. to diffuse information between non-intersecting pores in the structures of interest. In [Appendix \ref{appendix:Geod_and_Eucl_conv}] we perform a study to highlight the advantages of using dual convolutions.
    
    This work extends our previous work on 2D surrogate modelling using standard pixel-based CNNs \citep{krokos2021bayesian}, to 3D graph-based Graph Neural Networks. We have reported similar trends in prediction abilities for the 3D and 2D surrogate modelling approaches, including epistemic uncertainty quantification, which is performed via a Bayesian formulation of the GNN. 
    In particular, we have examined the ability of the proposed approach to predict maximum equivalent stress measures, namely the Von Mises stress, in two different settings :
    (i) a Monte-Carlo setting whereby repeated and/or fast structural calculations with random distributions of pores are to be performed and (ii) the surrogate modelling of parameterised structural problems with random microstructures, which may be of interest to develop methodologies of structural design and optimisation under (microscale) uncertainty. 

    \begin{itemize}
    \item
    For application (i), we have shown that relatively few offline simulations were required to train the GNN to an insightful level of accuracy. Typically, training the GNN with 50 direct numerical simulations yields predictions where 80\% of new outputs are predicted with less than 20\% relative error, whilst training it with 800 FE simulations gives predictions where 80\% of new outputs are predicted with less than 15\% relative error.
    % R1.Q4
    However, our investigations also show that in cases where a lower error threshold is required, then the training dataset needs to be considerably larger. In this case the applicability of the method is limited to applications whereby (a) the surrogate model needs to be called thousands of times, or (b) results are to be delivered in quasi-real-time. (a) typically arises when studying all possible relative arrangements of micro-structural features within a structural component of fixed geometry, whilst (b) may arise when performing non-destructive testing from CT scans (i.e. physics-based data augmentation). Of course, it is always possible to revert to a ``more online" strategy, whereby the NN is only used to predict likely hot-spots, typically using upper quartiles of the probabilistic predictions, that are then reanalysed as needed using concurrent multiscale finite element analyses (i.e. local FE analyses of highly stressed regions with boundary conditions computed from structural-scale analyses).
    % R1.Q5

    \item
    For application (ii), we have shown that introducing parameterised geometries vastly increases the complexity of the problem. Using global parameters makes the dataset sparser, leading to an exponential increase in training data requirements (the infamous curse of dimensionality). So, at the current stage of our understanding of the capabilities of deep learning algorithms, we recommend to either deploy this type of deep learning methodologies to predict stresses in random distributions of microscale pores over fixed macroscale geometries, or use a very small number of structural parameters.
    \end{itemize}
    
    Lastly, we have introduced a novel approach to enforce physics constraints at evaluation time. We choose to force the stress predictions delivered by the NN surrogate to satisfy the homogeneous Neumann conditions on the surface of the porous structure. This is done within a recursive Bayesian setting, similarly to probabilistic state estimation. The Bayesian GNN (BGNN) predictions are considered as prior for this task, and posterior predictions that conform to the imposed Neumann conditions are computed via an ensemble Kalman update. We have shown that without this online correction, downward bias is observed for maximum equivalent stresses, which is problematic. Indeed, maximum equivalent stresses, which are statistically the tails of predicted fields of equivalent stresses, are precisely what we are interested in in structural integrity assessment. Online physics-based corrections, in this context, may be interpreted as \say{free} data that, as we have shown, help lift those quantities of interest back up. It should be noted that this approach may be used complementary to active learning in the future, as we have shown that these biased predictions are correctly identified by the Bayesian Neural Network as uncertain outputs.

    Finally, we stress that we have not compared the efficiency of our Graph Neural Network to voxel-based CNN. We gave qualitative arguments for computational gain, namely the facts that (i) our NN operates on the surface of the porous structure instead of its volume and (ii) we may be able to operate on surface meshes with heterogeneous element size, hence paving the way to neural meta-models with adaptive spatial resolution. Yet, voxel-based CNNs can be (and are) highly optimised taking advantage of the regular grid that they operate on. Thus, a definitive CPU time comparison remained to be done to better quantify the advantage of geometric learning over voxel-based CNNs in the context of stress analysis for porous structures.

% \clearpage 
\section*{Acknowledgments} 
%% This is mandatory acknowledgement of the MSCA funding RAINBOW, leave as is
\begin{minipage}{0.5\textwidth}
    \includegraphics[width=\textwidth]{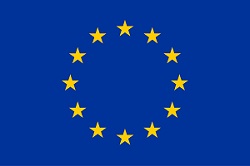}
\end{minipage}\hspace{10pt}
\begin{minipage}{0.5\textwidth}
    This project has received funding from the European Union’s Horizon 2020 research and innovation programme under the Marie Sklodowska-Curie grant agreement No. 764644.\\
    
    This paper only contains the author's views and the Research Executive Agency and the Commission are not responsible for any use that may be made of the information it contains.\\
\end{minipage}
\hfill \break
\hfill \break
We acknowledge the support of the Supercomputing Wales project, which is part-funded by the European Regional Development Fund (ERDF) via Welsh Government.

\hfill \break
We thank Synopsys for its support and specifically Dr Viet Bui Xuan for his constructive suggestions during the planning and development of this research work.

% References
\bibliography{references}

\begin{thebibliography}{}

\bibitem[Alnæs et~al., 2015]{fenics}
Alnæs, M., Blechta, J., Hake, J., Johansson, A., Kehlet, B., Logg, A., Richardson, C., Ring, J., Rognes, M., and Wells, G. (2015).
\newblock The fenics project version 1.5.
\newblock 3.

\bibitem[Ba et~al., 2016]{ba2016layer}
Ba, J.~L., Kiros, J.~R., and Hinton, G.~E. (2016).
\newblock Layer normalization.
\newblock arXiv preprint arXiv: 1607.06450.

\bibitem[Battaglia et~al., 2018]{battaglia2018relational}
Battaglia, P.~W., Hamrick, J.~B., Bapst, V., Sanchez-Gonzalez, A., Zambaldi, V., Malinowski, M., Tacchetti, A., Raposo, D., Santoro, A., Faulkner, R., Gulcehre, C., Song, F., Ballard, A., Gilmer, J., Dahl, G., Vaswani, A., Allen, K., Nash, C., Langston, V., Dyer, C., Heess, N., Wierstra, D., Kohli, P., Botvinick, M., Vinyals, O., Li, Y., and Pascanu, R. (2018).
\newblock Relational inductive biases, deep learning, and graph networks.
\newblock arXiv preprint arXiv: 1806.01261.

\bibitem[Bishop, 1995]{Bishop}
Bishop, C.~M. (1995).
\newblock {\em Neural Networks for Pattern Recognition}.
\newblock Oxford University Press, Inc., USA.

\bibitem[Blundell et~al., 2015]{blundell2015weight}
Blundell, C., Cornebise, J., Kavukcuoglu, K., and Wierstra, D. (2015).
\newblock Weight uncertainty in neural networks.
\newblock In {\em Proceedings of the 32nd International Conference on International Conference on Machine Learning - Volume 37}, ICML'15, page 1613–1622. JMLR.org.

\bibitem[Buda et~al., 2019]{BUDA2019218}
Buda, M., Saha, A., and Mazurowski, M.~A. (2019).
\newblock Association of genomic subtypes of lower-grade gliomas with shape features automatically extracted by a deep learning algorithm.
\newblock {\em Computers in Biology and Medicine}, 109:218--225.

\bibitem[Chakraborty et~al., 2022]{Ayan_2022}
Chakraborty, A., Anitescu, C., Zhuang, X., and Rabczuk, T. (2022).
\newblock Domain adaptation based transfer learning approach for solving pdes on complex geometries.
\newblock {\em Engineering with Computers}.

\bibitem[Deshpande et~al., 2023a]{deshpande2023magnet}
Deshpande, S., Bordas, S. P.~A., and Lengiewicz, J. (2023a).
\newblock Magnet: A graph u-net architecture for mesh-based simulations.
\newblock arXiv preprint arXiv.2211.00713.

\bibitem[Deshpande et~al., 2022]{deshpande2021fembased}
Deshpande, S., Lengiewicz, J., and Bordas, S.~P. (2022).
\newblock Probabilistic deep learning for real-time large deformation simulations.
\newblock {\em Computer Methods in Applied Mechanics and Engineering}, 398:115307.

\bibitem[Deshpande et~al., 2023b]{Deshpande_Sosa_Bordas_Lengiewicz_2023}
Deshpande, S., Sosa, R.~I., Bordas, S. P.~A., and Lengiewicz, J. (2023b).
\newblock Convolution, aggregation and attention based deep neural networks for accelerating simulations in mechanics.
\newblock {\em Frontiers in Materials}, 10.

\bibitem[Evensen, 1994]{Evensen1994_EnKF}
Evensen, G. (1994).
\newblock Sequential data assimilation with a nonlinear quasi-geostrophic model using monte carlo methods to forecast error statistics.
\newblock {\em Journal of Geophysical Research: Oceans}, 99(C5):10143--10162.

\bibitem[Gendre et~al., 2009]{gendre:hal-00437023}
Gendre, L., Allix, O., Gosselet, P., and Comte, F. (2009).
\newblock {Non-intrusive and exact global/local techniques for structural problems with local plasticity}.
\newblock {\em {Computational Mechanics}}, 44(2):233--245.

\bibitem[Geuzaine and Remacle, 2009]{gmsh}
Geuzaine, C. and Remacle, J.-F. (2009).
\newblock Gmsh: A 3-d finite element mesh generator with built-in pre- and post-processing facilities.
\newblock {\em International Journal for Numerical Methods in Engineering}, 79:1309 -- 1331.

\bibitem[Goetz et~al., 2022]{Goetz_2022}
Goetz, A., Durmaz, A., Müller, M., Thomas, A., Britz, D., Kerfriden, P., and Eberl, C. (2022).
\newblock Addressing materials’ microstructure diversity using transfer learning.
\newblock {\em npj Computational Materials}, 8:27.

\bibitem[Gong et~al., 2020]{gong2020geometrically}
Gong, S., Bahri, M., Bronstein, M.~M., and Zafeiriou, S. (2020).
\newblock Geometrically principled connections in graph neural networks.
\newblock In {\em 2020 IEEE/CVF Conference on Computer Vision and Pattern Recognition (CVPR)}, pages 11412--11421, Los Alamitos, CA, USA. IEEE Computer Society.

\bibitem[Graves, 2011]{Graves_2011}
Graves, A. (2011).
\newblock Practical variational inference for neural networks.
\newblock In {\em Proceedings of the 24th International Conference on Neural Information Processing Systems}, NIPS'11, page 2348–2356, Red Hook, NY, USA. Curran Associates Inc.

\bibitem[Guo and Buehler, 2020]{Guo_Buehler_2020}
Guo, K. and Buehler, M.~J. (2020).
\newblock A semi-supervised approach to architected materials design using graph neural networks.
\newblock {\em Extreme Mechanics Letters}, 41:101029.

\bibitem[Hanocka et~al., 2019]{Hanocka_2019}
Hanocka, R., Hertz, A., Fish, N., Giryes, R., Fleishman, S., and Cohen-Or, D. (2019).
\newblock Meshcnn: a network with an edge.
\newblock {\em ACM Transactions on Graphics (TOG)}, 38:1 -- 12.

\bibitem[He et~al., 2015]{he2015delving}
He, K., Zhang, X., Ren, S., and Sun, J. (2015).
\newblock Delving deep into rectifiers: Surpassing human-level performance on imagenet classification.
\newblock In {\em 2015 IEEE International Conference on Computer Vision (ICCV)}, pages 1026--1034.

\bibitem[He et~al., 2016]{he2015deep}
He, K., Zhang, X., Ren, S., and Sun, J. (2016).
\newblock Deep residual learning for image recognition.
\newblock In {\em 2016 IEEE Conference on Computer Vision and Pattern Recognition (CVPR)}, pages 770--778, Los Alamitos, CA, USA. IEEE Computer Society.

\bibitem[Hesthaven et~al., 2015]{Multiscale_Zhu}
Hesthaven, J., Zhang, S., and Zhu, X. (2015).
\newblock Reduced basis multiscale finite element methods for elliptic problems.
\newblock {\em SIAM Journal on Multiscale Modeling and Simulation}, 13:316--337.

\bibitem[Hinton et~al., 2012]{hinton2012improving}
Hinton, G.~E., Srivastava, N., Krizhevsky, A., Sutskever, I., and Salakhutdinov, R.~R. (2012).
\newblock Improving neural networks by preventing co-adaptation of feature detectors.
\newblock arXiv preprint arXiv: 1207.0580.

\bibitem[Hinton and van Camp, 1993]{vanCamp_1993}
Hinton, G.~E. and van Camp, D. (1993).
\newblock Keeping the neural networks simple by minimizing the description length of the weights.
\newblock In {\em Proceedings of the Sixth Annual Conference on Computational Learning Theory}, COLT '93, page 5–13, New York, NY, USA. Association for Computing Machinery.

\bibitem[Hoang et~al., 2016]{HOANG2016121}
Hoang, K., Kerfriden, P., and Bordas, S. (2016).
\newblock A fast, certified and “tuning free” two-field reduced basis method for the metamodelling of affinely-parametrised elasticity problems.
\newblock {\em Computer Methods in Applied Mechanics and Engineering}, 298:121--158.

\bibitem[Hochreiter et~al., 2001]{chapter-gradient-flow-2001}
Hochreiter, S., Bengio, Y., and Frasconi, P. (2001).
\newblock Gradient flow in recurrent nets: the difficulty of learning long-term dependencies.
\newblock In Kolen, J. and Kremer, S., editors, {\em Field Guide to Dynamical Recurrent Networks}. IEEE Press.

\bibitem[Jaegle et~al., 2022]{Jaegle_et_al_2022}
Jaegle, A., Borgeaud, S., Alayrac, J.-B., Doersch, C., Ionescu, C., Ding, D., Koppula, S., Zoran, D., Brock, A., Shelhamer, E., Hénaff, O., Botvinick, M.~M., Zisserman, A., Vinyals, O., and Carreira, J. (2022).
\newblock Perceiver io: A general architecture for structured inputs \& outputs.
\newblock arXiv preprint arXiv.2107.14795.

\bibitem[Jiang et~al., 2021]{Nie2020}
Jiang, H., Nie, Z., Yeo, R., Farimani, A.~B., and Kara, L.~B. (2021).
\newblock {StressGAN: A Generative Deep Learning Model for Two-Dimensional Stress Distribution Prediction}.
\newblock {\em Journal of Applied Mechanics}, 88(5).
\newblock 051005.

\bibitem[Kerfriden et~al., 2009]{Kerfriden_Allix_2009}
Kerfriden, P., Allix, O., and Gosselet, P. (2009).
\newblock A three-scale domain decomposition method for the 3d analysis of debonding in laminates.
\newblock {\em Computational Mechanics}, 44:343--362.

\bibitem[Khosla et~al., 2018]{3D_class_2018}
Khosla, M., Jamison, K., Kuceyeski, A., and Sabuncu, M. (2018).
\newblock {\em 3D Convolutional Neural Networks for Classification of Functional Connectomes: 4th International Workshop, DLMIA 2018, and 8th International Workshop, ML-CDS 2018, Held in Conjunction with MICCAI 2018, Granada, Spain, September 20, 2018, Proceedings}, pages 137--145.

\bibitem[Kim et~al., 2016]{kim2016deeplyrecursive}
Kim, J., Lee, J.~K., and Lee, K.~M. (2016).
\newblock Deeply-recursive convolutional network for image super-resolution.
\newblock In {\em 2016 IEEE Conference on Computer Vision and Pattern Recognition (CVPR)}, pages 1637--1645.

\bibitem[Kingma and Welling, 2014]{kingma2014autoencoding}
Kingma, D.~P. and Welling, M. (2014).
\newblock Auto-encoding variational bayes.
\newblock arXiv preprint arXiv: 1312.6114.

\bibitem[Krokos et~al., 2021]{krokos2021bayesian}
Krokos, V., Bui~Xuan, V., Bordas, S., Young, P., and Kerfriden, P. (2021).
\newblock A bayesian multiscale cnn framework to predict local stress fields in structures with microscale features.
\newblock {\em Computational Mechanics}, pages 1--34.

\bibitem[Lei et~al., 2021]{Lei2021PicassoAC}
Lei, H., Akhtar, N., and Mian, A. (2021).
\newblock Picasso: A cuda-based library for deep learning over 3d meshes.
\newblock In {\em Proceedings of the IEEE/CVF Conference on Computer Vision and Pattern Recognition}, pages 13854--13864.

\bibitem[Lim et~al., 2017]{lim2017enhanced}
Lim, B., Son, S., Kim, H., Nah, S., and Lee, K.~M. (2017).
\newblock Enhanced deep residual networks for single image super-resolution.
\newblock In {\em 2017 IEEE Conference on Computer Vision and Pattern Recognition Workshops (CVPRW)}, pages 1132--1140.

\bibitem[Lino et~al., 2021]{lino2021simulating}
Lino, M., Cantwell, C., Bharath, A.~A., and Fotiadis, S. (2021).
\newblock Simulating continuum mechanics with multi-scale graph neural networks.
\newblock arXiv preprint arXiv: 2106.04900.

\bibitem[Logg et~al., 2012]{LoggMardalEtAl2012}
Logg, A., Mardal, K.-A., Wells, G.~N., et~al. (2012).
\newblock {\em Automated Solution of Differential Equations by the Finite Element Method}.
\newblock Springer.

\bibitem[Lu et~al., 2019]{LU2019422}
Lu, H., Wang, H., Zhang, Q., Yoon, S.~W., and Won, D. (2019).
\newblock A 3d convolutional neural network for volumetric image semantic segmentation.
\newblock {\em Procedia Manufacturing}, 39:422--428.
\newblock 25th International Conference on Production Research Manufacturing Innovation: Cyber Physical Manufacturing August 9-14, 2019 | Chicago, Illinois (USA).

\bibitem[Lye et~al., 2020]{Lye2020DeepLO}
Lye, K.~O., Mishra, S., and Ray, D. (2020).
\newblock Deep learning observables in computational fluid dynamics.
\newblock {\em Journal of Computational Physics}, 410:109339.

\bibitem[Masci et~al., 2015]{masci2018geodesic}
Masci, J., Boscaini, D., Bronstein, M.~M., and Vandergheynst, P. (2015).
\newblock Geodesic convolutional neural networks on riemannian manifolds.
\newblock In {\em 2015 IEEE International Conference on Computer Vision Workshop (ICCVW)}, pages 832--840, Los Alamitos, CA, USA. IEEE Computer Society.

\bibitem[Mendizabal et~al., 2020]{Cotin1}
Mendizabal, A., Márquez-Neila, P., and Cotin, S. (2020).
\newblock Simulation of hyperelastic materials in real-time using deep learning.
\newblock {\em Medical Image Analysis}, 59:101569.

\bibitem[Mylonas et~al., 2022]{mylonas2021bayesian}
Mylonas, C., Tsialiamanis, G., Worden, K., and Chatzi, E.~N. (2022).
\newblock Bayesian graph neural networks for strain-based crack localization.
\newblock In Madarshahian, R. and Hemez, F., editors, {\em Data Science in Engineering, Volume 9}, pages 253--261, Cham. Springer International Publishing.

\bibitem[Nie et~al., 2019]{ResCNN}
Nie, Z., Jiang, H., and Kara, L.~B. (2019).
\newblock Stress field prediction in cantilevered structures using convolutional neural networks.
\newblock {\em Journal of Computing and Information Science in Engineering}, 20(1).

\bibitem[Oden et~al., 2006]{Oden_2006}
Oden, J., Prudhomme, S., Romkes, A., and Bauman, P. (2006).
\newblock Multiscale modeling of physical phenomena: Adaptive control of models.
\newblock {\em SIAM Journal on Scientific Computing}, 28(6):2359--2389.

\bibitem[Oden et~al., 1999]{ODEN19993}
Oden, J.~T., Vemaganti, K., and Moës, N. (1999).
\newblock Hierarchical modeling of heterogeneous solids.
\newblock {\em Computer Methods in Applied Mechanics and Engineering}, 172(1):3--25.

\bibitem[Paladim et~al., 2016]{Paladin_Kerfriden_2016}
Paladim, D., Almeida, J., Bordas, S., and Kerfriden, P. (2016).
\newblock Guaranteed error bounds in homogenisation: an optimum stochastic approach to preserve the numerical separation of scales.
\newblock {\em International Journal for Numerical Methods in Engineering}, 110.

\bibitem[Pereira-Alvarez et~al., 2021]{Pereira2021}
Pereira-Alvarez, P., Kerfriden, P., Ryckelynck, D., and ROBIN, V. (2021).
\newblock Real-time data assimilation in welding operations using thermal imaging and accelerated high-fidelity digital twinning.
\newblock {\em Mathematics}, 9:2263.

\bibitem[Perera et~al., 2022]{Perera_Guzzetti_Agrawal_2022}
Perera, R., Guzzetti, D., and Agrawal, V. (2022).
\newblock Graph neural networks for simulating crack coalescence and propagation in brittle materials.
\newblock {\em Computer Methods in Applied Mechanics and Engineering}, 395:115021.

\bibitem[Pfaff et~al., 2021]{pfaff2021learning}
Pfaff, T., Fortunato, M., Sanchez-Gonzalez, A., and Battaglia, P.~W. (2021).
\newblock Learning mesh-based simulation with graph networks.
\newblock International Conference on Learning Representations; arXiv preprint arXiv: 2010.03409.

\bibitem[Pilkey and Pilkey, 2008]{SIFs}
Pilkey, W. and Pilkey, D. (2008).
\newblock Peterson's stress concentration factors, third edition.
\newblock {\em Peterson's Stress Concentration Factors, Third Edition}, pages 1--522.

\bibitem[Qi et~al., 2017]{qi2017pointnet}
Qi, C.~R., Su, H., Mo, K., and Guibas, L.~J. (2017).
\newblock Pointnet: Deep learning on point sets for 3d classification and segmentation.
\newblock Conference on Computer Vision and Pattern Recognition (CVPR) 2017; arXiv preprint arXiv: 1612.00593.

\bibitem[Raghavan and Ghosh, 2004]{Ghosh_2004}
Raghavan, P. and Ghosh, S. (2004).
\newblock Concurrent multi-scale analysis of elastic composites by a multi-level computational model.
\newblock {\em Computer Methods in Applied Mechanics and Engineering}, 193(6):497--538.

\bibitem[Raissi et~al., 2019]{RAISSI2019686}
Raissi, M., Perdikaris, P., and Karniadakis, G. (2019).
\newblock Physics-informed neural networks: A deep learning framework for solving forward and inverse problems involving nonlinear partial differential equations.
\newblock {\em Journal of Computational Physics}, 378:686--707.

\bibitem[Raissi et~al., 2020]{Karniadakis_HiddenFluidDynamics}
Raissi, M., Yazdani, A., and Karniadakis, G.~E. (2020).
\newblock Hidden fluid mechanics: Learning velocity and pressure fields from flow visualizations.
\newblock {\em Science}, 367(6481):1026--1030.

\bibitem[Rao and Liu, 2020]{RAO2020109850}
Rao, C. and Liu, Y. (2020).
\newblock Three-dimensional convolutional neural network (3d-cnn) for heterogeneous material homogenization.
\newblock {\em Computational Materials Science}, 184:109850.

\bibitem[Rocha et~al., 2020]{ROCHA2020103995}
Rocha, I., Kerfriden, P., and {van der Meer}, F. (2020).
\newblock Micromechanics-based surrogate models for the response of composites: A critical comparison between a classical mesoscale constitutive model, hyper-reduction and neural networks.
\newblock {\em European Journal of Mechanics - A/Solids}, 82:103995.

\bibitem[Ryckelynck, 2009]{Ryckelynck2009}
Ryckelynck, D. (2009).
\newblock Hyper-reduction of mechanical models involving internal variables.
\newblock {\em International Journal for Numerical Methods in Engineering}, 77:75 -- 89.

\bibitem[Sanchez-Gonzalez et~al., 2020]{sanchezgonzalez2020learning}
Sanchez-Gonzalez, A., Godwin, J., Pfaff, T., Ying, R., Leskovec, J., and Battaglia, P. (2020).
\newblock Learning to simulate complex physics with graph networks.
\newblock In III, H.~D. and Singh, A., editors, {\em Proceedings of the 37th International Conference on Machine Learning}, volume 119 of {\em Proceedings of Machine Learning Research}, pages 8459--8468. PMLR.

\bibitem[Sanchez-Gonzalez et~al., 2018]{sanchez2018graph}
Sanchez-Gonzalez, A., Heess, N., Springenberg, J.~T., Merel, J., Riedmiller, M., Hadsell, R., and Battaglia, P. (2018).
\newblock Graph networks as learnable physics engines for inference and control.
\newblock In {\em International Conference on Machine Learning}, pages 4470--4479. PMLR.

\bibitem[Schlömer, 2021]{schlomer_nico_2021_5591953}
Schlömer, N. (2021).
\newblock pygmsh: A python frontend for gmsh.

\bibitem[Schult et~al., 2020]{schult2020dualconvmeshnet}
Schult, J., Engelmann, F., Kontogianni, T., and Leibe, B. (2020).
\newblock Dualconvmesh-net: Joint geodesic and euclidean convolutions on 3d meshes.
\newblock In {\em 2020 IEEE/CVF Conference on Computer Vision and Pattern Recognition (CVPR)}, pages 8609--8619.

\bibitem[Senior et~al., 2019]{DeepMind_Protein_2019}
Senior, A.~W., Evans, R., Jumper, J., Kirkpatrick, J., Sifre, L., Green, T., Qin, C., Žídek, A., Nelson, A. W.~R., Bridgland, A., Penedones, H., Petersen, S., Simonyan, K., Crossan, S., Kohli, P., Jones, D.~T., Silver, D., Kavukcuoglu, K., and Hassabis, D. (2019).
\newblock Protein structure prediction using multiple deep neural networks in the 13th critical assessment of protein structure prediction (casp13).
\newblock {\em Proteins: Structure, Function, and Bioinformatics}, 87(12):1141--1148.

\bibitem[Senior et~al., 2020]{DeepMind_Protein_2020}
Senior, A.~W., Evans, R., Jumper, J., Kirkpatrick, J., Sifre, L., Green, T., Qin, C., Žídek, A., Nelson, A. W.~R., Bridgland, A., Penedones, H., Petersen, S., Simonyan, K., Crossan, S., Kohli, P., Jones, D.~T., Silver, D., Kavukcuoglu, K., and Hassabis, D. (2020).
\newblock Improved protein structure prediction using potentials from deep learning.
\newblock {\em Nature}, 577(7792):706—710.

\bibitem[Sullivan and Kaszynski, 2019]{sullivan2019pyvista}
Sullivan, C.~B. and Kaszynski, A. (2019).
\newblock {PyVista}: 3d plotting and mesh analysis through a streamlined interface for the visualization toolkit ({VTK}).
\newblock {\em Journal of Open Source Software}, 4(37):1450.

\bibitem[Sun et~al., 2020]{sun2020predicting}
Sun, Y., Hanhan, I., Sangid, M.~D., and Lin, G. (2020).
\newblock Predicting mechanical properties from microstructure images in fiber-reinforced polymers using convolutional neural networks.
\newblock arXiv: 2010.03675.

\bibitem[Sussillo and Abbott, 2015]{sussillo2015random}
Sussillo, D. and Abbott, L.~F. (2015).
\newblock Random walk initialization for training very deep feedforward networks.
\newblock arXiv preprint arXiv: 1412.6558.

\bibitem[Thomas et~al., 2020]{ma13153298}
Thomas, A., Durmaz, A.~R., Straub, T., and Eberl, C. (2020).
\newblock Automated quantitative analyses of fatigue-induced surface damage by deep learning.
\newblock {\em Materials}, 13(15).

\bibitem[Vlassis et~al., 2020]{Vlassis2020}
Vlassis, N.~N., Ma, R., and Sun, W. (2020).
\newblock Geometric deep learning for computational mechanics part i: anisotropic hyperelasticity.
\newblock {\em Computer Methods in Applied Mechanics and Engineering}, 371:113299.

\bibitem[Vu et~al., 2020]{VU2020117328}
Vu, H., Kim, H.-C., Jung, M., and Lee, J.-H. (2020).
\newblock fmri volume classification using a 3d convolutional neural network robust to shifted and scaled neuronal activations.
\newblock {\em NeuroImage}, 223:117328.

\bibitem[Zagoruyko and Komodakis, 2016]{zagoruyko2017wide}
Zagoruyko, S. and Komodakis, N. (2016).
\newblock Wide residual networks.
\newblock In Richard C.~Wilson, E. R.~H. and Smith, W. A.~P., editors, {\em Proceedings of the British Machine Vision Conference (BMVC)}, pages 87.1--87.12. BMVA Press.

\bibitem[Zohdi and Wriggers, 2005]{zohdiwrigger}
Zohdi, T. and Wriggers, P. (2005).
\newblock {\em An Introduction to Computational Micromechanics}, volume~20.

\bibitem[Évariste Sanchez-Palencia, 1987]{sanchezpalencia74}
Évariste Sanchez-Palencia (1987).
\newblock {\em General introduction to asymptotic methods}, volume 272.

\end{thebibliography}
\clearpage

% Appendices
\begin{appendices}
\section{Accuracy} \label{appendix:accuracy_algorithm}

Below we can see the algorithm that we use to calculate the accuracy of the GNN prediction [Algorithm \ref{alg:acc}].

\begin{algorithm}[h]
    \caption{Compute accuracy}\label{alg:acc}
    \begin{algorithmic}[1]
        % \Function{acc}{$datapoints$, $prediction$, $ground\_truth$, $ROI$, $threshold$ = 0.1}
        \Function{acc}{$datapoints$, $threshold$ = 0.1}
            \State
            \State $N = \textbf{length}(datapoints)$ 
            \State $accepted = \textbf{zeros}(N)$
            \State
            \For{$patch \textbf{ in } datapoints$}\ 
                \State
                \State $S_{\text{NN}} = patch.\text{prediction}$\         \Comment{get the predicted stress tensor for the current patch}
                \State $S_{\text{FE}} = patch.\text{ground\_truth}$\     \Comment{get the real stress tensor for the current patch}
                \State
                \State $S_{\text{NN}\_\text{ROI}} = S_{\text{NN}}[patch.\text{ROI}]$\  \         \Comment{extract the predicted stress values present in the ROI}
                \State $S_{\text{FE}\_\text{ROI}} = S_{\text{FE}}[patch.\text{ROI}]$\     \Comment{extract the real stress values present in the ROI}
                \State
                \State $y_{\text{NN}} = \textbf{VM}(S_{\text{NN}\_\text{ROI}})$\ \Comment{calculate the Von Mises stress for the predicted stress tensor}
                \State $y_{\text{FE}} = \textbf{VM}(S_{\text{FE}\_\text{ROI}})$\ \Comment{calculate the Von Mises stress for the real stress tensor}
                \State
                \State $y_{\text{NN\_max}} = \textbf{max}(y_{\text{NN}})$\ \Comment{get the maximum Von Mises stress for the predicted stress tensor}
                \State $y_{\text{FE\_max}} = \textbf{max}(y_{\text{FE}})$\ \Comment{get the maximum Von Mises stress for the real stress tensor}
                \State
                \State $error$ = $|y_{\text{NN\_max}}-y_{\text{FE\_max}}|/y_{\text{FE\_max}}$\ \Comment{calculate the relative error}
                \State
                \If {$error \leq threshold$}\ \Comment{decide if the error is acceptable}
                    \State $accepted[patch] = 1$
                \EndIf
            \EndFor\label{accendfor}
            \State
            \State $accuracy$ = $\textbf{sum}(accepted)/N$
            \State
            \State \textbf{return} $accuracy$
        \EndFunction
        \end{algorithmic}
    \end{algorithm}

\section{Data pre-processing} \label{appendix:preprocessing}

    Data pre-processing is a standard technique used in data driven methodologies. A common practice is to pre-process the input data usually with a simple linear rescaling \citep{Bishop}. In this work we perform rescaling to both have smoother convergence of the optimiser and to restrict the space of possible inputs. We divide the input, and the output for consistence reasons, of the NN with the maximum absolute component of the stress tensor. This rescales the input data to a range $[-1, 1]$. This rescaling alleviates the differences in scales across the data points which helps the optimiser to avoid bad local minima and converge to a stable solution. Additionally, scaling the input in the case of linear elasticity caries a deeper physical meaning. Scaling the macroscale stress field with a scaling factor, $k$, results in scaling the microscale stress tensor with the same scaling factor, $k$. By taking advantage of this linear relationship we can effectively train our NN in a range of values from -1 to 1 and then evaluate it to an input of any range. This is achieved by dividing the new input with the scaling factor $k$, to map it to a range from -1 to 1. To retrieve the real output, microscale tensor, the output of the NN needs to be multiplied with the scaling factor $k$.
    
%-----------------------------------------------------------------------------------------

\section{Volume and surface mesh} \label{appendix:Volume_vs_Surface}

\begin{figure}[htb]
        \centering
        \begin{subfigure}{.5\textwidth}
          \centering
          \includegraphics[width=0.5\linewidth]{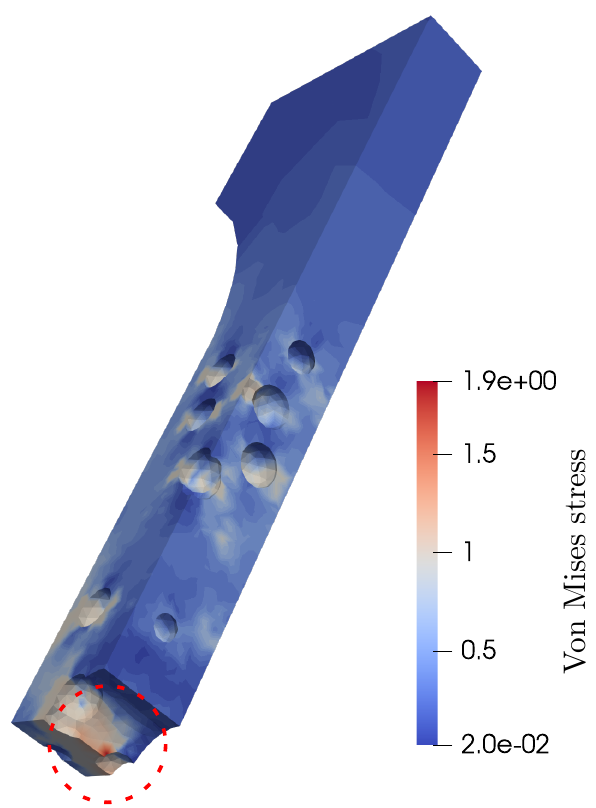}
          \caption{Volume Stress}
          \label{fig:Volume:a}
        \end{subfigure}%
        \begin{subfigure}{.5\textwidth}
          \centering
          \includegraphics[width=0.5\linewidth]{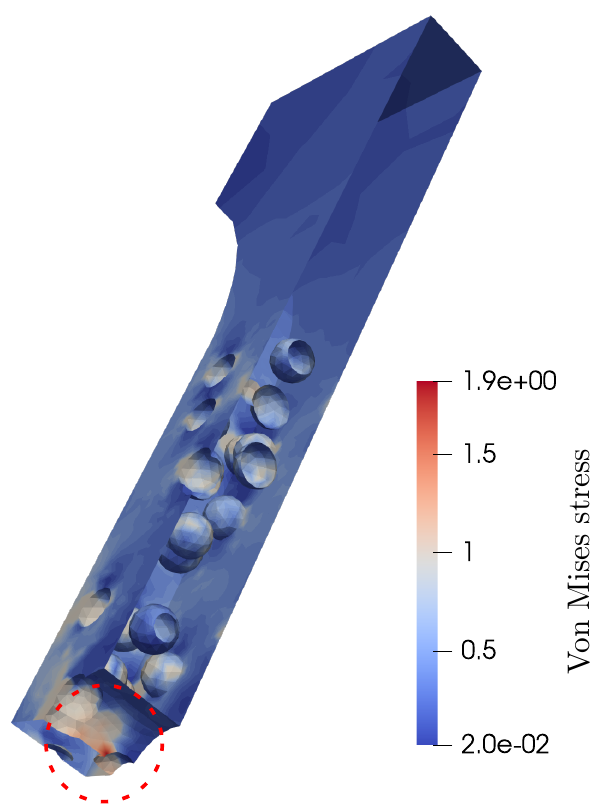}
          \caption{Surface Stress}
          \label{fig:Surface:b}
        \end{subfigure}
        \caption{Two structures corresponding to one quarter of a random realisation of the dogbone geometry. In the diagram on the left (a) we see the Von Mises stress distribution on the volume mesh. In the diagram on the right (b) we see the Von Mises stress distribution on the surface mesh. We can see, in the red circles, that the maximum values in both structures are in the same location.}
        \label{fig:volume_2_surface}
    \end{figure}

    As discussed in section [3.2] of the main paper, the input and output of the model is the surface and not the volume stress. In [Fig. \ref{fig:volume_2_surface}] we see an example of a surface stress extracted from the volume mesh. We can observe that the maximum value of the Von Mises stress is the same in both cases and the location where the maximum Von Mises stress occurs is the same as well.    

%-------------------------------------------------------------------------------------

\section{GNN parameters and architectural choices} \label{appendix:GNN_parameters}

    We identified some key parameters for the training of the GNN and performed several tests to identify their optimum values. The parameters that we examined is the number of filters in the MLPs, the number of GN Blocks, the number of maximum neighbours per node, the existence of a skip connection between the input and the output of the network and finally the type of encoder. For these tests we used 500 patches as the training set and 100 patches for the test set.

    \subsection{Skip Connection} \label{appendix:Skip Connection}
    
\begin{figure}[htb]
            \begin{center}
                 \includegraphics[width=.5\linewidth]{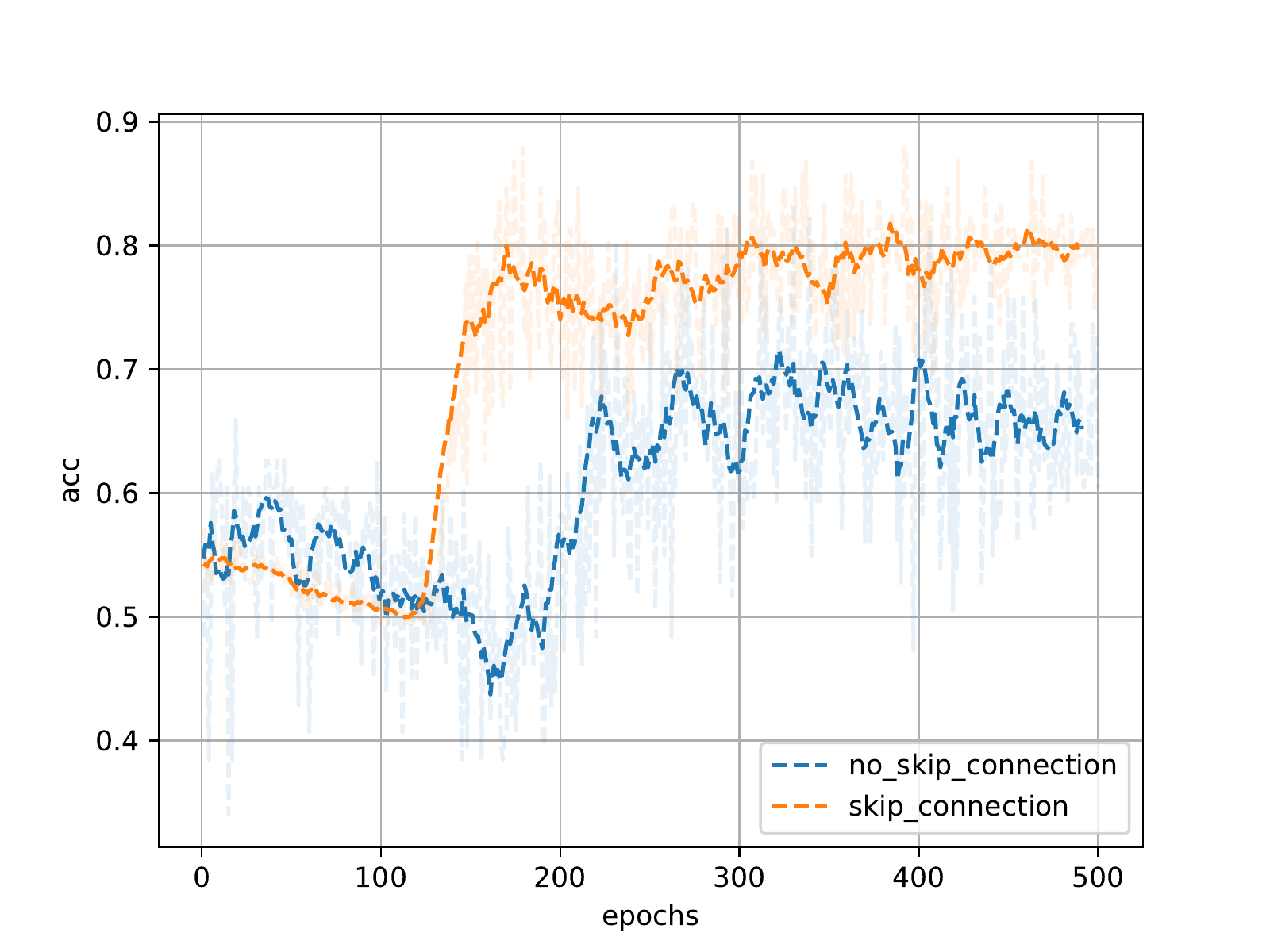}
                \caption{In the diagram we see accuracy curves for the test dataset defined using the maximum Von Mises stress. The yellow line corresponds to a GNN trained with a skip connection to add the input macroscale stress to the output of the GNN and the blue line corresponds to a GNN trained without this skip connection. We observe that the skip connection results in smoother training and higher accuracy.}
                \centering
                \label{fig:skip}
            \end{center}
        \end{figure}

        We suggest that using a skip connection to add the input macroscale stress tensor to the output of the network will improve GNN's performance. The reason behind this is that the GNN will learn how the microscale stress field deviates from the macroscale stress field instead of learning the microscale stress field from zero which we believe that it is easier and it should also improve the generalization ability of the network. In order to validate our assumption we train 2 identical GNNs with the exact same parameters and training set but one of them has a skip connection and the other does not. Both networks have MLPs with 128 filters, 10 GN blocks and each node has at most 10 neighbours. The results can be found in [Fig \ref{fig:skip}] where we can observe that the GNN trained with the skip connection has smoother convergence compared to the GNN trained without the skip connection. Also, the skip connection improved accuracy from 65\% to 80\%. We conclude that indeed the skip connection helped not only to have more stable training but also to improve the accuracy and thus we decide to include it in our architecture.
        
    \clearpage 
    \subsection{Number of Filters} \label{appendix:Number of Filters}
    
\begin{figure}[htb]
            \begin{center}
                 \includegraphics[width=.5\linewidth]{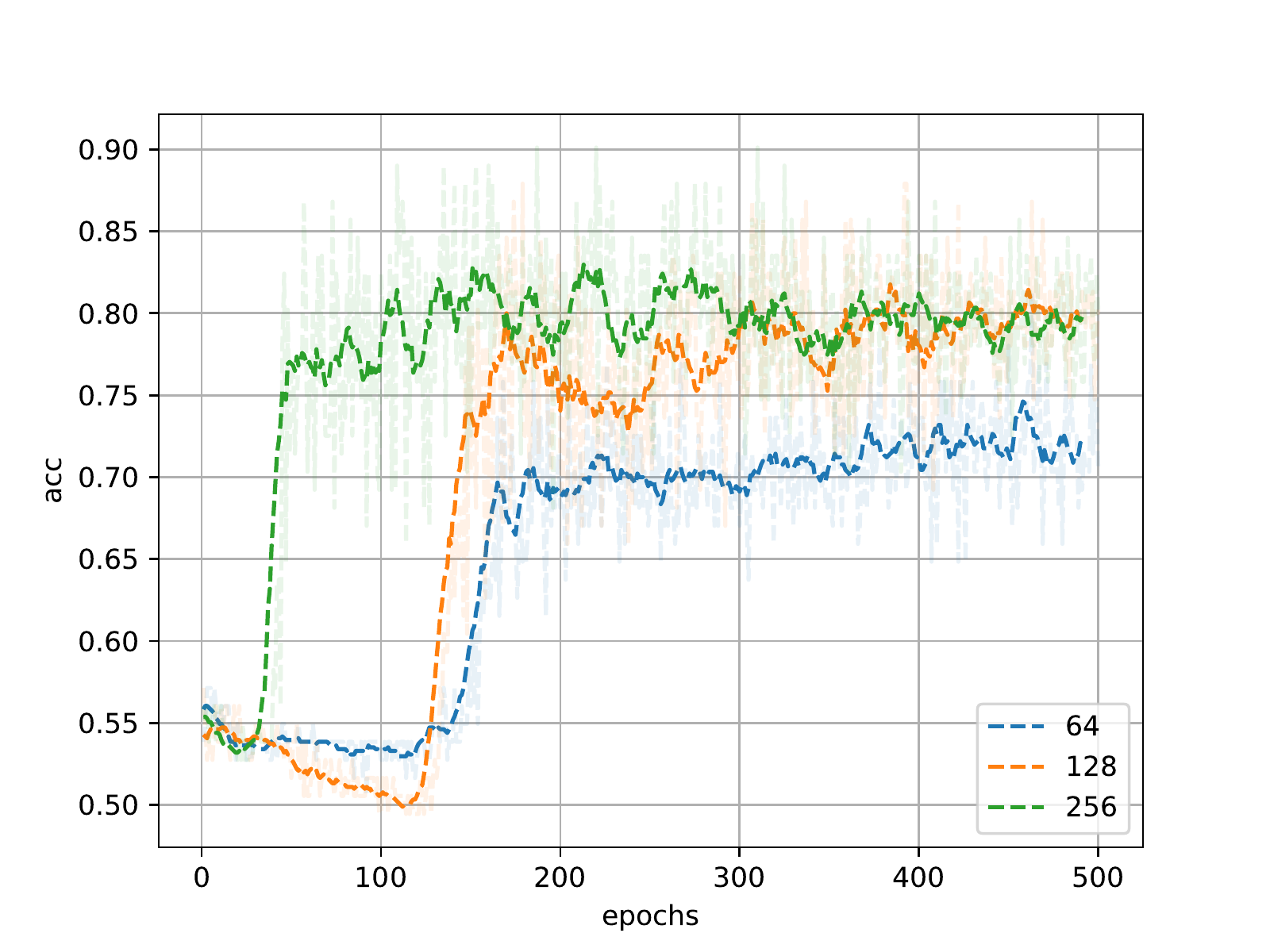}
                \caption{In the diagram we see accuracy curves for the test dataset defined using the maximum Von Mises stress. Coloured lines correspond to networks trained with different number of filters, namely 64, 128 and 256. We observe that 128 and 256 filters result in the same accuracy 80\% where 64 filters result in a lower accuracy of 72\%.}
                \centering
                \label{fig:filters}
            \end{center}
        \end{figure}

        An important parameter that heavily influences not only the accuracy of the GNN but also the training time and memory requirements is the number of filters in the MLPs. In order to identify the minimum number of filters that result in optimum accuracy we trained 3 GNNs with 64, 128 and 256 filters in the MLPs and we kept all the other parameters the same. Specifically, all networks have the skip connection described in [\ref{appendix:Skip Connection}], 10 GN blocks and each node has at most 10 neighbours. The results can be found in [Fig \ref{fig:filters}] where we observe that 128 and 256 filters result in the same accuracy which is higher than the accuracy for the 64 filters, 80\% compared to 72\%. Additionally, the training time for the GNN with the 256 filters is 12.7 hours and the maximum required memory is 15.4 GB while for the 128 filter GNN the training was completed in 5.8 hours and the maximum required memory was 7.7 GB. By choosing to use 128 filters in the MLPs of the GNN we achieve the same accuracy as in the 256 filters GNN but with a 54\% decrease in the training time and a 50\% decrease in memory requirements. 

    \clearpage 
    \subsection{Independent Decoder} \label{appendix:Independent Decoder}
    
\begin{figure}[htb]
            \begin{center}
                 \includegraphics[width=.5\linewidth]{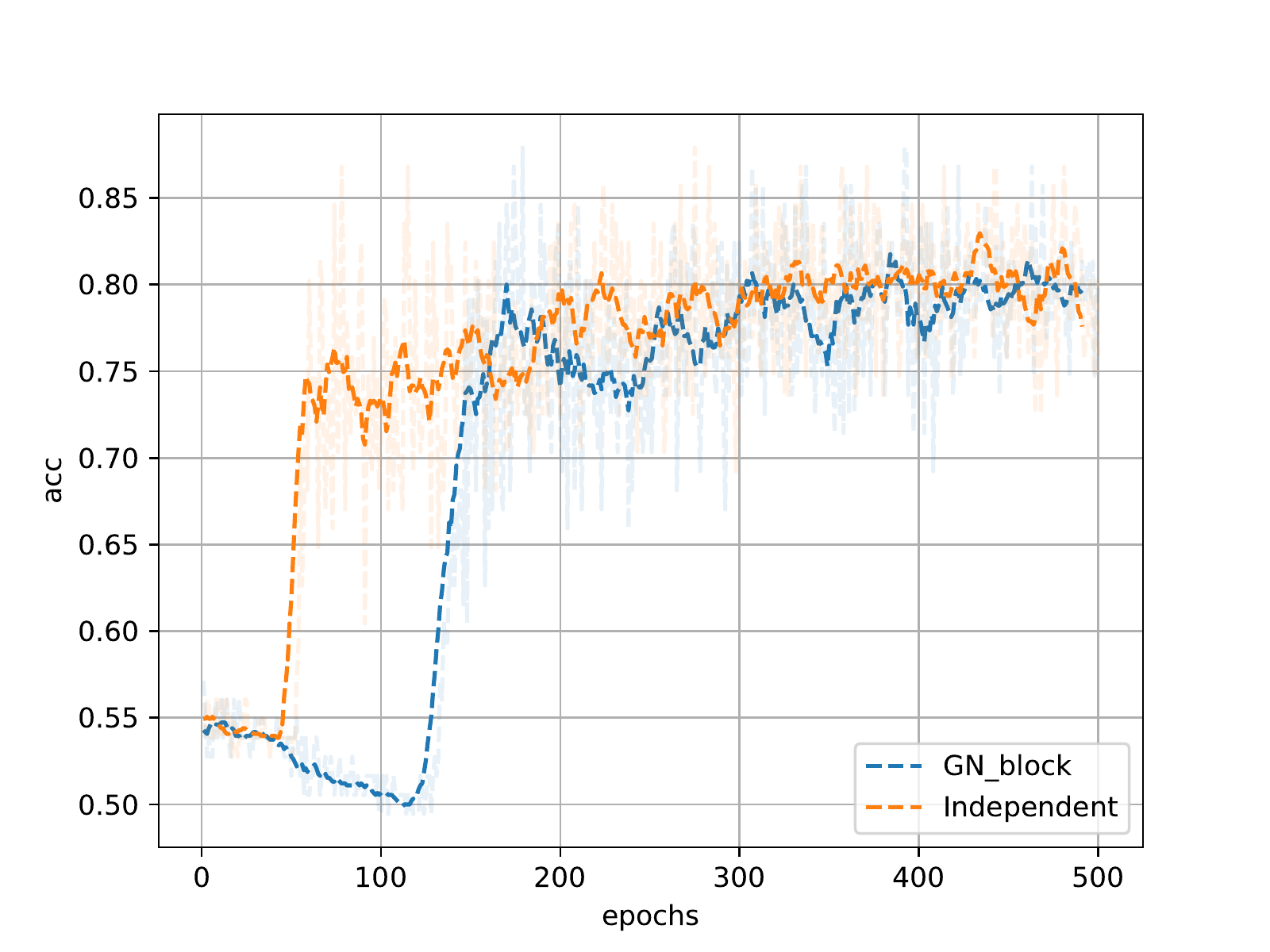}
                \caption{In the diagram we see accuracy curves for the test dataset defined using the maximum Von Mises stress. The yellow line corresponds to a GNN trained with an independent encoder and the blue line corresponds to a GNN trained with a GN block as an encoder. We observe that both GNNs have similar accuracy curves with the same final accuracy but the GNN with the independent encoder starts increasing its accuracy sooner, epoch 50, compared to the other one, epoch 120.}
                \centering
                \label{fig:Encoder}
            \end{center}
        \end{figure}

        As already discussed in section [3.6] of the main paper, in the first layer of the GNN we choose to encode the edge and node features independently into the latent dimension as suggested by \citep{sanchezgonzalez2020learning, pfaff2021learning, mylonas2021bayesian}. We want to test if this choice results in an improved GNN and thus we perform an experiment between 2 GNNs with the same parameters where one uses a GN block to encode the inputs into the latent dimension and the other encodes them independently. Both of the GNNs have 10 GN blocks, 128 filters in the MLPs, the skip connection described in [\ref{appendix:Skip Connection}] and each node has at most 10 neighbours. The results can be found in [Fig \ref{fig:Encoder}] where we observe that both GNNs have similar accuracy curves and they both have a final accuracy of 80\%. Nevertheless, we can observe that in the independent encoder case the accuracy starts increasing sooner, at epoch 50, where in the GN block case it starts increasing later, at epoch 120. Lastly, the independent encoder version involves slightly less calculations and results in a 5\% decrease in training time. Consequently, we choose to use the independent encoder GNN.
        
    \clearpage 
    \subsection{Number of GN blocks} \label{appendix:Number of GN blocks}
    
\begin{figure}[htb]
            \begin{center}
                 \includegraphics[width=.5\linewidth]{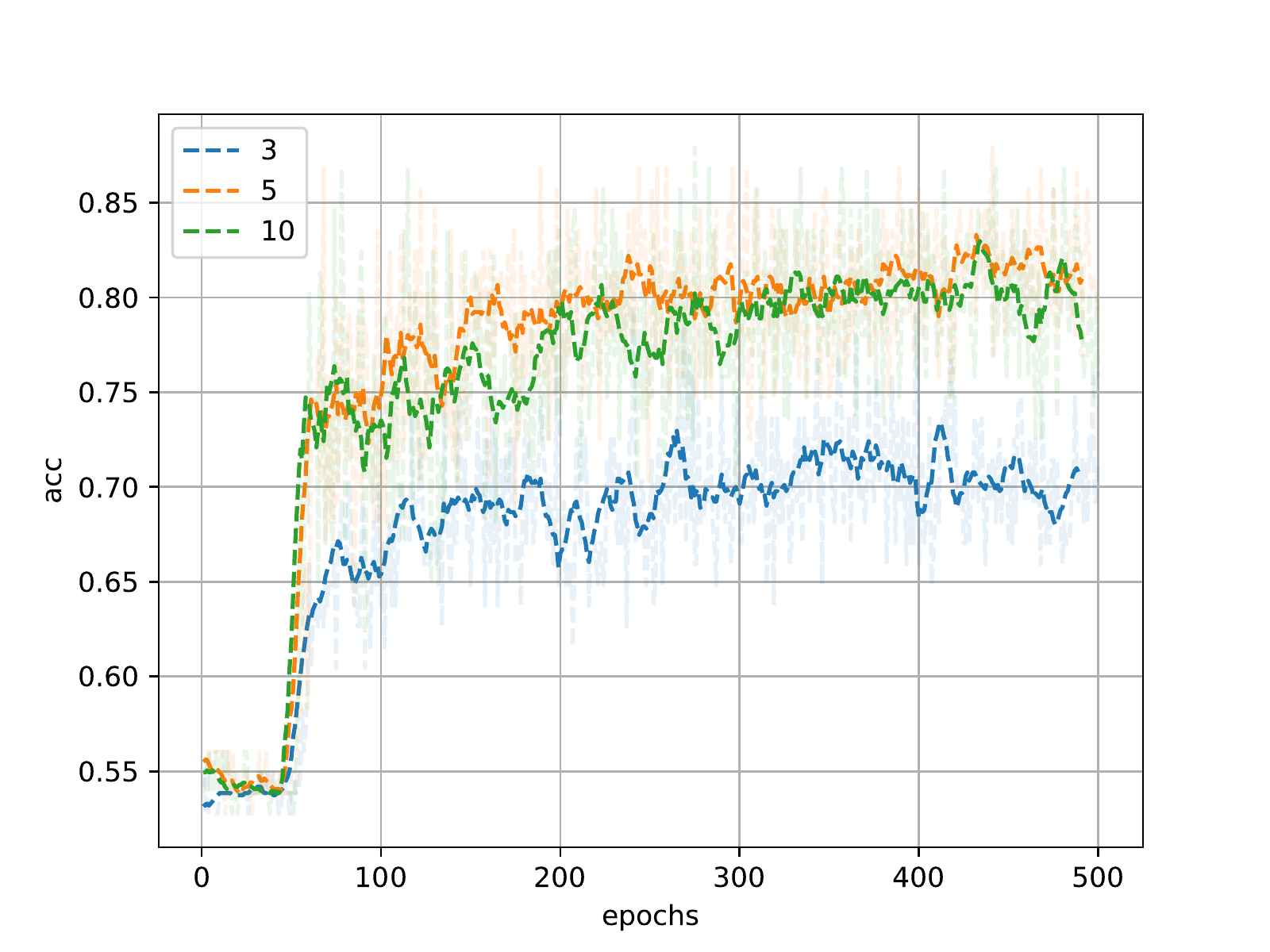}
                \caption{In the diagram we see accuracy curves for the test dataset defined using the maximum Von Mises stress. Coloured lines correspond to networks trained with different number of GN blocks, namely 3, 5 and 10. We observe that the GNN with 3 GN blocks has a decreased accuracy compared to the rest. Additionally, we can see that the GNNs trained with 5 and 10 GN blocks have almost exactly the same behaviour.}
                \centering
                \label{fig:NPS}
            \end{center}
        \end{figure}

        Another important parameter that affects both the memory requirement and the training time is the number of residual GN blocks. Because we are using residual connections we do not expect a decrease in accuracy as we add more GN blocks but we are expecting that there should be a saturation point where the GNN does not benefit anymore from the GN blocks but it performs unnecessary calculations resulting in increased computational cost. To validate this assumption and determine the most proper number of GN blocks we train 3 GNNs with the same parameters but different number of GN blocks. All the GNNs have 128 filters in the MLPs, the skip connection described in [\ref{appendix:Skip Connection}], independent encoder and each node has at most 10 neighbours. The results can be found in [Fig \ref{fig:NPS}] where we observe that the GNN with only 3 GN blocks presents a decrease of 10\% in the test accuracy compared to the other two. Also we can see that the GNNs with 5 and 10 GN blocks have similar test accuracy curves and they both reach a test accuracy of 80\%. We decide to opt for the GNN with the 5 GN blocks that presents optimum accuracy without additional computational cost. This results in a training time of 3.1 hours and a maximum memory of 4.3 GB which is a decrease of 45\% in training time and 44\% in memory requirements compared to the GNN with the 10 GN blocks. 
        
    \clearpage 
    \subsection{Number of neighbours} \label{appendix:Number of neighbours}
    
\begin{figure}[htb]
            \begin{center}
                 \includegraphics[width=.5\linewidth]{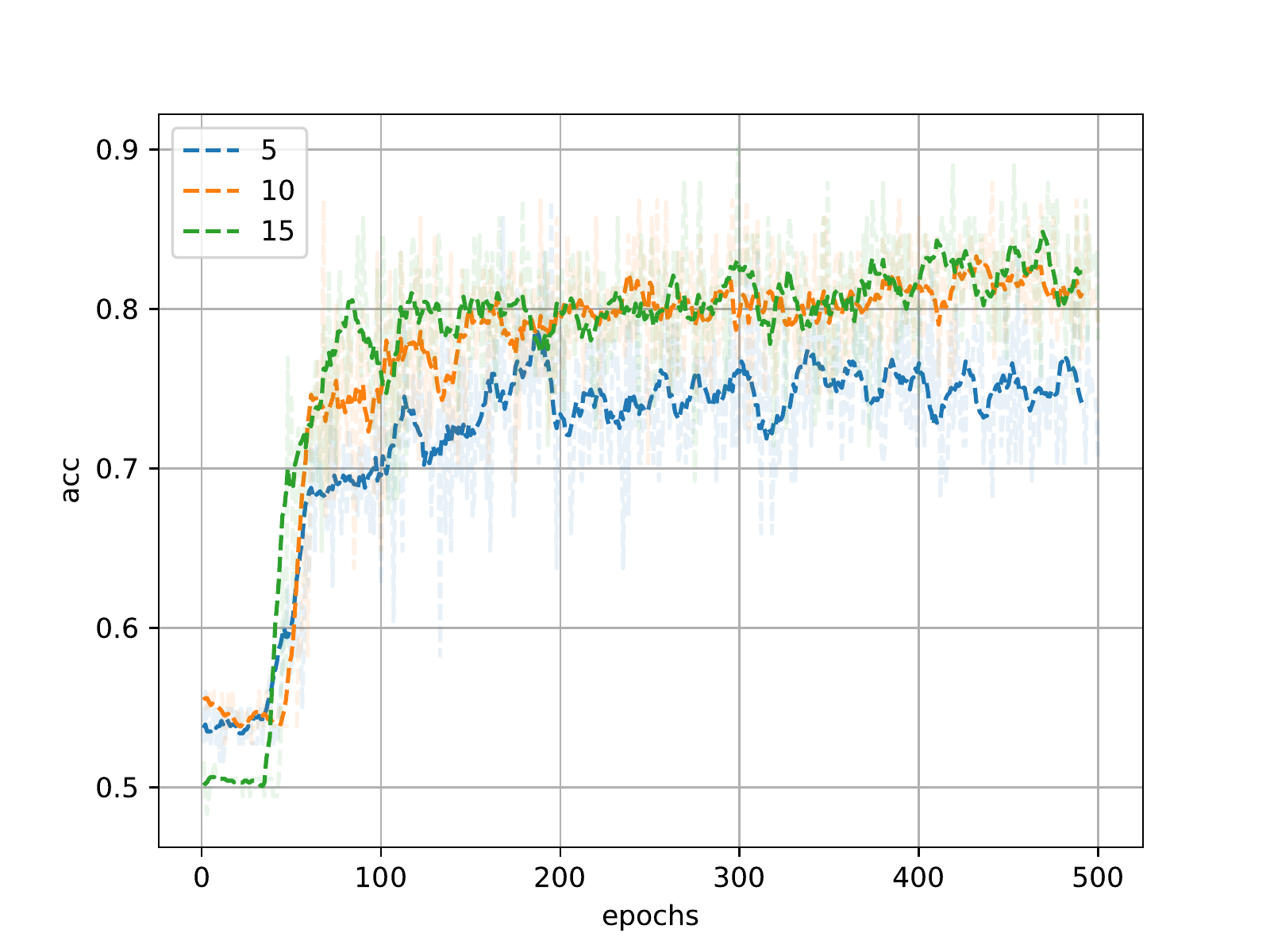}
                \caption{In the diagram we see accuracy curves for the test dataset defined using the maximum Von Mises stress. Coloured lines correspond to networks trained with different number of maximum neighbours per node, namely 5, 10 and 15. We observe that 10 and 15 neighbours result in the same accuracy 81\% where 5 neighbours result in a lower accuracy 75\%.}
                \centering
                \label{fig:neighbours_appendix}
            \end{center}
        \end{figure}

        We want to study the effect of the maximum number of neighbours on the GNN training. A small number of maximum neighbours will result in a graph where disconnected areas of the FE mesh do not share an edge and thus cannot exchange information, for instance microscale and macroscale features. We train 3 GNNs with the same parameters apart from the maximum number of neighbours that have values 5, 10 and 15. All the GNNs have 5 GN blocks, 128 filters in the MLPs, the skip connection described in [\ref{appendix:Skip Connection}] and independent encoder. The results can be found in [Fig \ref{fig:neighbours_appendix}] where we observe that 10 and 15 neighbours result in the same accuracy which is higher than the accuracy for the 5 neighbours, 81\% compared to 75\%. Additionally, the training time for the GNN with the 15 neighbours is 27\% higher and the memory requirements 31\% higher compared to the GNN with 10 neighbours. We conclude that in our case we do not have a reason to use more than 10 neighbours although in denser meshes this number could be different.
        
    \clearpage 
    \subsection{Geodesic and Euclidean Convolutions} \label{appendix:Geod_and_Eucl_conv}    
    
\begin{figure}[htb]
            \centering
            \begin{subfigure}{.5\textwidth}
              \centering
              \includegraphics[width=\linewidth]{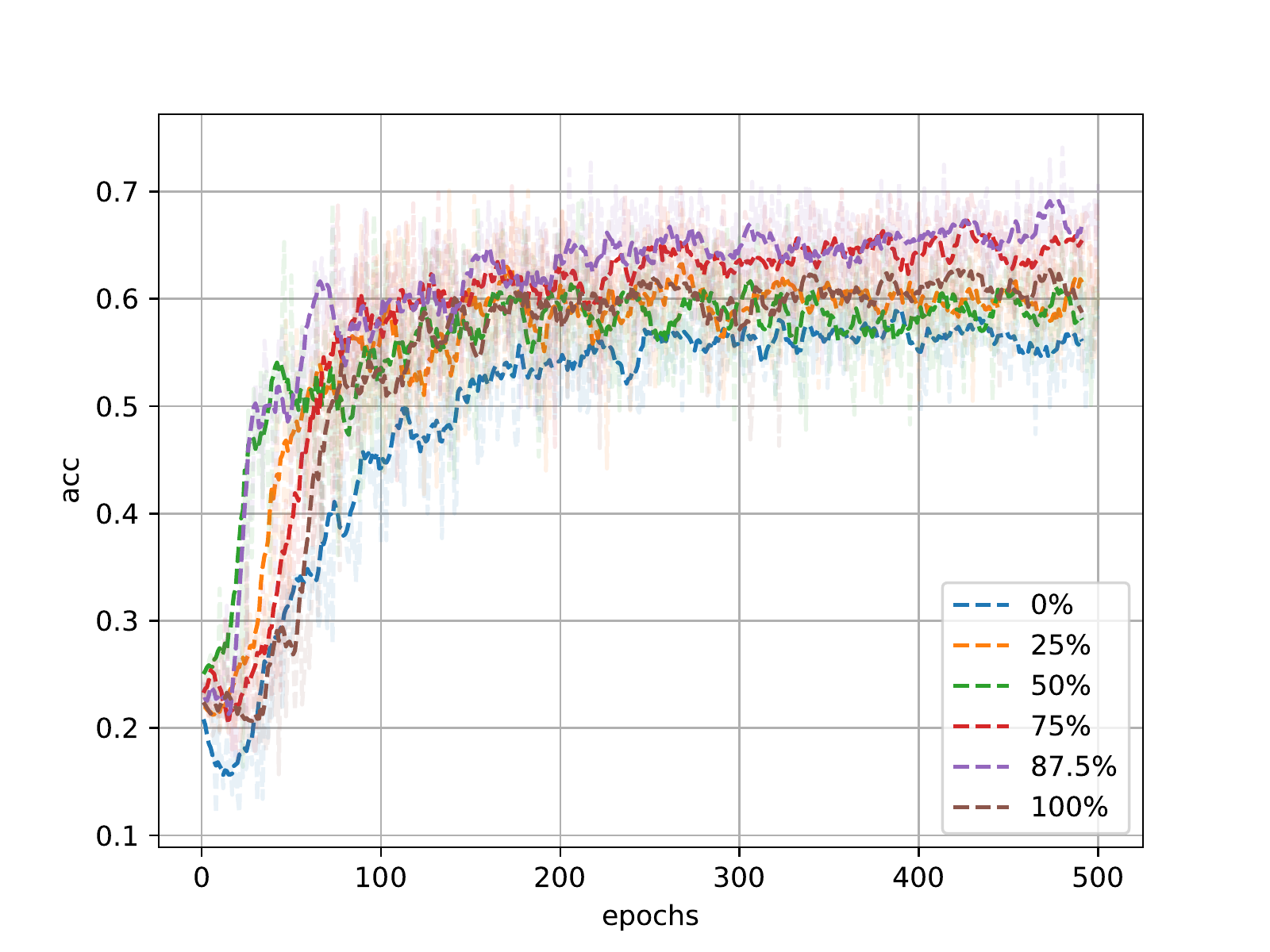}
              \caption{Accuracy}
              \label{fig:DualConv_study:a}
            \end{subfigure}%
            \begin{subfigure}{.5\textwidth}
              \centering
              \includegraphics[width=\linewidth]{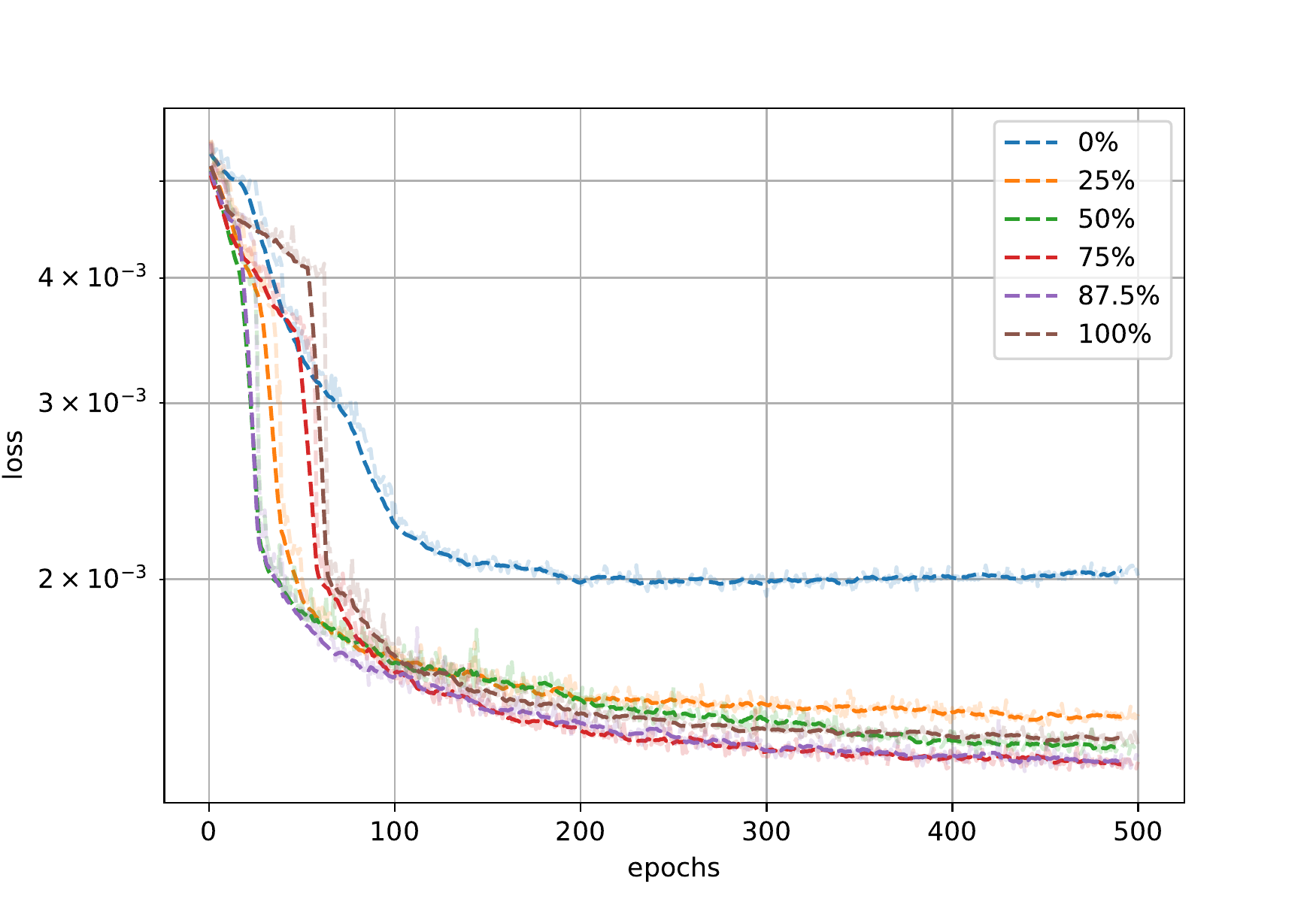}
              \caption{Loss}
              \label{fig:DualConv_study:b}
            \end{subfigure}
            \caption{In the diagram on the left (a) we see accuracy curves defined using the maximum Von Mises stress and a 15\% threshold for the relative error. Specifically, we see the accuracy as function of the training epochs. In the diagram on the right (b) we see the loss function as a function of the epochs. For both diagrams, coloured lines correspond to GNNs trained with different Geodesic to total (Euclidean + Geodesic) filter ratio, namely 0\% (only Euclidean), 25\%, 50\%, 75\%, 87.5\% and 100\% (only Geodesic).}
            \label{fig:DualConv_study}
        \end{figure}

    We study the effect of the joint convolutions by training 6 GNNs with different number of Geodesic and Euclidean filters. We perform the study on the dogbone dataset. We keep the total number of filters constant to 128 but we change the ratio between Geodesic and total (Euclidean + Geodesic) filters. We investigate the case where the ratio is 0\% (only Euclidean), 25\%, 50\% and 75\%, 87.5\% and 100\% (only Geodesic). All the GNNs have 5 GN blocks, residual connection, independent encoder, 128 filters and a maximum of 20 neighbours per node. The results can be found in [Fig \ref{fig:DualConv_study}]. We can observe that both for the accuracy and the loss the worst case is when the ratio is 0, so no Geodesic convolutions are present. Both the accuracy and the loss improve (accuracy increases and loss decreases) as we increase the ratio until the value 75\%. After that no further improvement in performance is observed, both the accuracy and loss curves for ratio values 75\% and 87.5\% are the same. On the contrary when the ratio becomes 1 there is a substantial drop in performance, which is expected since no Euclidean convolutions are performed. We conclude that a 75\% ratio is the most beneficial for this case.
        
    \clearpage
    \subsection{Full stress field VS maximum stress} \label{appendix:stress_vs_max}
    
\begin{figure}[htb]
        \begin{center}
            \includegraphics[width=.7\linewidth]{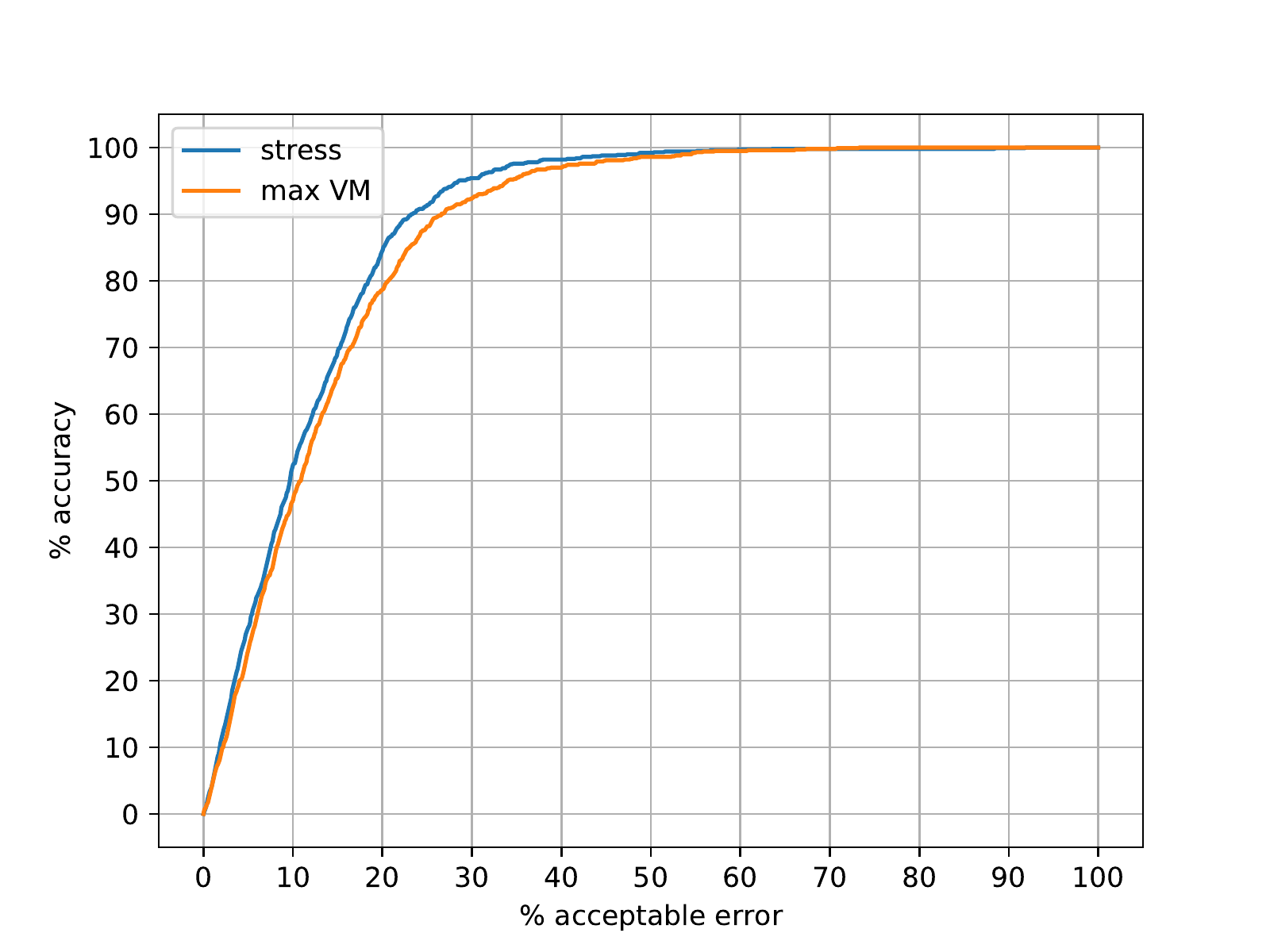}
            \caption{Accuracy curves as a function of the threshold value used for the relative error. The accuracy curves are defined using the maximum Von Mises stress. The blue line corresponds to a GNN that predicts the full microscale stress distribution first and then from this the maximum Von Mises stress in the ROI of the patch. The orange line corresponds to a GNN that directly predicts the maximum Von Mises in the ROI of the patch.}
            \centering
            \label{fig:acc_maxVM_vs_stress}
        \end{center}
    \end{figure}

    We investigate the option to directly predict the maximum Von Mises stress on the surface of the dogbone structure. To this end, we train 2 identical GNNs on patches extracted from 100 FE simulations and we evaluate them in a different set of patches extracted from 100 FE simulations. One of the GNNs directly predicts the maximum Von Mises stress and the other the full stress tensor. Both GNNs have 5 GN blocks, independent encoder, 128 filters (32 Euclidean and 96 Geodesic) and a maximum of 20 neighbours per node. The results can be found in [Fig \ref{fig:acc_maxVM_vs_stress}]. We observe that the 2 GNNs have similar accuracy, but the one predicting the full stress has slightly higher values. We conclude that predicting the full stress is beneficial since it not only results in better prediction of the maximum Von Mises stress but it also provides us with the full stress tensor that can be used for instance for the online bias corrections.
    
    \clearpage
    \subsection{Patches VS full structure} \label{appendix:full_vs_patches}
        
\begin{figure}[htb]
        \centering
        \begin{subfigure}{.5\textwidth}
          \centering
          \includegraphics[width=\linewidth]{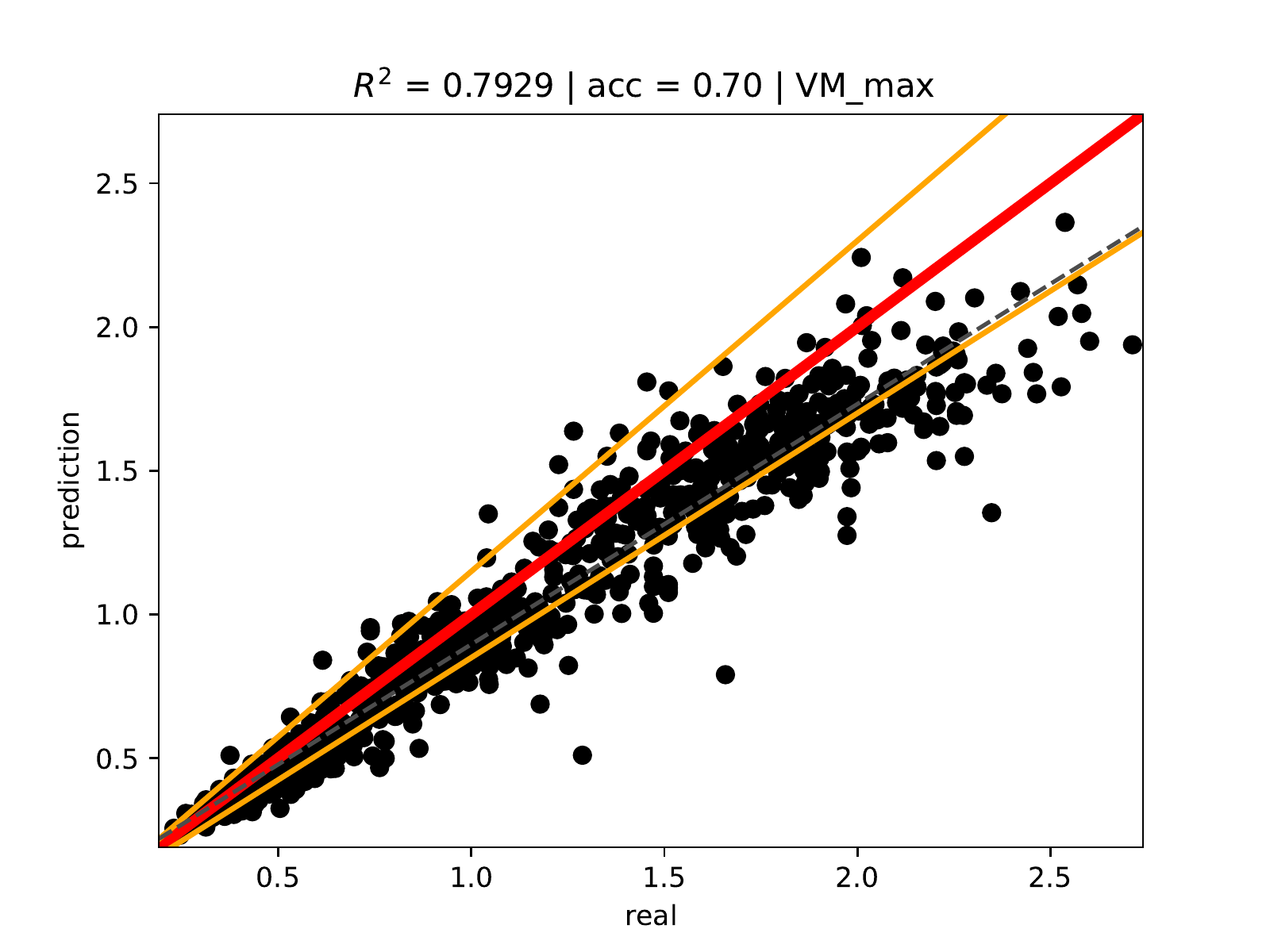}
          \caption{patches}
          \label{fig:VM_max_patches:a}
        \end{subfigure}%
        \begin{subfigure}{.5\textwidth}
          \centering
          \includegraphics[width=\linewidth]{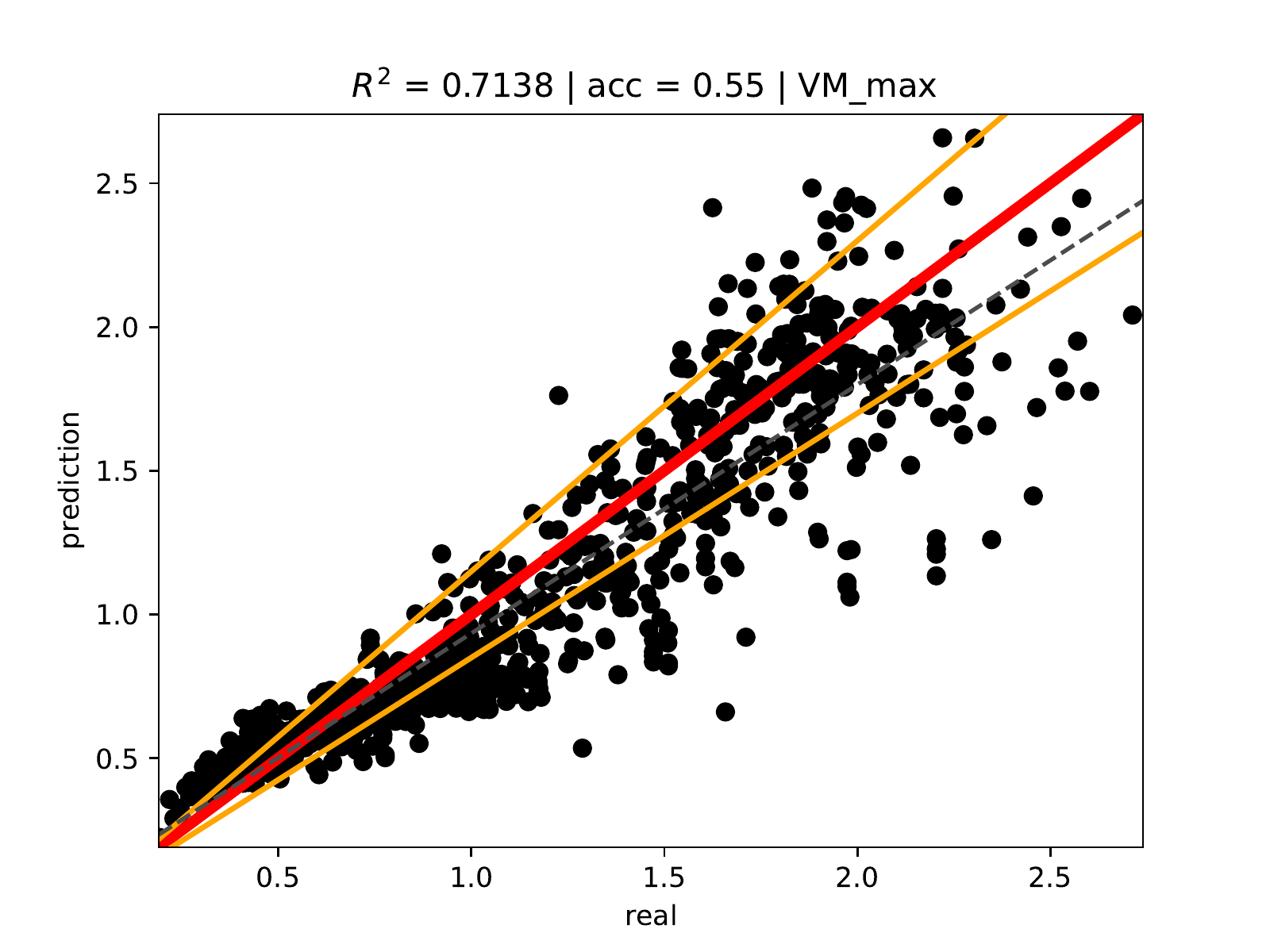}
          \caption{full}
          \label{fig:VM_max_patches:b}
        \end{subfigure}
        \caption{In both diagrams the x-axis corresponds to the maximum Von Mises stress in the ROI of the patches calculated by FE simulations and the y-axis to the maximum Von Mises stress in the ROI of the patches predicted by the GNN. The yellow lines correspond to the 15\% relative error threshold. The accuracy is defined as the percentage of points with relative error less than 15\%. The image on the left (a) corresponds to a GNN trained with patches extracted by 400 FE simulations and the image on the right (b) to a GNN trained directly on the same 400 FE simulations. We can observe that the GNN trained on patches outperforms the GNN trained on the full structures both in terms of accuracy and coefficient of determination.}
        \label{fig:VM_max_patches_vs_full}
    \end{figure}

\begin{figure}[htb]
        \begin{center}
            \includegraphics[width=\linewidth]{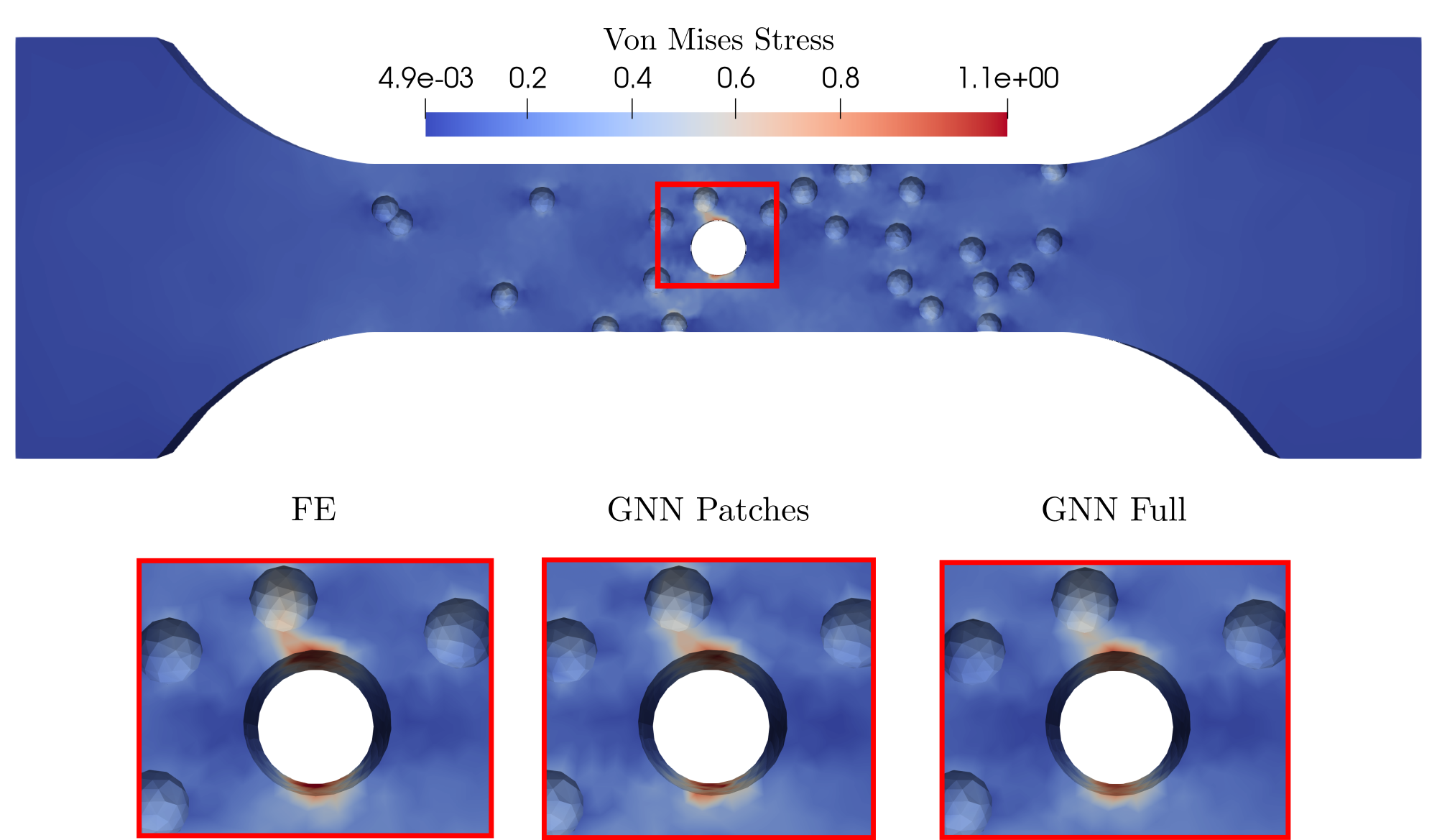}
            \caption{Von Mises stress on the surface of a dogbone structure (top). On the bottom of the image we have zoomed in the high stress area. From left to right we see the FE result, the prediction of the GNN trained with patches and the GNN prediction of the GNN trained with full structures. We observe that the GNN that is trained with patches is in better agreement with the FE results compared to the one trained on the full structures.}
            \centering
            \label{fig:VM_distribution_patches_vs_full}
        \end{center}
    \end{figure}

    Lastly, we want to evaluate to what degree our choice to train the GNN using patches of the geometry affects the quality of the results. This choice stems from the fact that defects only locally affect the macroscale stress field and distant areas do not offer significant information. On one hand, training on patches has the benefit of making the GNN unaware of the specific structure and it encourages it to focus on predicting how the microscale features affect the global stress field, thus leading to improved generalisation. On the other hand, our choice for the size of ROI and patch has to be careful so that the network has all the necessary information to make predictions in the ROI. Choosing to train the GNN using the entire structure alleviates us from this problem. We perform an experiment to compare the two strategies. We train two GNNs with the same parameters, namely 5 GN blocks, residual connection, independent encoder, 128 filters (32 Euclidean and 96 Geodesic) and a maximum of 20 neighbours per node. The first one is trained using patches from 400 FE simulations (10 patches from each simulation) and the second one is trained directly on the 400 FE simulation results. We evaluate both GNNs on a separate set of patches extracted from 100 FE simulations. The results can be found in  [Fig \ref{fig:VM_max_patches_vs_full}]. We can observe that for the GNN that was trained on patches the maximum Von Mises in the ROI of the patches is less scattered around the $y=x$ line, true value. The coefficient of determination for the patch case is 0.79 compared to 0.71 for the full case and the accuracy significantly drops from 70\% for the patch case to 55\% for full dogbone case. Thus we conclude that our choice to train the GNN with patches indeed leads to improved generalisation. We also compare the Von Mises stress distribution on the surface of a dogbone predicted by the GNN trained with patches with the GNN trained with the entire structure. In [Fig \ref{fig:VM_distribution_patches_vs_full}] we see that the GNN that was trained using patches more accurately predicts the high stress area compared to the GNN that was trained with the full structure.
    
%-------------------------------------------------------------------------------------

\section{Ensemble Kalman Method: Observation matrix free implementation} \label{appendix:EnkF}
    
    \label{appendix:matrix_free_implementation}
    
        We can simplify the calculations involved in the Kalman update procedure by avoiding to explicitly define the observation matrix. Instead we can define a function that will provide the noise free value of the data. This function is called observation function and is of the form 
        
        \begin{equation}
            h(\mathbf{x}) = \mathbf{Hx}
        \end{equation}
        
        \noindent
        The posterior can be written as 
        
        \begin{subequations}
            \begin{gather}
                \mathbf{X^{\star}} = \mathbf{X} + \frac{1}{N-1} \mathbf{A}(\mathbf{HA})^T \mathbf{P}^{-1}(\mathbf{D}-\mathbf{HX})\\
                \mathbf{HX} = h(\mathbf{X})\\
                \mathbf{A} = \mathbf{X} - \frac{1}{N} \mathbf{X} \\
                \mathbf{HA} = \mathbf{HX} - \frac{1}{N} \mathbf{HX} \\
                \mathbf{P} =  \frac{1}{N-1} \mathbf{HA} (\mathbf{HA})^T + \mathbf{\Sigma}_{\epsilon}
            \end{gather}
        \end{subequations}

%-------------------------------------------------------------------------------------

\section{Dogbone scaled error examples} \label{appendix:dogbone_examples}

Below we can compare the GNN prediction with the FEA results in 5 cases. The scaled error seen in the next figures is defined as the absolute difference between the microscale Von Mises stress as calculated by FEA and as predicted by the GNN, divided by the maximum microscale Von Mises stress as calculated by FEA.

\begin{figure}[h!]
    \begin{center}
        \includegraphics[width=.565\linewidth]{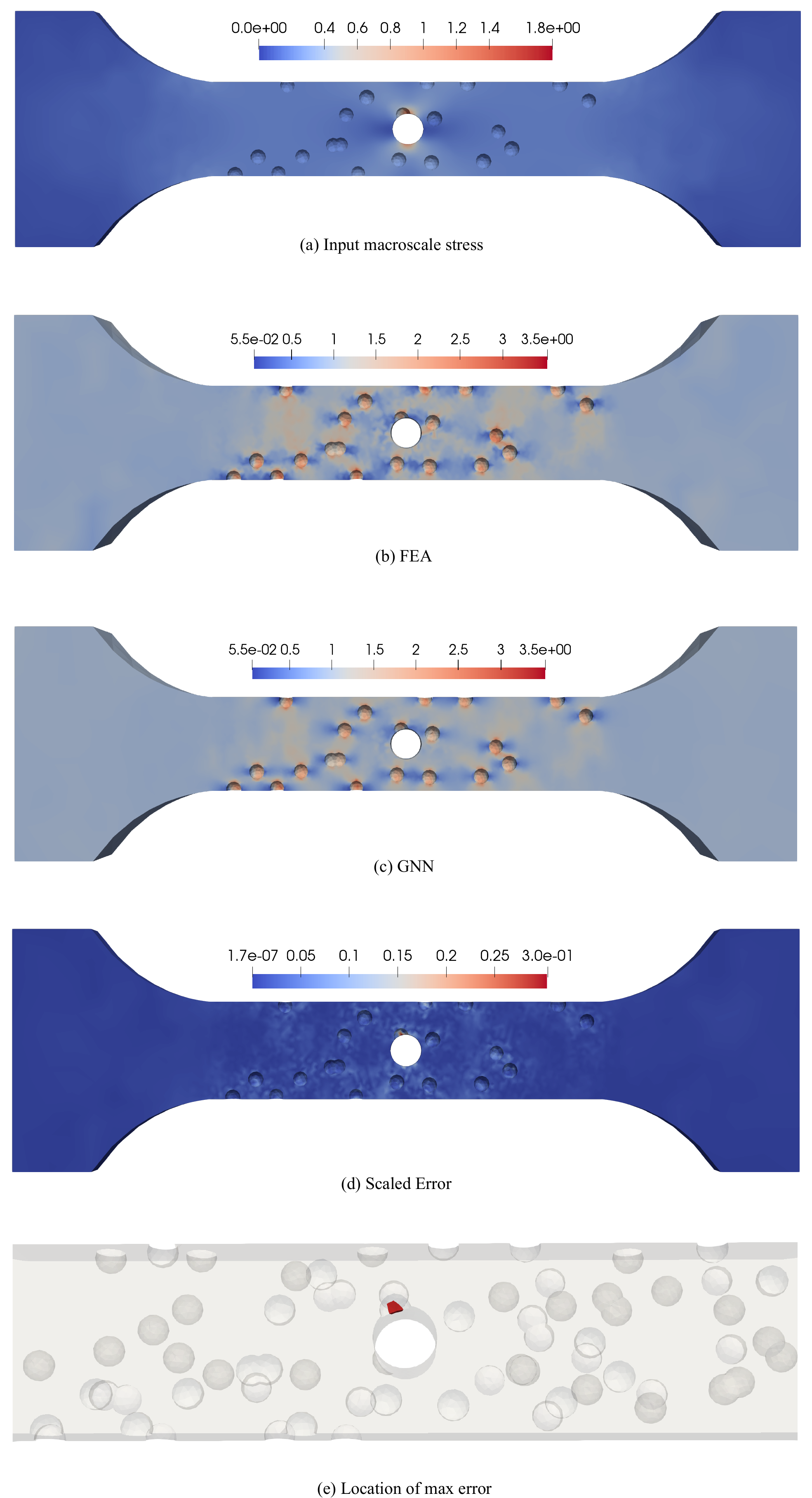}
        \caption{
        Comparison between the microscale Von Mises stress distribution as calculated by FEA (b) and the one predicted by the GNN (c). We can also see the input macroscale Von Mises stress interpolated on the microscale mesh (a) and the scaled error distribution (d). Lastly, the maximum absolute error location is visible in (e). The GNN result is reconstructed from the union of multiple patch predictions where only the ROI is extracted.}
        \centering
    \end{center}
\end{figure}

\begin{figure}[h!]
    \begin{center}
          \includegraphics[width=\linewidth]{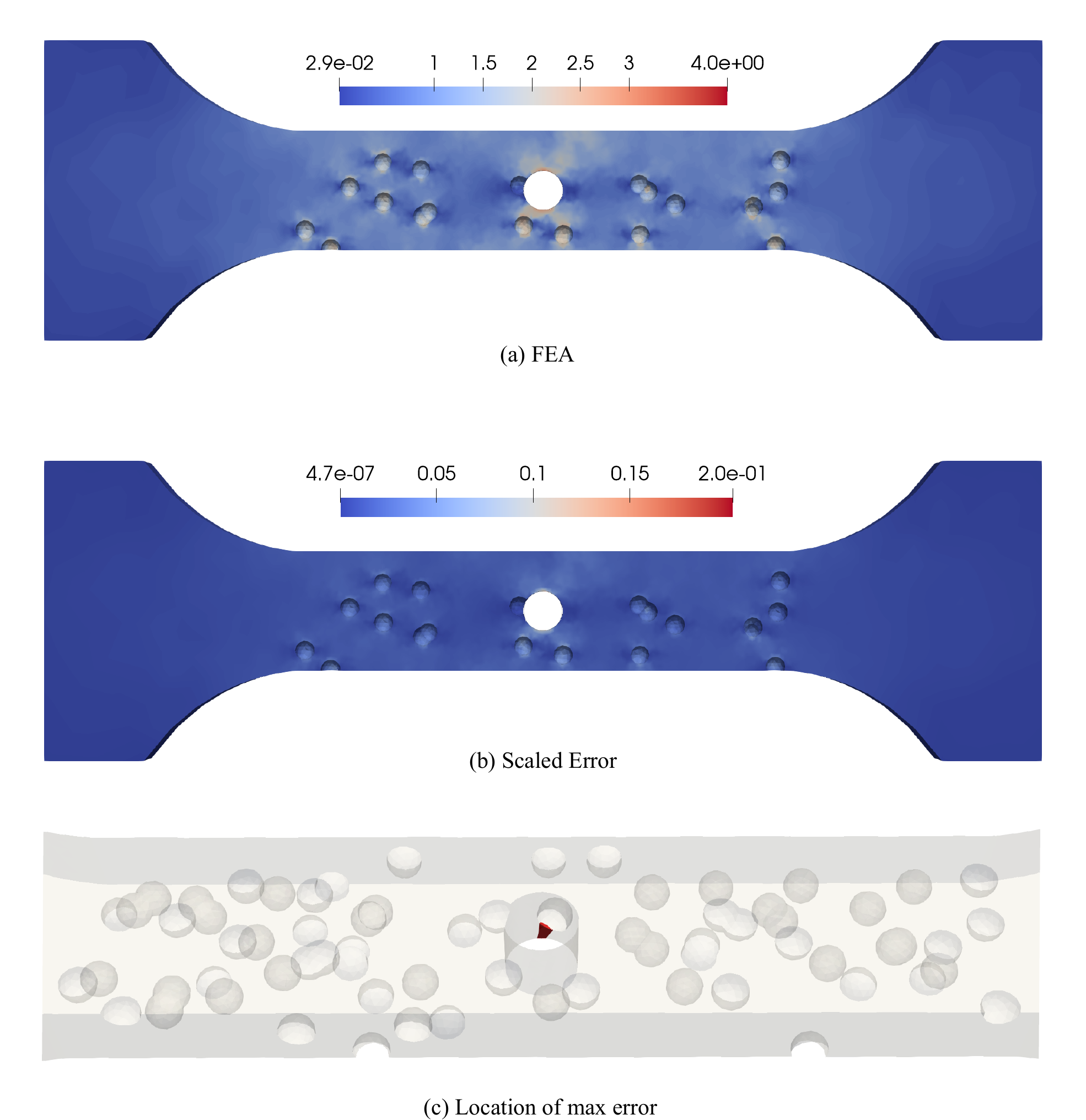}
        \caption{
        In subfigure (a) we can see the  microscale Von Mises stress distribution as calculated by FEA. In subfigure (b) we can see the scaled error distribution. Lastly, we can also see the maximum absolute error location in (c).}
        \centering
    \end{center}
\end{figure}

\begin{figure}[h!]
    \begin{center}
         \includegraphics[width=\linewidth]{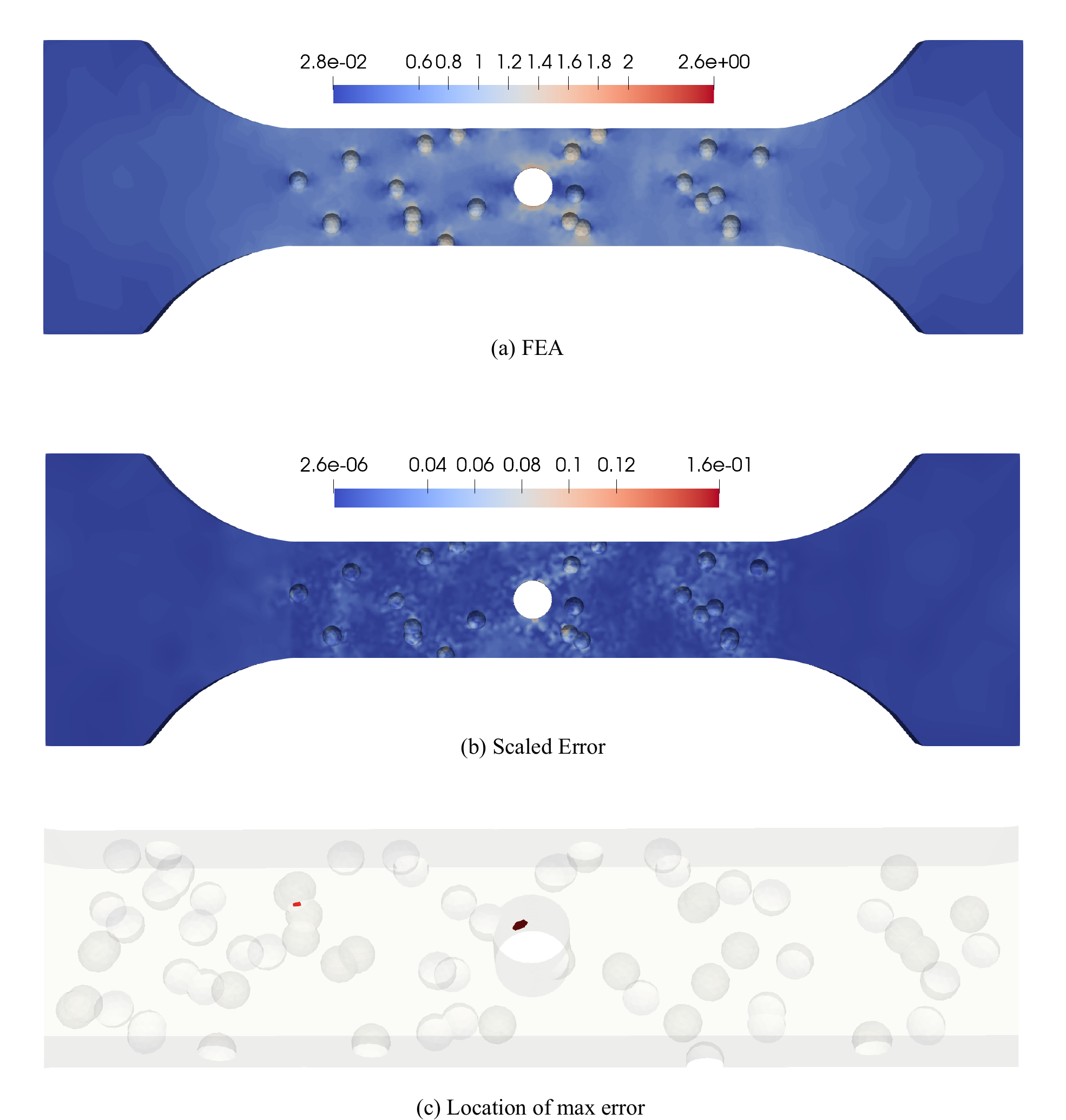}
        \caption{In subfigure (a) we can see the  microscale Von Mises stress distribution as calculated by FEA. In subfigure (b) we can see the scaled error distribution. Lastly, we can also see the location of the two highest absolute error values in (c).}
        \centering
    \end{center}
\end{figure}

\begin{figure}[h!]
    \begin{center}
        \includegraphics[width=\linewidth]{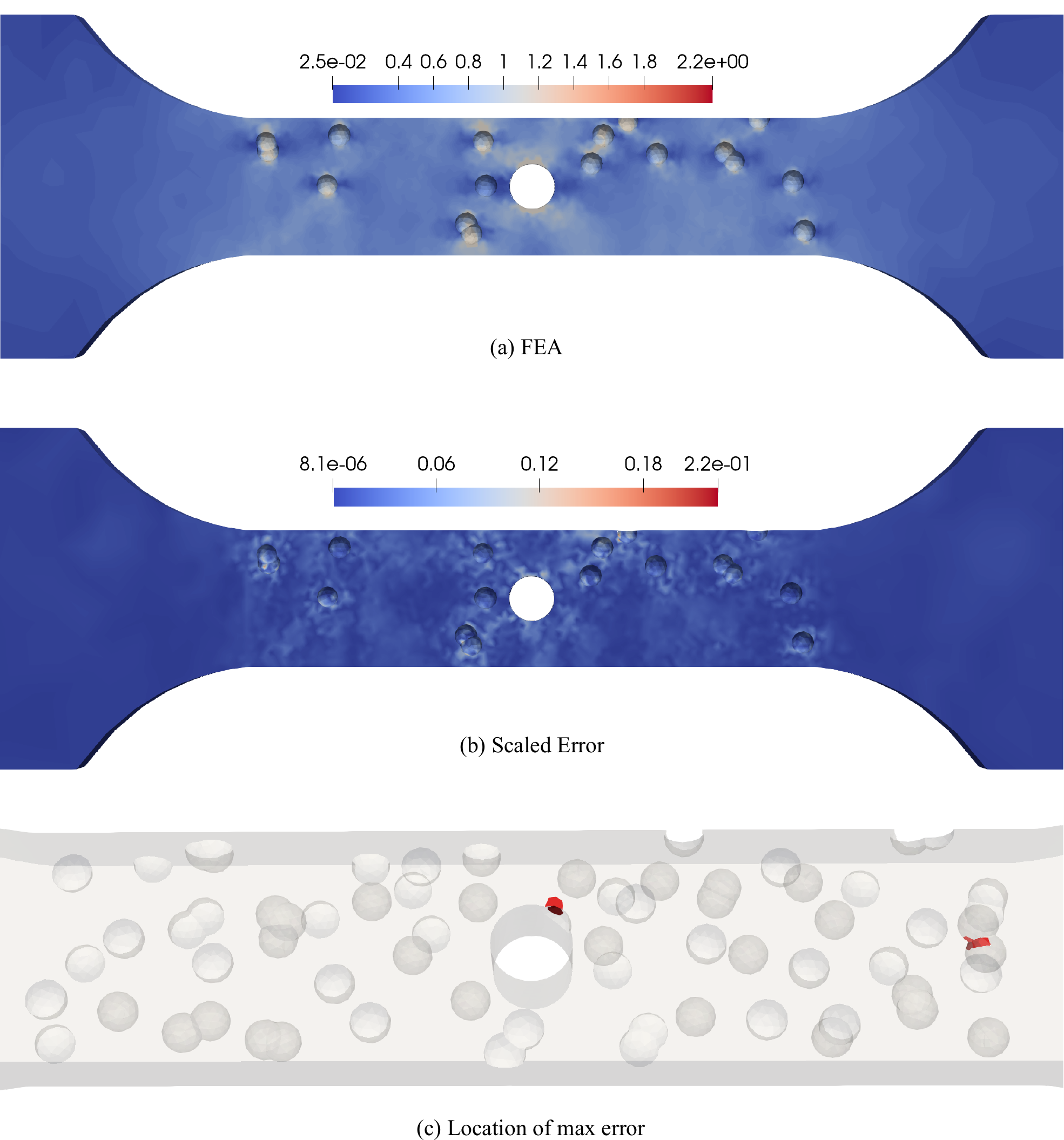}
        \caption{In subfigure (a) we can see the  microscale Von Mises stress distribution as calculated by FEA. In subfigure (b) we can see the scaled error distribution. Lastly, we can also see the location of the two highest absolute error values in (c).}
        \centering
    \end{center}
\end{figure}

\begin{figure}[h!]
    \begin{center}
         \includegraphics[width=\linewidth]{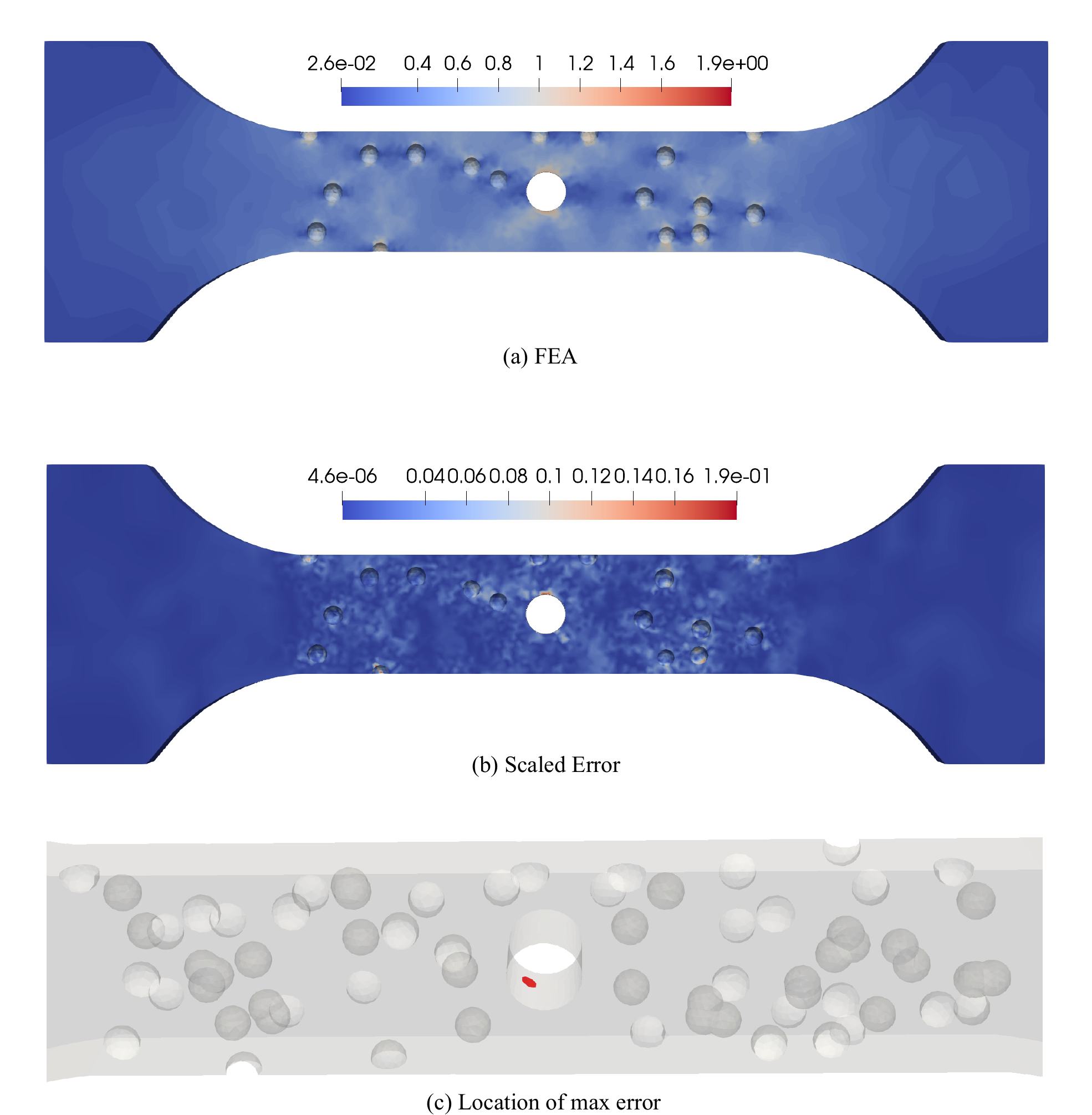}
        \caption{In subfigure (a) we can see the  microscale Von Mises stress distribution as calculated by FEA. In subfigure (b) we can see the scaled error distribution. Lastly, we can also see the maximum absolute error location in (c).}
        \centering
    \end{center}
\end{figure}

%-------------------------------------------------------------------------------------

\clearpage

\section{Evaluate GNN performance using stress triaxiality} \label{appendix:triaxiality}

In this section we evaluate the performance of the GNN using the triaxiality instead of the Von Mises stress. The triaxiality reflects the ratio between the isotropic (or spherical) part of the stress tensor, $\pmb{\sigma}^{sph}$, and the anisotropic (or deviatoric) part of the stress tensor, $\mathbf{s}$. The triaxiality, $\eta$, can be calculated as:

\begin{equation}
    \eta = \frac{\sqrt{2}}{3} \frac{ \pmb{\sigma}^{sph} } { \norm{\mathbf{s}} } 
    \label{eq:triaxiality}
\end{equation}

\noindent
where the isotropic and anisotropic stress parts can be expressed as

\begin{equation}
    \pmb{\sigma}^{sph}  = \sqrt{3}\sigma_m
\end{equation}

\begin{equation}
    \norm{\mathbf{s}} = \sqrt{2J_2} 
\end{equation}

\noindent
where $\sigma_m$  and $J_2$ read as, respectively

\begin{equation}
    \sigma_m  = \frac{1}{3}\text{tr}(\pmb{\sigma})
\end{equation}

\begin{equation}
    J_2 = \frac{1}{2} [\text{tr} (\pmb{\sigma^2}) - 
    \frac{1}{3} \text{tr} (\pmb{\sigma})^2]
\end{equation}

\noindent
where $\pmb{\sigma}$ is the stress tensor in a coordinate system aligned with the principal axes of the stress state.
\newline

\begin{figure}[h!]
    \begin{center}
         \includegraphics[width=\linewidth]{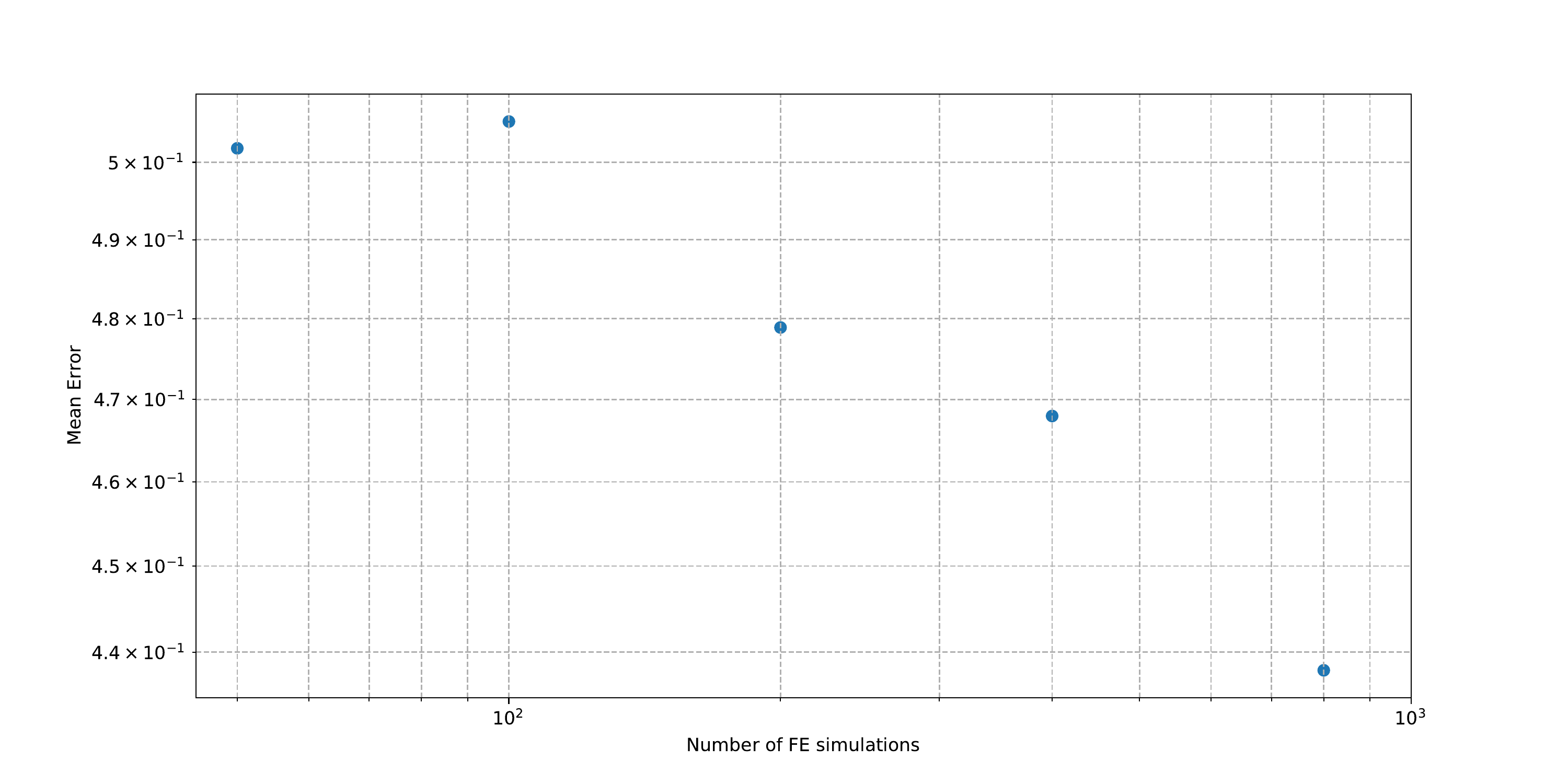}
        \caption{In this diagram we see the mean error over all the patches between the triaxiality stress in the ROI of the patches as calculated by FEA and as predicted from the GNN, $\text{mean}(error_\text{triaxiality})$. The x axis corresponds to the number of data used to train the GNNs. The figure is plotted in a log-log scale.}
        \centering
        \label{fig:SingleCylinder_convergennce_triaxiality}
    \end{center}
\end{figure}

\begin{figure}[h]
    \centering
    \begin{subfigure}{.5\textwidth}
      \centering
      \includegraphics[width=\linewidth]{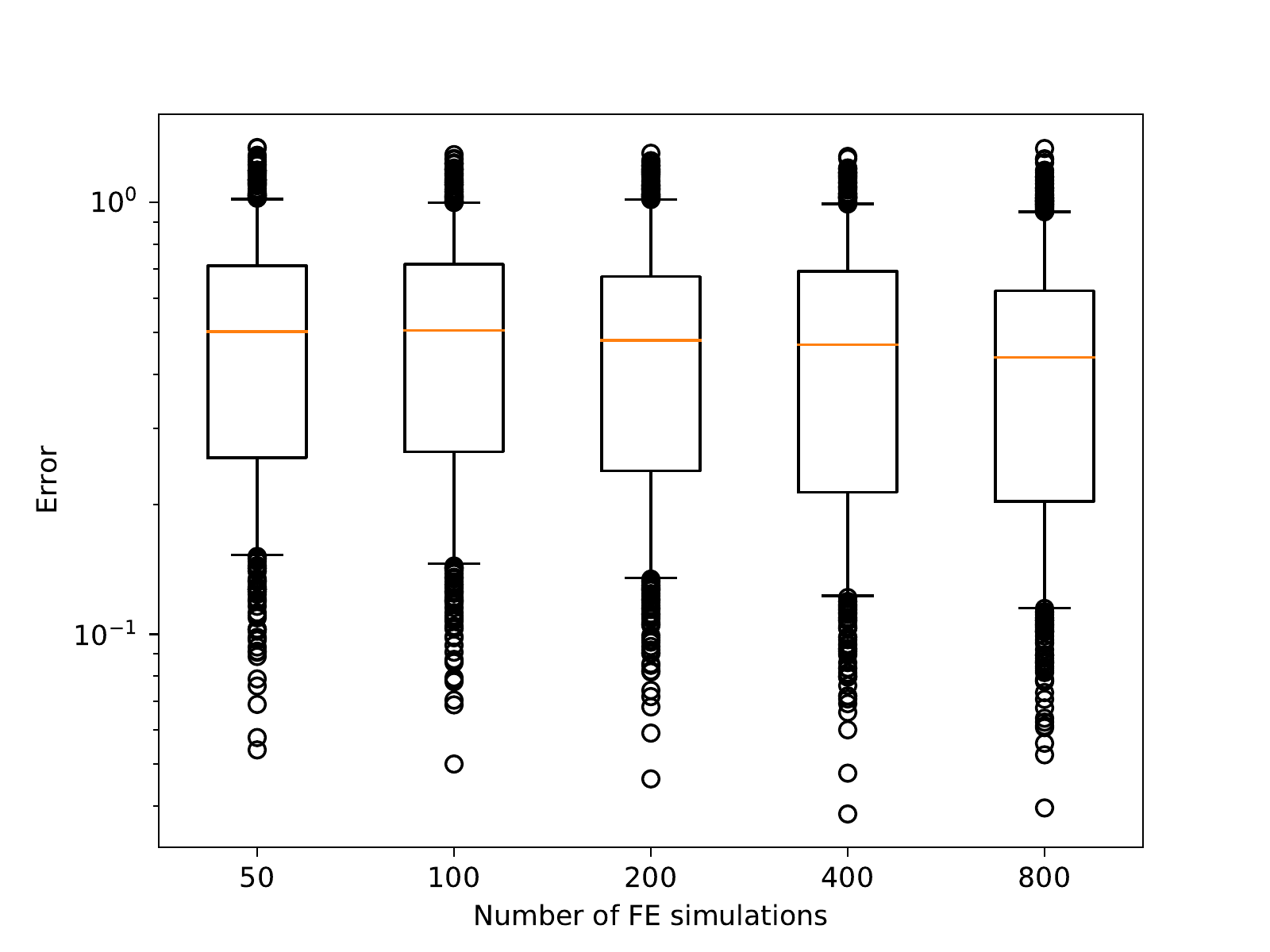}
      \caption{Box plot with outliers}
      \label{fig:SingleCylinder_relative_error_boxplots_triaxiality:a}
    \end{subfigure}%
    \begin{subfigure}{.5\textwidth}
      \centering
      \includegraphics[width=\linewidth]{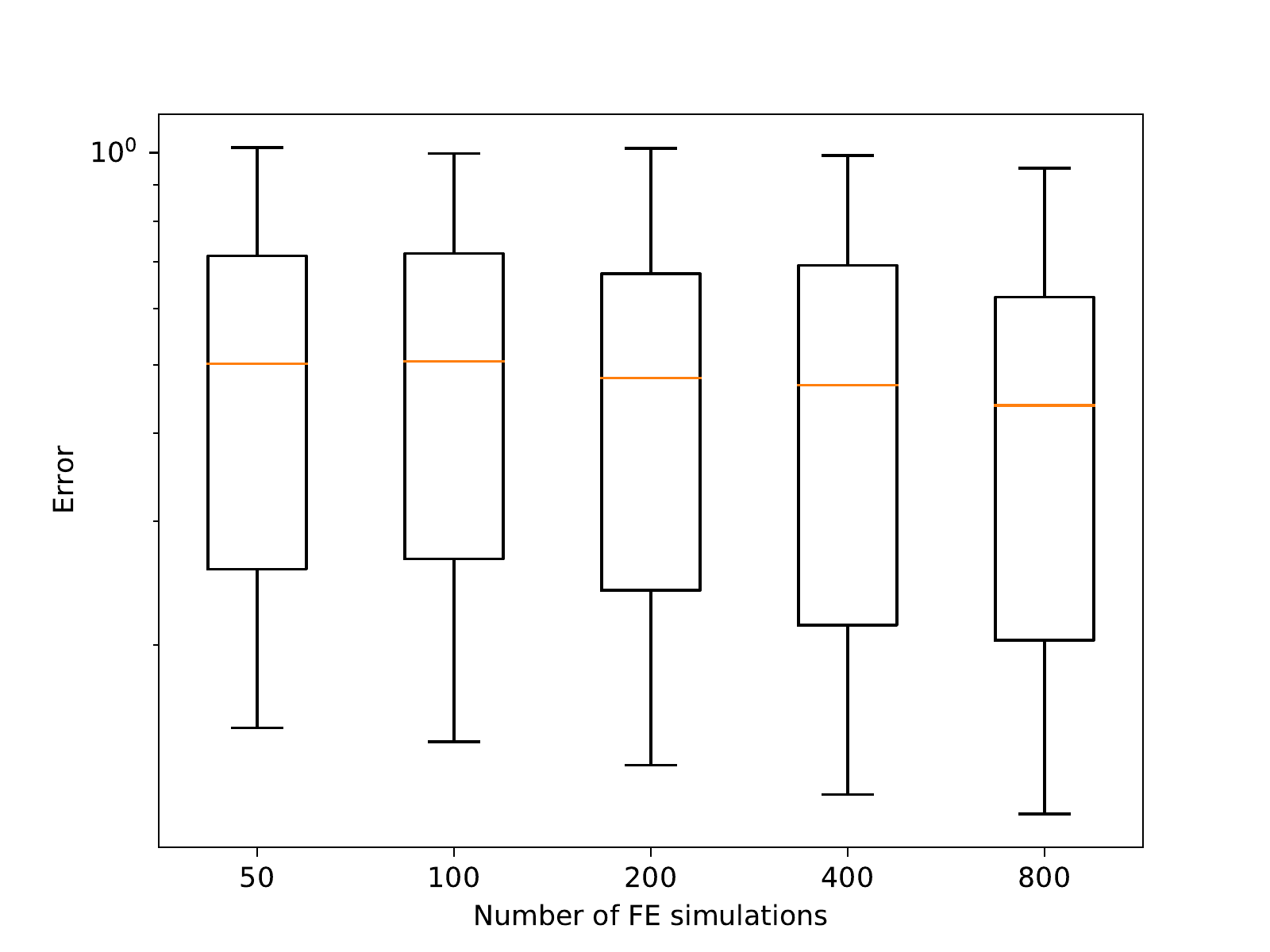}
      \caption{Box plot without outliers}
      \label{fig:SingleCylinder_relative_error_boxplots_triaxiality:b}
    \end{subfigure}
    \caption{In the diagram on the left (a) we see 5 box plots for the error between the triaxiality stress in the ROI of the patches as calculated by FEA and as predicted from the GNN, $error_\text{triaxiality}$. Each box plot is defined by 5 numbers. The 2 whiskers at the top and bottom correspond to the 95th and 5th percentiles respectively. The top and bottom limits of the box correspond to the 75th and 25th percentiles respectively. Finally, the yellow line corresponds to the mean of the data. Each box plot corresponds to a GNN trained with different number of data. In the diagram on the right (b) we can see the same box plots without the outliers so that the trend is clearly visible. Each box plot corresponds to 1,000 points.}
    \label{fig:SingleCylinder_relative_error_boxplots_triaxiality}
\end{figure}

We conduct a similar study as the one conducted in [section 5.3.3] of the main work (but using the triaxiality instead of the Von Mises stress) and we have reached the same conclusions. We report the error in the triaxiality stress in the ROI of the patches between the FEA calculations and the GNN predictions. Specifically, we are interested in the maximum absolute error of the log triaxiality stress, defined for every ROI as
% $\text{max}(|\text{log10(}|triaxiality_\text{FEA}|) - \text{log10(}|triaxiality_\text{NN}|)|)$.
$error_\text{triaxiality} = \text{max}(|log\_triaxiality_\text{FEA} - log\_triaxiality_\text{NN}|)$, where $log\_triaxiality_\text{FEA} = \text{log10}(|triaxiality_\text{FEA}|)$ and $log\_triaxiality_\text{NN} = \text{log10}(|triaxiality_\text{NN}|)$.
From [Fig \ref{fig:SingleCylinder_convergennce_triaxiality}] we can see that the mean error decreases as more data are added in the training set. 
We also give some insights for the distribution of the error across the patches instead of just showing the mean value. To this end we provide box plots for the relative error [Fig \ref{fig:SingleCylinder_relative_error_boxplots_triaxiality}]. We can see that as more data are added, not only the mean error but also the error values that correspond to the high error points decrease. 

%-------------------------------------------------------------------------------------

\end{appendices}

\end{document}